\documentclass[11pt, a4paper]{article}
\usepackage[T1]{fontenc}
\usepackage{a4wide, url}
\usepackage[english]{babel}
\usepackage{graphicx}
\usepackage{array}
\usepackage[longnamesfirst]{natbib}
\usepackage[textfont=sc,labelfont=bf]{caption}
\usepackage{setspace}
\usepackage{multirow}
\usepackage{amsthm}
\usepackage{amsmath}
\usepackage{amsfonts}
\usepackage{amssymb}
\usepackage{pgf} 
\usepackage{bm}
\usepackage{paralist}
\usepackage{pdflscape}
\usepackage{xfrac}
\newcolumntype{P}[1]{>{\centering\arraybackslash}p{#1}}

\DeclareMathAlphabet\mathbfcal{OMS}{cmsy}{b}{n}

\newcommand{\mbf}{\mathbf}
\newcommand{\beq}{\begin{equation}}
\newcommand{\eeq}{\end{equation}}
\newcommand{\bea}{\begin{eqnarray}}
\newcommand{\eea}{\end{eqnarray}}
\newcommand{\ba}{\begin{array}}
\newcommand{\ea}{\end{array}}
\newcommand{\bit}{\begin{itemize}}
\newcommand{\eit}{\end{itemize}}
\newcommand{\ben}{\begin{enumerate}} 
\newcommand{\een}{\end{enumerate}}
\newcommand{\bpm}{\begin{pmatrix}}
\newcommand{\epm}{\end{pmatrix}}
\newcommand{\bbm}{\begin{bmatrix}}
\newcommand{\ebm}{\end{bmatrix}}

\renewcommand{\l}{\left}
\renewcommand{\r}{\right}

\newcommand{\E}[0]{\mathsf{E}}
\newcommand{\Var}[0]{\mathsf{Var}}

\newcommand{\nn}{\nonumber}
\newcommand{\wh}{\widehat}
\newcommand{\wt}{\widetilde}

\newtheorem{prop}{Proposition}
\newtheorem{rem}{Remark}
\newtheorem{lem}{Lemma}

\title{\textsc{\LARGE Common factors, trends, and cycles\\ in large datasets }}
\date{ }

\defcitealias{Claudia1}{Gilbert et al., 2015}
\defcitealias{Claudia2}{Lengermann et al., 2017}

\begin{document}
\maketitle

\begin{center}\vspace{-1.5cm}
\begin{tabular}{cp{1cm}c}
\large Matteo Barigozzi && \large Matteo Luciani\\[.1cm] 
\small London School of Economics &&\small Federal Reserve Board\\[.1cm] 
\footnotesize  m.barigozzi@lse.ac.uk &&\footnotesize matteo.luciani@frb.gov\\[1cm] 
\end{tabular}

\small \today\\[1.5cm]
\end{center}

\begin{abstract}
This paper considers a non-stationary dynamic factor model for large datasets to disentangle long-run from short-run co-movements. We first propose a new Quasi Maximum Likelihood estimator of the model based on the Kalman Smoother and the Expectation Maximisation algorithm. The asymptotic properties of the estimator are discussed. Then, we show how to separate trends and cycles in the factors by mean of eigenanalysis of the estimated non-stationary factors. Finally, we employ our methodology on a panel of US quarterly macroeconomic indicators to estimate aggregate real output, or Gross Domestic Output, and the output gap.\\

\vspace{0.5cm}

\end{abstract}

\renewcommand{\thefootnote}{$\ast$} 
\thispagestyle{empty}

\footnotetext{We thank for helpful comment the participants to the conferences:  ``Inference in Large Econometric Models'', CIREQ, Montr\' eal, May 2017; ``Big Data in Dynamic Predictive Econometric Modelling'', University of Pennsylvania, Philadelphia, May 2017; ``Computing in Economics and Finance'', Fordham University, 
New York City, June 2017; and to the seminars at: Federal Reserve Board, September 2016; Warwick Business School, February 2017; Department of Statistics, Universidad Carlos III, Madrid, May 2017. We would like also to thank: Stephanie Aaronson, Gianni Amisano, Massimo Franchi, Marco Lippi, Filippo Pellegrino, Ivan Petrella, Lucrezia Reichlin, John Roberts, and Esther Ruiz.\\

Disclaimer: the views expressed in this paper are those of the authors and do not necessarily reflect the views and policies of the Board of Governors or the Federal Reserve System.  } 

\renewcommand{\thefootnote}{\arabic{footnote}}

\newpage

\section{Introduction}
This paper is about two stylized facts of macroeconomic time series: co-movements and non-stationarity \citep{lippireichlin94}. More precisely, this paper is about disentangling long-run co-movements (common trends) from short-run co-movements (common cycles) in a large dataset of non-stationary US macroeconomic indicators.

Since the seminal work of \citet{beveridgenelson1981}, the issue of decomposing GDP into a trend and a cycle has been a central question in both time series econometrics and policy analysis. This is not surprising, as long-run trends are mainly influenced by supply-side factors, while short-run cycles are mainly associated with demand-side factors, and therefore different estimates of the trend and of the cycle can lead to different policy recommendations. Given the relevance of the issue, in the last 30 years, many papers have suggested different ways to obtain a Trend-Cycle (TC) decomposition of GDP. Roughly speaking, those works can be grouped under two main approaches: one based on univariate methods \citep[e.g.][]{watson86,lippireichlin94b,mnz03,dungeyetal15}, and another using multivariate, but low-dimensional, time series techniques \citep[e.g.][]{stockwatson88JASA,lippireichlin94,gonzalogranger,garratetal06,crealetal10}. 

In this paper we use a novel approach to decompose GDP into a trend and cycle based on large datasets. We first disentangle common and idiosyncratic dynamics by using a Non-Stationary Approximate Dynamic Factor Model (DFM), and then we disentangle common trends from common cycles by applying a non-parametric TC decomposition to the latent common factors. Our methodology builds on four points: first, focusing on a high-dimensional setting is crucial, as only in a high-dimensional setting it is possible to disentangle common from idiosyncratic dynamics in a consistent way \citep{FHLR00,baing02, stockwatson02JASA} --- i.e., we can separate macroeconomic fluctuations from sectoral dynamics and measurement error only in a high-dimensional setting. Second, assuming the existence of a factor structure is a realistic and convenient way to represent co-movements in large macroeconomic datasets. Third, considering non-stationary data is necessary to account for the presence of common trends or, equivalently, cointegration \citep{bai04,baing04,BLL1,BLL2}. And fourth, by using a non-parametric TC decomposition we do not have to make assumptions on the law of motion of either the trend, or the cycle. Our approach is deliberately reduced form, and therefore our empirical analysis is conducted ``without pretending to have too much a priori economic theory'' \citep{SS77}, thus letting the data speak as freely as possible.
 
The first contribution of this paper is methodological. Namely, we propose a Quasi Maximum Likelihood estimator of the non-stationary DFM based on the Expectation Maximisation (EM) algorithm combined with the Kalman Filter and the Kalman Smoother estimators of the factors. The theoretical properties of this approach in the large stationary DFM case have been studied in \citet{DGRfilter,DGRqml}, and here we extend their results to the non-stationary case by proving consistency and by providing rates of convergence for the factors and the parameters of the model. Compared to the non-stationary principal component estimator \citep{baing04}, the estimator proposed in this paper is more efficient, and it is more flexible in that, thanks to the use of the Kalman Filter, it allows us to explicitly model the idiosyncratic dynamics, and to impose economically meaningful restrictions.

The second contribution of this paper is to show how to isolate common trends and common cycles in large macroeconomic datasets. In detail, we use a non-parametric approach that identifies the common trends as those linear combinations of the factors obtained by the leading eigenvectors of the long-run covariance matrix (\citealp{bai04,penaponcela04}), and the common cycles as deviations from the long-run equilibria, which coincide with the space orthogonal to that of the common trends --- i.e., the cointegration space \citep{ZRY}. Because our approach is non-parametric, we are not imposing any particular form to the trend, which is not constrained to be a random walk, or to the cycle. This is what differentiates our approach from the standard state-space, which normally is applied on a handful of variables and where the trend and the cycle dynamics are explicitly specified and jointly estimated with the parameters of the model (\citealp{harvey90}). 

Our final contributions are empirical. Specifically, we employ our methodology to analyse a large panel of US quarterly macroeconomic time series with the goal of estimating the cyclical position of the economy and the observation error. With the expression ``estimating the observation error,'' we mean estimating aggregate real output. With the expression ``estimating the cyclical position of the economy,'' we mean decomposing aggregate real output into potential output and output gap. To the best of our knowledge, \citet{charlesjohn} and \citet{ADNSS} are the only works that, so far, have used (small) factor models to estimate aggregate real output. On the other hand, a few papers have used low-dimensional factor models to estimate the cyclical position of the economy \citep[e.g.][]{charlesjohn,jarocinskilenza}, and a few more to estimate long-run trends \citep[e.g.][]{ADP16}. Finally, \citet{aastveittrovik14} and \citet{morleywong2017} have used a high-dimensional setting for estimating the output gap by means of a factor model and a large Bayesian VAR, respectively. However, in both works the variables are transformed to stationarity prior to model estimation.

The first part of our empirical analysis is about estimating aggregate real output, to which we refer as Gross Domestic Output (GDO). We first show that our model naturally produces an estimate of GDO as that part of GDP/GDI that is driven by the macroeconomic (common) shocks. We then compare our estimate of GDO with ``the average of GDP and GDI'' released by the Bureau of Economic Analysis, and with  ``GDPplus''  proposed by \citet{ADNSS} and released by the Philadelphia Fed. Our results show that these three measures are very similar, which is not surprising, as they are attempting to estimate the same thing. However, we estimate that since 2010 quarterly annualized GDO growth was on average \sfrac{1}{2} of a percentage point higher than estimated by the BEA or the Philadelphia Fed, thus pointing out that --- based on the commonality in the data --- the US economy grew at a faster pace than measured by national account statistics.

The second part of our empirical analysis is about estimating the output gap. To this end, we use the above-mentioned TC decomposition in order to separate long-run from short-run co-movements, and in particular we focus on the decomposition derived for GDO. We compare our estimate with the one produced by the Congressional Budget Office (CBO), which estimates potential output as that level of output consistent with current technologies and normal utilisation of capital and labour, and the output gap as the residual part of output. Although these two estimates are obtained in completely different ways, in practice they look very similar. The two estimates are comparable for most of the sample considered, but from the late nineties to the financial crisis, when our measure suggests that a greater part of the produced output was driven by transitory factors. In particular, according to our estimate between 2001:Q1 and 2005:Q4 the output gap was on average 2\sfrac{1}{2} percentage points higher than estimated by the CBO.

The rest of this paper is structured as follows.~In Section \ref{sec:Representation} we discuss representation of large non-stationary panels of time series. In this section we first present the non-stationary dynamic factor model, and we define the concepts of commonality --- i.e., the common factors. Then we discuss how to disentangle long-run co-movements from short-run co-movements --- i.e., we define what common trends and common cycles are. In Section \ref{sec:Estimation} we discuss estimation. We first introduce in Section \ref{sec:StaticFM} the {\it static} representation of the DFM, which is just a convenient way to approach estimation of the dynamic model presented in Section \ref{sec:Representation}. We then present in Section \ref{sec:ML} our  estimator, we discuss its properties, and we compare it with existing methods. Finally, in Section \ref{sec:TC2} we present the non-parametric TC decomposition that we use in the empirical section. Then, Section \ref{sec:emp} presents the empirical analysis. This section is split in two, with the first part presenting our estimate of GDO (Section \ref{sec_gdo}), and the second part presenting our estimate of the output gap (Section \ref{sec_ogap}). To conclude, in Section \ref{sec:Conclusions} we discuss our findings and the advantages and limitations of our methodology, and we propose directions for further research. In the Appendix we report all technical proofs and the description of the data used and their transformation.

\subsection*{Notation}
A vector $\mbf z_t$ is $I(1)$ if the higher-order of integration among all its components is 1, thus under this definition some components of $\mbf z_t$ can be stationary. 
Eigenvalues are always considered as ordered from the largest to the smallest, so for a given set of eigenvalues $\{\mu_j\}_{j=1}^m$, we have $\mu_1\ge\mu_2\ge\ldots\ge\mu_{m-1}\ge \mu_m$. Therefore, the spectral norm of $\bm A$ is defined as $\Vert \bm A\Vert^2=\mu_1^{A'A}$. The $j$-th largest eigenvalue of a spectral density matrix at frequency $\omega$ is denoted as $\mu_j(\omega)$. The generic $(i,j)$-th entry of a matrix $\bm A$ is denoted as $[\bm A]_{ij}$. We denote by $L$ the lag operator, such that $L^k y_t =y_{t-k}$, for any $k\in\mathbb Z$ and we use the notation $\Delta y_t:=(1-L)y_t$. Finally, we let $M,M_0,M_1\ldots$ denote generic positive and finite constants that do not depend on the panel dimensions $n$ or $T$, and whose value may change from line to line. 

\section{Representation of non-stationary panels of time series}\label{sec:Representation}

Let us assume to observe a vector of $n$ time series $\{\mbf y_t =(y_{1t}\cdots y_{nt})':t=1,\ldots ,T\}$ such that
\beq\label{eq:trend}
y_{it} = \mathcal D_{it} + x_{it},
\eeq
where $\mathcal D_{it}$ is a deterministic component --- e.g., a linear trend --- and $\mbf x_t=(x_{1t}\cdots x_{nt})'$ is such that $\mbf x_t\sim I(1)$. We also assume that $\E[x_{it}]= 0$, for any $i$ and $t$, therefore, $\mbf x_{t}$ contains all the stochastic trends but no deterministic component. Throughout, the spectral density matrix of $\Delta \mbf x_t$ is assumed to exist. 

In a high-dimensional setting, it is reasonable to assume that there are common trends and common cycles, but also idiosyncratic terms. Thus, for each variable $x_{it}$ we write
\beq\label{eq:TC}
x_{it} =  {\mathcal T}_{it} + {\mathcal C}_{it} + \xi_{it},
\eeq
where $\mathcal T_{it}\sim I(1)$ is the trend component, $\mathcal C_{it}\sim I(0)$ is the cycle component, and $\xi_{it}$ is the idiosyncratic component, which is allowed to be either $I(1)$ (in presence of idiosyncratic trends) or $I(0)$ (e.g. measurement errors). The trend and the cycle  are capturing the common dynamics across series, and thus constitute the common component defined as $\chi_{it}={\mathcal T}_{it} + {\mathcal C}_{it}$. Hence, \eqref{eq:TC} is also written as
\beq
x_{it} =  \chi_{it} + \xi_{it}.\label{eq:commonidio}
\eeq
We define the vectors of common and idiosyncratic components as $\bm\chi_t=(\chi_{1t}\cdots \chi_{nt})'$ and $\bm\xi_t=(\xi_{1t}\cdots \xi_{nt})'$, respectively. Finally, notice that consistently with the data considered in this paper: (i) some (but not all) components of $\mbf x_t$ are allowed to be stationary, and (ii)  the deterministic components $\mathcal D_{it}$ are not common to all series --- i.e., there are no common deterministic trends.  

We assume that the co-movements in $\bm \chi_t$ are driven by  $q$ ``structural'' shocks, with  $q\ll n$, which are collected in a  weak white noise vector process $\mbf u_t=(u_{1t}\cdots u_{qt})'$. Then, for a given $q$, we decompose each element of $\mbf x_t$ as
\begin{align}
 x_{it}&=\bm b_i'(L) \bm f_t+\xi_{it},\label{eq:gdfm0}\\
\Delta\bm f_t &= \mbf C(L)\mbf u_t, \label{eq:dynf}
\end{align}
where from \eqref{eq:commonidio} the common component is given by $\chi_{it}=\bm b_i'(L) \bm f_t$ and the following properties hold:
\ben
\item [A1.] $\mbf u_t\stackrel{w.n.}{\sim} (\mbf 0_q,\mbf I_q)$, with $q$ is independent of $n$;
\item [A2.] $\E[u_{jt} \xi_{is}]=0$, for any $j=1,\ldots q$, $i=1,\ldots, n$, and $s,t=1,\ldots,T$;
\item [A3.] $\mbf B(L)=(\bm b_1'(L)\cdots\bm b_n'(L))'$ is an $n\times q$ one-sided, matrix polynomial matrix of finite order $s$, $\bm f_t\sim I(1)$ of dimension $q$;
\item [A4.] $\mbf C(L)=(\bm c_1'(L)\cdots\bm c_q'(L))'$ is a $q\times q$ one-sided, infinite matrix polynomial with square-summable coefficients and such that $\mathsf {rk}(\mbf C(1))=(q-d)$ with $0<d< q$;
\item [A5.] 
the $q$-th largest eigenvalue $\mu_q^{\Delta \chi}(\omega)$ of the spectral density matrix of $\Delta \bm\chi_t$ is such that
$$
M_1\le\lim\!\!\!\inf_{n\to\infty} n^{-1}{\mu_q^{\Delta \chi}(\omega)}\le \lim\!\!\sup_{n\to\infty} n^{-1}{\mu_q^{\Delta \chi}(\omega)}\le M_2,\qquad \omega\mbox{-a.e.}\in [-\pi,\pi],
$$  
while the largest eigenvalue $\mu_{1}^{\Delta \xi}(\omega)$ of the spectral density matrix of $\Delta \bm\xi_t$ is such that
$$
M_3\le \lim\!\!\!\inf_{n\to\infty}\mu_{1}^{\Delta \xi}(\omega)\le \lim\!\!\sup_{n\to\infty}\mu_{1}^{\Delta \xi}(\omega)\le M_4, \qquad \omega\mbox{-a.e.}\in [-\pi,\pi].
$$ 
\een
Equations \eqref{eq:gdfm0} and \eqref{eq:dynf} together with properties A1-A5 define a  Non-Stationary Approximate Dynamic Factor Model (DFM).  In the case of stationary time series our model is a special case of the Generalised Dynamic Factor Model originally proposed by \citet{FHLR00}.

Condition A5 is crucial and it allows for identification of the common component by defining it according to its spectral properties. An explanation for A5 in the time domain is provided by \citet{hallinlippi13} who show that this condition is equivalent to defining the common and idiosyncratic component by asking that for any dynamic aggregation scheme given by an $n$-dimensional vector of weights $\mbf a_k$ such that $\sum_{k\in\mathbb Z}\mbf a_k'\mbf a_k=1$, the following holds 
\begin{align}
0< \lim_{n\to\infty}\Var\l(\frac 1n\sum_{k=-\infty}^\infty \mbf a_k'\Delta \bm\chi_{t-k}\r)\le M\;\mbox{ and }\;\lim_{n\to\infty} \Var\l(\frac 1n\sum_{k=-\infty}^{\infty} \mbf a_k'\Delta \bm\xi_{t-k}\r)= 0.\label{example}
\end{align}

The following asymptotic conditions for the eigenvalues $\mu_{i}(\omega)$ of the spectral density of $\Delta \mbf x_t$ are a direct consequence of A4, A5, and Weyl's inequality:
\ben
\item [B1.] for $\omega\mbox{-a.e.}\in [-\pi,\pi]$ the following holds:\\
$M_1\le \lim\inf_{n\to\infty}n^{-1}{\mu_q(\omega)}\le\lim\sup_{n\to\infty}n^{-1}{\mu_q(\omega)} \le M_2$,\\ $M_3\le \lim\inf_{n\to\infty}\mu_{q+1}(\omega)\le \lim\sup_{n\to\infty}\mu_{q+1}(\omega)\le M_4$;
\item [B2.] for $\omega = 0$ the following holds:\\
$M_1\le \lim\inf_{n\to\infty}n^{-1}{\mu_{q-d}(0)}\le\lim\sup_{n\to\infty}n^{-1}{\mu_{q-d}(0)} \le M_2$, \\ $M_3\le \lim\inf_{n\to\infty}\mu_{q-d+1}(0)\le \lim\sup_{n\to\infty}\mu_{q-d+1}(0)\le M_4$.
\een 
By means of B1 the number of shocks $q$ can then be identified (\citealp{hallinliska07}, \citealp{onatski09}).
Similarly, by means of B2 the number of common trends, $(q-d)$, can be identified (\citealp{BLL2}). In particular, from the intuition given in \eqref{example} and because of B1 and B2, it is clear that the DFM is identifiable only in the limit $n\to\infty$.

Condition A4 allows for the presence of $(q-d)$ common trends in the factors $\bm f_t$. In line with our empirical results in Section \ref{sec:emp} we rule out the degenerate cases $d=0$ or $d=q$. This implies that the vector $\bm f_t$ admits a VECM representation with $d$ cointegration relations (\citealp{englegranger1987}), as well as the factor representation (\citealp{EP1994}):
\beq\label{eq:EP}
\bm f_t = \bm\Psi \bm\tau_t + \bm\gamma_t,
\eeq
where $\bm\Psi$ is $q\times (q-d)$ and $\bm \tau_t$ is the vector of $(q-d)$ common trends with components $\tau_{jt}\sim I(1)$ for $j=1,\ldots, (q-d)$, while $\bm\gamma_t$ is a $q$-dimensional stationary vector.\footnote{Notice that in general all factors are non-stationary, unless some \textit{ad hoc} zero-constraint is imposed on the elements of $\mbf C(1)$. On the other hand if we were to ask for one of the factors to be stationary then the corresponding row of $\bm\Psi$ must be set to zero. However, we do not consider this case further since it could easily be included in our framework by imposing the appropriate identifying assumptions.} Notice that \eqref{eq:EP} is different from the common trends representation (or multivariate Beveridge-Nelson decomposition) of \citet{stockwatson88JASA} in that the trend $\bm\tau_t$ is not constrained to be a  vector random walk, a property advocated for by many authors (e.g. \citealp{lippireichlin94}).

For a given choice of $\bm\Psi$, the $(q-d)$ common trends can then be obtained by linear projection onto the space spanned by the columns of $\bm \Psi$: 
\beq\nn
\bm \tau_t = (\bm\Psi '\bm\Psi )^{-1}\bm\Psi '\bm f_t=\bm\Psi '\bm f_t.
\eeq
where the second equality holds because, without loss of generality, we can always assume the identifying constraint $\bm\Psi'\bm\Psi=\mbf I_{(q-d)}$.

Different choices of $\bm\Psi$ lead to different definitions of common trends. Here we opt for a non-parametric approach and we identify the elements of $\bm\tau_t$ as the first $(q-d)$ principal components of $\bm f_t$, as proposed by \citet{bai04} and \citet{penaponcela04} (see Section \ref{sec:TC2} for details on estimation). Given this definition, the columns of  $\bm\Psi$ are orthonormal and therefore there exists a $q\times d$ matrix $\bm\Psi_{\perp}$ such that $\bm\Psi_{\perp}'\bm\Psi_{\perp}=\mbf I_d$ and $\bm\Psi_{\perp}'\bm\Psi=\mbf 0_{d\times (q-d)}$. Now, consider the $d$-dimensional process obtained by projecting $\bm f_t$ onto the space orthogonal to the common trends
\beq\nn
\bm c_t = (\bm\Psi_{\perp}'\bm\Psi_{\perp})^{-1}\bm\Psi_\perp'\bm f_t=\bm\Psi_\perp'\bm f_t=\bm\Psi_\perp'\bm \gamma_t.
\eeq 
It is straightforward to see that ${\bm c}_t\sim I(0)$, that its components are $d$ common cycles in the sense of \citet{vahidengle93}, and that the columns of $\bm\Psi_{\perp}$ are a basis of the cointegration space of $\bm f_t$, thus these common cycles represent deviations from long-run equilibria --- see also e.g. \citet{johansen91} and \citet{kasa1992} for similar definitions.\footnote{Other TC decompositions based on a different definitions of cycles than the one used here are in \citet{gonzalogranger} and \citet{GN01}.} 

According to our definition, common trends and common cycles are orthogonal by construction, and we have the TC decomposition of the factors:
\begin{align}\label{eq:tc1}
\bm f_t =  \bm\Psi \bm\Psi '\bm f_t + \bm\Psi_\perp \bm\Psi_\perp '\bm f_t= \bm\Psi \bm\tau_t + \bm\Psi_\perp \bm c_t,
\end{align}
and therefore, by combining \eqref{eq:trend}, \eqref{eq:gdfm0} and \eqref{eq:tc1}, we have the TC decomposition of the data:
\beq\label{eq:tc2}
y_{it} = \mathcal D_{it}+ \bm b_i'(L) \bm\Psi\bm\tau_t + \bm b_i'(L) \bm\Psi_\perp \bm c_t + \xi_{it}= \mathcal D_{it}+\mathcal T_{it} +\mathcal C_{it}+\xi_{it}.
\eeq

\section{Estimation} \label{sec:Estimation}
In order to estimate \eqref{eq:tc2}, we need to estimate the factors, $\bm f_t$ and their TC decomposition. We opt for a two-step approach, where we first extract the common factors and then we estimate their TC decomposition. In particular, we first introduce a convenient re-parametrization of the DFM based on its \textit{static} state-space representation (Section \ref{sec:StaticFM}), which is then used for retrieving the factors space by means of the EM algorithm (Section \ref{sec:ML}). Then, in a second step we use principal component analysis for extracting common trends and cycles (Section \ref{sec:TC2}). Notice that compared to the classical state-space approach (e.g. \citealp{charlesjohn}) or from the Bayesian approach (e.g. \citealp{jarocinskilenza}) in which the trend and the cycle are estimated in one-step together with the parameters of the models, our approach has the advantage that it does not require us to specify a law of motion for the trend and the cycles. 

For simplicity of exposition we assume in this section that there is no deterministic component and we refer to Section \ref{sec:emp} and to \ref{sec:data} for the treatment of these terms in practice.

\subsection{The \textit{static} representation of dynamic factor models}\label{sec:StaticFM}
Consider the state-space form of the DFM in \eqref{eq:gdfm0}-\eqref{eq:dynf} (\citealp{stockwatson05}; \citealp{FGLR09}):
\begin{align}
x_{it} &= \bm\lambda_i'\mbf F_t + \xi_{it},\label{eq:static}\\
\Delta\mbf F_t &= \mbf D(L)\mbf u_t\label{eq:static2},
\end{align}
where from \eqref{eq:commonidio} the common component is now given by $\chi_{it}=\bm\lambda_i'\mbf F_t$ and $\mbf u_t$ is the same as in \eqref{eq:dynf}. We assume that A1, A2 and A5 still hold and in addition we require:
\ben
\item [C1.] $\mbf D(L)=(\bm d_1'(L)\cdots\bm d_r'(L))'$ is an $r\times q$ one-sided, infinite matrix polynomial with square-summable coefficients  and such that $\mathsf {rk}(\mbf D(1))=(q-d)$ with $0<d< q$;
\item [C2.] $\bm\Lambda=(\bm\lambda_1 \cdots\bm\lambda_n)'$ is an $n\times r$ loadings matrix such that $\lim_{n\to\infty}\Vert n^{-1}\bm\Lambda'\bm\Lambda-\mbf I_r\Vert=0$ and $\vert[\bm\Lambda]_{ij}\vert<M$, for any $i=1,\ldots, n$ and $j=1,\ldots, r$;
\item [C3.] $\mbf F_t\sim I(1)$ of dimension $r$, with $\E[\Delta\mbf F_t\Delta\mbf F_t']$ positive definite.
\een
Condition C1 is equivalent to A4 in that it requires the existence of $(q-d)$ common trends driving the common component. Conditions C2 and C3 are standard in the literature and imply that the eigenvalues of the covariance of  $\Delta\bm\chi_t$ diverge as $n\to\infty$ at a rate $n$ (\citealp{stockwatson02JASA,baing02,FLM13}). Finally, from A5 we immediately have that the largest eigenvalue of the covariance of $\Delta\bm\xi_t$ is finite for any $n$. Given the way $\mbf F_t$ and $\bm f_t$ are loaded by the data, hereafter we call $\mbf F_t$ \textit{static} factors and $\bm f_t$ \textit{dynamic} factors. 

Let us stress once more the fact that here the DFM and the related TC decomposition are our focus, while the \textit{static} representation is just a convenient way to approach estimation of the dynamic model. In particular, for \eqref{eq:static}-\eqref{eq:static2} to be equivalent to \eqref{eq:gdfm0}-\eqref{eq:dynf} we need the following restrictions to hold:
\ben
\item [R1.] there exists  an invertible $r\times r$ matrix $\mbf K$ such that $\mbf F_t=\mbf K (\bm f_t'\cdots \bm f_{t-s}')'$ and $\bm\lambda_i' = (\bm b_{i0}'\cdots \bm b_{is}')\mbf K^{-1}$, for any $i=1,\ldots, n$, where $\bm b_{ik}$, for $k=0,\ldots, s$, are the coefficients of $\bm b_i(L)$ defined in A3;
\item [R2.] the dimension of $\mbf F_t$ is $r=q(s+1)$;
\item [R3.] the cointegration rank of $\mbf F_t$ is $d$.
\een
Let us consider each restriction in detail. Restriction R1 implies that the spectral density of $\Delta\mbf F_t$ has reduced rank $q$. In the following, we impose this restriction when estimating the model but we do not attempt to identify $\mbf K$. 

Restriction R2 offers an alternative way to determine $r$ with respect to the typical methods available in the literature based  on the behavior of the eigenvalues of the covariance matrix of $\Delta\mbf x_t$ and therefore on C2, C3, and A5 (e.g. \citealp{baing02}). Specifically, by virtue of restriction R2, once we set $q$ using B1, we can choose $r$ such that the share of variance explained by the \textit{static} factors $\mbf F_t$ coincides with the share of variance explained by the $q$ \textit{dynamic} factors $\bm f_t$ --- see also \citet{dagostinogiannone12}.

Finally, restriction R3 tells us that the autoregressive representation for \eqref{eq:static2} is a VECM with $d$ cointegration relations (a proof is in \ref{sec:proofsRepresentation}). Moreover, since the vector $\mbf F_t$ is singular, the autoregressive representation has a finite order (\citealp{BLL1}). 
However, in the next section we do not estimate a VECM, rather we estimate an unrestricted VAR in the levels (\citealp{simsstockwatson}). We use the knowledge of the cointegration rank to determine the dimension of the common cycles space (see Section \ref{sec:TC2}). 

Summing up, by not fully imposing R1 and  R3 when estimating the factors, we opt for simplicity of estimation versus complexity of a more realistic representation, which implies that the model considered is deliberately mis-specified.~The effects of such mis-specification will appear clear in Section \ref{sec:TC2}, when we consider TC decompositions of $\mbf F_t$ as opposed to those of $\bm f_t$.

\subsection{Estimating the space of factors and loadings} \label{sec:ML}
We consider the following state-space form of \eqref{eq:static}-\eqref{eq:static2} in which we assume a VAR(2) for the \textit{static} factors as in the empirical analysis of Section \ref{sec:emp}:
\begin{align}
x_{it} &= \bm\lambda_i'\mbf F_t + \xi_{it},\label{eq:SS1}\\
\mbf F_t&= \mbf A_1\mbf F_{t-1}+\mbf A_2\mbf F_{t-2}+\mbf H\mbf u_t,\label{eq:SS2}\\
\xi_{it}&=\rho_i\xi_{it-1}+e_{it}.\label{eq:SS3}
\end{align}
We estimate \eqref{eq:SS1}-\eqref{eq:SS3} via the EM algorithm (\citealp{DLR77}), combined with the Kalman Filter (KF) and the Kalman Smoother (KS) estimators of the factors (\citealp{AM79,harvey90}). 
In the stationary, low-dimensional --- i.e., finite $n$ --- setting, estimation of a factor model by means of the EM algorithm can be found in \citet{shumwaystoffer82} and \citet{watsonengle83}, while the asymptotic properties of this factors' estimator are studied by \citet{DGRfilter,DGRqml} under the joint limit $n,T\to\infty$.\footnote{For recent applications of this approach see e.g. \citet{reiswatson10,banburamodugno14,juvenalpetrella2015,smokinggun,CGM16}.} In the non-stationary  case, applications of the EM algorithm can be found in \citet{quahsargent93} and \citet{SAZ13} in a low-dimensional setting. Here, we study the theoretical properties in the non-stationary case when $n,T\to\infty$.
 
In order to run the KF-KS it is necessary to make some additional assumptions on the idiosyncratic component. Let $\mbf R$ be the covariance matrix of the vector $\bm e_t=(e_{1t}\cdots e_{nt})'$ of the idiosyncratic innovations in \eqref{eq:SS3}, then we assume:
\ben
\item [D1.] $\rho_i=1$ if  $\xi_{it}\sim I(1)$ or $\rho_i=0$ if  $\xi_{it}\sim I(0)$;
\item [D2.] $\bm e_{t}\stackrel{w.n.}{\sim}  \mathcal N(\mbf 0_n, \mbf R)$, with $[\mbf R]_{ii}>0$ and $[\mbf R]_{ij}=0$ for any $i\neq j$ and $i,j=1,\ldots, n$;
\item [D3.] $\mbf u_t\stackrel{w.n.}{\sim} \mathcal N(\mbf 0_q, \mbf I_q)$.
\een
It is clear from D1, D2 and \eqref{eq:SS3} that if some idiosyncratic components are $I(1)$, we can still consider a factor model for $\mbf x_t$ with stationary errors in \eqref{eq:SS1} by adding additional latent states with unit loadings and evolving as random walks. Notice that the dimension of the parameter space does not increase by increasing the number of $I(1)$ idiosyncratic components. On the other hand modeling the dynamics of $I(0)$ idiosyncratic components would increase the complexity of the estimation problem. For this reason, in D1 we choose to leave the dynamics of the stationary idiosyncratic components unspecified --- see Section \ref{sec:emp} for practical implementation of this assumption. Assumptions D1-D3 define a mis-specified approximating model of the true DFM and in this sense our EM approach delivers  Quasi Maximum Likelihood (QML) estimators. The effect of these mis-specifications are discussed at the end of this section, but before discussing them we present the asymptotic properties of the estimated factors and loadings. 

We collect all unknown parameters of the model into the vector $$\bm\Theta:=(\mbox{vec}(\bm\Lambda)' \;\mbox{vec}(\mbf A_1)'\;\mbox{vec}(\mbf A_2)'\; \mbox{vec}(\mbf H)'\;\mbox{diag}(\mbf R)')'.$$ We denote by $Q$ the dimension of ${\bm\Theta}$, then we assume that the true values of the parameters satisfy:
\ben
\item [D4.] $\bm\Theta \in \mbox{int} (\Omega)$, with $\Omega\subseteq \mathbb R^Q$ and compact.
\een
This condition is standard in QML theory and ensures existence of the true values of the parameters.

The EM algorithm is based on the iteration of two steps. In the E-step, for a given estimator of the parameters $\wh{\bm\Theta}_k$, we compute the expected likelihood conditional on all observed data $\{\mbf x_1,\ldots,\mbf x_T\}$. This is in turn a function of the first and second conditional moments of the \textit{static} factors, which are computed by means of the KS when using $\wh{\bm\Theta}_k$. 

Note that, under the assumption of normality, as in D2 and D3, and for a given value of the parameters $\bm\Theta$, the KF-KS give the conditional expectations: 
\begin{align}\nn
&{\mbf F}_{t|t-1} := \E_{{\bm\Theta}}[\mbf F_t |\mbf x_1,\ldots,\mbf x_{t-1}],\qquad
{\mbf F}_{t|t} := \E_{{\bm\Theta}}[\mbf F_t |\mbf x_1,\ldots,\mbf x_t],\qquad
&{\mbf F}_{t|T} := \E_{{\bm\Theta}}[\mbf F_t |\mbf x_1,\ldots,\mbf x_T],
\end{align}
with the associated covariance matrices denoted as ${\mbf P}_{t|t-1}$, ${\mbf P}_{t|t}$, and ${\mbf P}_{t|T}$, respectively. These are therefore optimal estimators of the \textit{static} factors since they minimize the associated Mean-Square-Error (MSE) for a given value of the parameters.

In the M-step a new estimator of the parameters $\wh{\bm\Theta}_{k+1}$ is computed by maximizing the expected likelihood. At convergence of the EM algorithm, say at iteration $k^*$, we obtain the estimator of the parameters, which we denote by $\wh{\bm\Theta}:= \wh{\bm\Theta}_{k^*}$. The estimator of the factors is then obtained by running the KS a last times using $\wh{\bm\Theta}$ and it is denoted by $\wh{\mbf F}_{t}:=\E_{\wh{\bm\Theta}}[\mbf F_t |\mbf x_1,\ldots,\mbf x_T]$. The estimated common and idiosyncratic components are then given by $\wh{\chi}_{it}=\wh{\bm \lambda}_i'\wh{\mbf F}_{t}$ and $\wh{\xi}_{it}=x_{it}-\wh{\chi}_{it}$. Details of the EM algorithm, as well as closed form expressions for all the estimators, are in \ref{sec:proofsDetails}. 

To initialise the EM algorithm we use as initial estimator of the loadings the $r$ leading eigenvectors of the covariance of $\Delta \mbf x_t$, from which we have an estimator of the \textit{static} factors as the integrated principal components of $\Delta \mbf x_t$ (\citealp{baing04}). This factors' estimator is in turn used to: (i) initialize the KF, together with a diffuse prior for the factors' covariance (\citealp{koopman97,KD00}) and (ii) estimate the VAR parameters (\citealp{BLL2}). Define as $\mbf V$ the $n\times r$ matrix having as columns the $r$ leading normalised eigenvectors of the covariance of $\Delta \bm \chi_t$, then the following identifying assumptions are convenient for proving consistency:
\ben
\item [E1.] $\bm\Lambda= \sqrt n\, \mbf V$ with $[\bm \Lambda]_{1j}>0$ for all $j=1,\ldots, r$; 
\item [E2.] $\mbf F_t=n^{-1/2}\,\mbf V'\bm\chi_t$ with $\mbf F_0=\mbf 0_r$. 
\een
Since the \textit{static} factors have no economic meaning, these identifying assumptions are perfectly valid and --- together with assumption C2 on the loadings scale --- they rule out any indeterminacy in the estimators used to initialize the EM algorithm --- see \citet{DGRfilter} for similar assumptions. 

We have the following consistency result.

\begin{prop}\label{prop:EMcons} Let A1, A2, A5, C1, C2, C3, D1, D2, D3, D4, E1, and E2 hold and let $\bar t(T)>0$ be such that 
\beq
\lim\!\!\sup_{T\to\infty} T e^{-\bar t(T)}\le M.\label{eq:exprate}
\eeq
Define $\mbf F_t^\dag:=(\bm f_t'\cdots \bm f_{t-s}')'$ and $\bm\lambda_i^\dag:=(\bm b_{i0}'\cdots \bm b_{is}')'$. Then, there exists an invertible $r\times r$  matrix $\mbf K$ such that, as $n,T\to\infty$, for all $\, \bar t (T) \le t\le T$ and any given $i=1,\ldots n$,
\begin{align}
& \sqrt T\,\Vert\wh{\bm \lambda}_i-\mbf K^{-1'}{\bm\lambda_i^\dag} \Vert=O_p(1),\label{eq:consEM2}\\
&\min(\sqrt n, \sqrt T)\,\Vert\wh{\mbf F}_{t}-\mbf K \mbf F_t^\dag\Vert=O_p(1), \label{eq:consEM3}\\
&\min(\sqrt n, \sqrt T)\,\vert\wh{\chi}_{it}-\chi_{it}\vert=O_p(1). \label{eq:consEM4}
\end{align} 
\end{prop}
\noindent
Proposition 1 states that under the assumptions presented before, we can consistently estimate the common component, as well as the spaces spanned by the \textit{dynamic} factors $\bm f_t$ and the corresponding \textit{dynamic} loadings which are the coefficients of $\bm b_i(L)$ defined  A3.  

Our proof, which is presented in detail in \ref{sec:proofsProposition}, is based on the same approach followed by \citet{PR15} in the one-factor case, and it is made of two main parts which we summarize here.\medskip

\begin{figure}[t!]\caption{Conditional Mean Squared Errors}\label{fig_Ptt}
\centering
\includegraphics[width=.6\textwidth,trim= 0cm 0cm 0cm 0cm,clip]{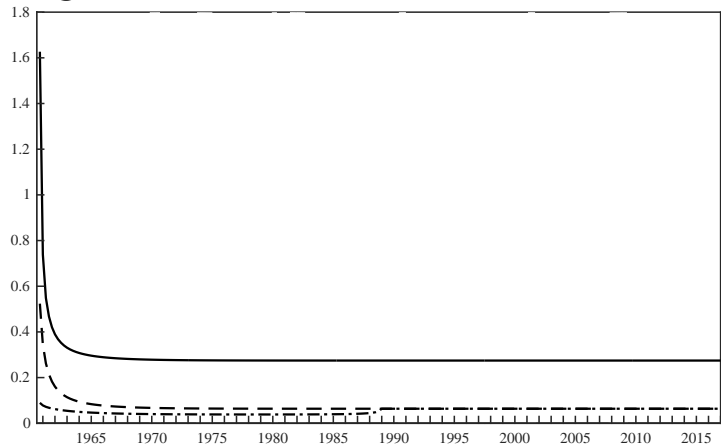}
\begin{tabular}{p{.6\textwidth}} \scriptsize
This figure reports $\mathsf{tr}(\mbf P_{t|s})$ when using $\wh{\bm\Theta}$, computed for the data analyzed in Section \ref{sec:emp},
where: $s=t-1$ is the one-step-ahead conditional MSE (solid line); $s=t$ is KF conditional MSE (dashed line); $s=T$ is the KS conditional MSE (dashed-dotted line).
\end{tabular}
\end{figure}

\noindent
\textbf{Population results}.
We first show that, when the parameters are known the one-step-ahead factors' MSE, $\mbf P_{t|t-1}$, converges to a steady state, while both the MSEs of the KF, $\mbf P_{t|t}$, and of the KS, $\mbf P_{t|T}$, tend to zero as $n\to\infty$ (Lemmas \ref{lem:steady1} and \ref{lem:steady3}). Notice that this is true also when initializing with a diffuse prior since this has an effect only for a finite number of initial periods, say $t_0$ (\citealp{koopman97}). In particular, convergence to the steady state is exponentially fast \citep{AM79}, hence our result holds for any $t\ge \bar t(T)>t_0$, where  $\bar t(T)$ satisfies condition \eqref{eq:exprate}, which asymptotically requires $\bar t(T)=O(\log T)$. In practice, though, the steady state is reached very quickly as shown in Figure \ref{fig_Ptt}, where we report the trace of $\mbf P_{t|t-1}$ (solid line), $\mbf P_{t|t}$ (dashed line) and $\mbf P_{t|T}$ (dashed-dotted line), computed for the data analysed in Section \ref{sec:emp}.  \medskip

\noindent
\textbf{Estimation results}.
In the second step of the proof, consistency of the KF and KS estimators of the \textit{static} factors when using estimated parameters is proved (Lemma \ref{lem:steady_est1}). This is done by taking into account an additional parameter estimation error which has two components: (i) the error of the QML estimator of the parameters for the case of known factors, say $\wh{\bm\Theta}^*$ (Lemma \ref{lem:eststar}) and (ii) the error due to the numerical approximation of $\wh{\bm\Theta}$ to $\wh{\bm\Theta}^*$ which is related to the stopping rule of the EM algorithm (\citealp{MR94}, and Lemma \ref{lem:convEM}). In particular, the latter error is shown to be negligible with respect to the former one. Therefore the rate of convergence of the loadings estimated via the EM algorithm is the same that one would obtain by QML estimation, were the true factors observable, and moreover, because of assumption D2 the loadings are estimated equation by equation, thus such error depends only on $T$. Results similar to \eqref{eq:consEM2} hold also for all other estimated parameters in $\wh{\bm\Theta}$. On the other hand the rate of convergence for the estimated \textit{static} factors is standard in the literature. \medskip

The results in Proposition \ref{prop:EMcons} extend those by \citet{DGRfilter,DGRqml} to the non-stationary case.
A major difference between the EM algorithm in levels proposed in this paper, and the EM algorithm in first differences proposed by \citet{DGRqml}, is relative to the way idiosyncratic components are modelled. Indeed, while by considering first differences it is implicitly assumed that all idiosyncratic components have a unit root, in our case we can distinguish between stationary and non-stationary idiosyncratic components --- i.e., we can allow for idiosyncratic trends only in some variables. This is not a minor difference, as it has substantial implications for the properties of the estimators. 

First of all we model non-stationary idiosyncratic components as additional latent states rather than differencing them, thus improving efficiency (see also Remark 2 below).  
Second, when $\xi_{it}\sim I(0)$, under D1 and D2 the QML estimator of the loadings of the $i$-th variable is obtained by minimizing the sample variance of $\xi_{it}$. In this case this is not the same as differencing before estimation, since in that case the loadings would be estimated by minimizing the sample variance of $\Delta\xi_{it}$. The resulting common component of the $i$-th variable has therefore different empirical properties: compared to our non-stationary approach, the common component estimated in first differences is likely to provide a better fit of the first differenced data, but not necessarily of the levels. Conversely, the common component obtained with our approach is likely to provide a better fit of the levels thus capturing better the lower frequencies --- and so the long-run trends --- and resulting in a smoother estimator, which however might have a worse fit of the differenced data.

We conclude this section by briefly discussing the possible mis-specifications introduced by assumptions D1, D2 and D3. In particular, we assume the vector of idiosyncratic shocks $\bm e_t$ to be i.i.d.~Gaussian, thus imposing four restrictions on: (1) the cross-sectional dependence; (2) the variances; (3) the serial dependence;  (4) the distribution. Let us consider the implications for the properties of the estimators of each of these restrictions --- see also \citet{DGRfilter} for a similar discussion.

\begin{rem}\upshape{
If the idiosyncratic components have some cross-sectional dependence, as allowed by A5, then the state-space form of the model is mis-specified, however by inspecting the proofs we see that, as long as we use an invertible estimator of $\mbf R$, consistency is not affected as long as $n\to\infty$. As a consequence of this asymptotic argument, we do not attempt here to model the off-diagonal terms of $\mbf R$.

This is better illustrated by a simple example showing the properties of the KF (an analogous argument holds for the KS). Denote as $\mbf P$ the steady state of $\mbf P_{t|t-1}$ then it can be shown that $\mbf P=\mbf H\mbf H'$ (Lemma \ref{lem:steady1}). Consider the case in which the parameters are given, $\bm\xi_{t}\sim I(0)$, and $r=q$, so that $\mbf P$ is invertible, then for $t\ge \bar t(T)$ the KF estimator is such that
\begin{align}
\mbf F_{t|t}&= \mbf F_{t|t-1} + \mbf P \bm\Lambda'(\bm\Lambda\mbf P\bm\Lambda'+\mbf R)^{-1}(\mbf x_t-\bm\Lambda\mbf F_{t|t-1})\nn\\
&=\mbf F_{t|t-1} +(\bm\Lambda'\mbf R^{-1}\bm\Lambda+\mbf P^{-1})^{-1}\bm\Lambda'\mbf R^{-1}(\mbf x_t-\bm\Lambda\mbf F_{t|t-1})\nn\\
&= (\bm\Lambda'\mbf R^{-1}\bm\Lambda)^{-1}\bm\Lambda'\mbf R^{-1}\mbf x_t + O(n^{-1})\nn\\
&=\mbf F_t + (\bm\Lambda'\mbf R^{-1}\bm\Lambda)^{-1}\bm\Lambda'\mbf R^{-1}\bm\xi_t+ O(n^{-1})\nn\\
&=\mbf F_t +O_p(n^{-1/2}),\nn
\end{align}
where we used (in order) the Woodbury formula, assumption C2, the definition of $\mbf x_t$ in \eqref{eq:SS1}, and assumption A5. Clearly consistency of the KF does not depend on the specific assumption for $\mbf R$, as long as it is invertible. However for finite $n$ the KF depends on $\mbf R$ and modeling also its out of diagonal terms could in principle improve its efficiency (e.g. \citealp{bailiao16}). }
\end{rem}

\begin{rem}\upshape{
From the example in Remark 1 it is clear that for finite $n$ the KF estimator is a weighted average of the data where the heteroskedasticity of the idiosyncratic components is accounted for. Again the same argument holds also for the KS. In this respect the KF-KS approach is analogous to the generalized principal component estimator, which is however derived in a stationary setting and without explicitly addressing the dynamics of the data (\citealp{choi12}).  
}
\end{rem}

\begin{rem}
\upshape{If the idiosyncratic components are autocorrelated, then, unless we model them explicitly as additional latent states, optimality is lost, in particular the loadings' estimators are still consistent but not efficient. By means of D1 we partially solve the problem at least for the series with $I(1)$ idiosyncratic components. 
}
\end{rem}

\begin{rem}\upshape{
If the idiosyncratic components are non-Gaussian then the estimator is not optimal being only the best linear estimator. Nevertheless, it has to be noticed that typical macroeconomic data show little deviations from normality, so we are minimally concerned by the restrictions imposed by this assumption.
}
\end{rem}
Summing up, regardless of these mis-specifications even though we might not have the most efficient estimator, we are likely to have gains in efficiency with respect to those estimators obtained by integrating the principal components of first differences of the data (\citealp{baing04}).~Indeed, principal components are optimal only in the case of serially and cross-sectionally i.i.d.~Gaussian idiosyncratic components (\citealp{lawleymaxwell71,tippingbishop99}), and such conditions clearly do not hold in a time series context, especially when non-stationarities are present and the cross-sectional dimension is large. On the contrary, our approach explicitly takes into account the autocorrelation in the factors and in the idiosyncratic components as well as their heteroscedasticity, and, as discussed above, it delivers consistent estimates even when some degree of cross-sectional dependence is present but not modelled. 

\subsection{Trend and cycles}\label{sec:TC2}
We now turn to estimation of common trends and common cycles. Notice that since we do not fully impose R1, the \textit{dynamic} factors $\bm f_t$ are not identified and instead we have to deal with a TC decomposition of the \textit{static} factors $\mbf F_t$, which can be carried out analogously to the one described in Section \ref{sec:Representation} for $\bm f_t$. Because of assumption C1 and restriction R3, for given values of $q$ and $d$, the vector $\mbf F_t$ admits the factor representation: 
\begin{align}\nn
{\mbf F}_t &= \bm\Phi_1\mbf T_t + \bm\Gamma_t,
\end{align}
where $\bm\Gamma_t\sim I(0)$, $\bm\Phi_1$ is $r\times (q-d)$ and $\mbf T_t$ is the vector of $(q-d)$ common trends with components $T_{jt}\sim I(1)$ for $j=1,\ldots, (q-d)$. Hence, in general the common trends admit the MA representation:
\[
\Delta \mbf T_t = \bm{\mathcal B}(L) \bm \eta_t,
\]
where $\bm\eta_t\stackrel{w.n.}{\sim} (\mbf 0_{q-d},\bm\Sigma_{\eta})$ with $\bm\Sigma_{\eta}$ positive definite and $\bm{\mathcal B}(L)$ is a $(q-d)\times (q-d)$ one-sided, infinite matrix polynomial with square-summable coefficients and $\mathsf{rk}(\bm{\mathcal B}(1))=(q-d)$. 

As a consequence of the results by \citet{penaponcela97} and Proposition \ref{prop:EMcons} above, given the estimated factors $\wh{\mbf F}_t$, it is clear that, as $n,T\to\infty$,
\beq\label{eq:PPGK}
\wh{\mbf S}:=\frac 1{T^2}\sum_{t=1}^{T} \wh{\mbf F}_t\wh{\mbf F}_{t}'\stackrel{}{\Rightarrow}
 \bm\Phi_1\,\bm{\mathcal B}(1)\,\bm\Sigma_{\eta}^{1/2}\l(\int_0^1\bm{\mathcal W}(u)\bm{\mathcal W}(u)'\mathrm d u\r)\bm\Sigma_{\eta}^{1/2}\,\bm{\mathcal B}(1)' \,\bm\Phi_1', 
\eeq
where convergence is in the sense of weak convergence of the associated probability measures and $\{\bm{\mathcal W}(u),\ 0\le u\le 1\}$ is a $(q-d)$-dimensional standard Wiener process. 
Hence, by virtue of \eqref{eq:PPGK}, we can estimate the common trends $\mbf T_t$ as the first $(q-d)$ principal components of the estimated \textit{static} factors $\wh{\mbf F}_t$ (\citealp{bai04,penaponcela04}). Specifically, we denote by $(\wh{\bm \Phi}_1\;\wh{\bm \Phi}_0)$ the $r\times r$ matrix with columns given by the normalized eigenvectors of $\wh{\mbf S}$, ordered according to the decreasing value of the corresponding eigenvalues, and such that $\wh{\bm \Phi}_1$ is $r\times (q-d)$ and $\wh{\bm \Phi}_0$ is $r\times (r-q+d)$. This leads to the estimator of common trends as the projection:
\begin{align}
\wh{\mbf T}_t = \wh{\bm\Phi}_1'\wh{\mbf F}_t. \nn
\end{align}
As for the common cycles, notice first that, by projecting $\wh{\mbf F}_t$ onto the columns of $\wh{\bm \Phi}_0$, we obtain the $(r-q+d)$-dimensional process 
\beq\nn
\wh{\mbf G}_t=\wh{\bm\Phi}_{0}'\wh{\mbf F}_t,
\eeq
which, by construction, is orthogonal to $\wh{\mbf T}_t$. Moreover, $\wh{\mbf G}_t$ is stationary since it belongs to the cointegration space of $\wh{\mbf F}_t$ \citep{ZRY}. However, by R3 we know that the cointegration space must have dimension $d$, but we do not impose R3 when estimating the \textit{static} factors. Thus, we face the problem of identifying $d$ cycles from the higher-dimensional stationary process $\wh{\mbf G}_t$. 

In order to identify the common cycles we then look for the $d$-dimensional projection of $\wh{\mbf G}_t$ with maximum spectral density. In the empirical analysis of Section \ref{sec:emp}, we consider the VAR(2):
\beq\label{eq:VARG}
\wh{\mbf G}_t = \bm{\mathcal A}_1\wh{\mbf G}_{t-1}+\bm{\mathcal A}_2\wh{\mbf G}_{t-2}+ \bm v_t,
\eeq
where $\bm v_t\stackrel{w.n.}{\sim} (\mbf 0_{r-q+d},\bm\Sigma_v)$ and $\det(\mbf I_{r-q+d}- \bm{\mathcal A}_1z-\bm{\mathcal A}_2z^2)\neq 0$ for $|z|\le 1$. Once we estimate \eqref{eq:VARG} we have its residuals $\wh{\bm v}_t$ and their covariance matrix $\wh{\bm\Sigma}_v$.
Denote as $\wh{\bm{\mathcal H}}$ the $(r-q+d)\times d$ matrix having as columns the leading $d$ normalized eigenvectors of $\wh{\bm\Sigma}_v$. We then define the estimated cycle component as the $d$-dimensional projection:
\beq
\wh{\mbf C}_t = \wh{\bm{\mathcal H}}'\wh{\mbf G}_t.\nn
\eeq
The estimated TC decomposition is then given by
\begin{align}
\wh{\mbf F}_t &= \wh{\bm\Phi}_1  \wh{\bm\Phi}_1'\wh{\mbf F}_t+ \wh{\bm\Phi}_{0} \wh{\bm\Phi}_{0}'\wh{\mbf F}_t\nn\\
&=\wh{\bm\Phi}_1 \wh{\mbf T}_t + \wh{\bm\Phi}_{0}\wh{\mbf G}_t \nn\\
&=\wh{\bm\Phi}_1 \wh{\mbf T}_t + \wh{\bm\Phi}_{0}\wh{\bm{\mathcal H}}\wh{\bm{\mathcal H}}'\wh{\mbf G}_t +  \wh{\bm\Phi}_{0}\wh{\bm{\mathcal H}}_\perp\wh{\bm{\mathcal H}}'_\perp\wh{\mbf G}_t \nn\\
&=\wh{\bm\Phi}_1 \wh{\mbf T}_t + \wh{\bm\Phi}_{0}\wh{\bm{\mathcal H}}\wh{\mbf C}_t + \wh{\bm\Phi}_{0}(\wh{\mbf G}_t-\wh{\bm{\mathcal H}}\wh{\mbf C}_t), \label{eq:FTCW}
\end{align}
where $\wh{\bm{\mathcal H}}_\perp$ is $(r-q+d)\times (r-q)$ and such that $\wh{\bm{\mathcal H}}_\perp'\wh{\bm{\mathcal H}}=\mbf 0_{(r-q)\times d}$. The last term on the right-hand-side of \eqref{eq:FTCW} appears due to the mis-specification caused by not fully imposing R1 and R3 and in particular it has covariance of rank $(r-q)$ and since $r>q$ it is  in general not zero. 

\begin{figure}[t!]\caption{Spectral Densities of Common Trends and Common Cycles}\label{fig_TC}
\centering \smallskip \noindent
\begin{tabular}{cc}
\includegraphics[width=.6\textwidth,trim= 0cm 0cm 0cm 0cm,clip]{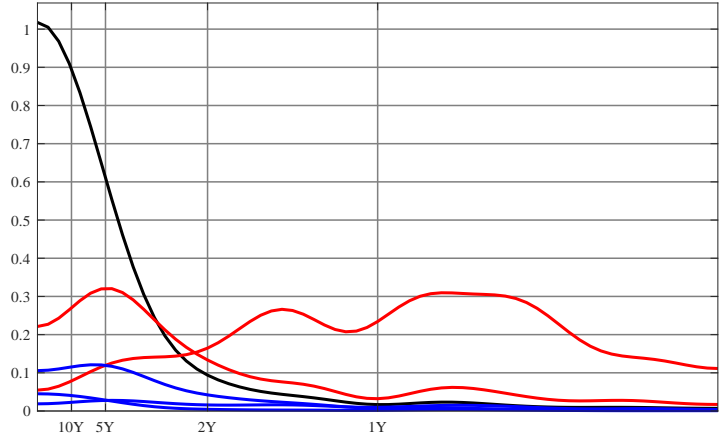} 
\end{tabular}
\begin{tabular}{p{.6\textwidth}} \scriptsize
This figure reports for the data analyzed in Section \ref{sec:emp} the spectral densities of the common trend $\Delta\wh{\mbf T}_t$ (black line), the common cycles $\Delta\wh{\mbf C}_t$ (red lines), and the residual cycles $(\Delta\wh{\mbf G}_t-\wh{\bm{\mathcal H}}\Delta\wh{\mbf C}_t)$ (blue lines). On the horizontal axis we report periods $\tau_j$ measured in years and corresponding to frequencies $\omega_j=\frac{2\pi}{4\tau_j}$ (the data considered is quarterly).
\end{tabular}
\end{figure}

To appreciate the meaning and the appropriateness of decomposition \eqref{eq:FTCW}, in Figure \ref{fig_TC} we show the spectral densities of the first differences of the three components of $\wh{\mbf F}_t$ for the data analyzed in Section \ref{sec:emp}, where $r=6$, $q=3$, and $d=2$. As expected the estimated common trend $\wh{\mbf T}_t$ (black line) contributes most at the lowest frequencies --- i.e., lower than $\frac \pi {10}$ --- which correspond to periods higher than five years. Once we remove the common trend, of the remaining five processes $\wh{\mbf G}_t$, the two estimated common cycles $\wh{\mbf C}_t$ (red lines) capture most of the variation for almost all frequencies: one cycle dominates at periods longer than two years --- i.e., frequencies lower than $\frac\pi 4$ --- and the other cycle dominates at periods shorter than two years --- i.e., frequencies higher than $\frac \pi 4$. With respect to those two cycles, the residual three cycles $(\wh{\mbf G}_t-\wh{\bm{\mathcal H}}\wh{\mbf C}_t)$ (blue lines) give a negligible contributions to the total variation.
Given this empirical result, the extra term in \eqref{eq:FTCW} can be neglected and treated as a mis-specification error.

Finally, from \eqref{eq:FTCW}, the estimated TC decomposition of the data immediately follows:
\begin{align}
x_{it}
&=  \wh{\bm\lambda}_i' \wh{\bm\Phi}_1 \wh{\mbf T}_t +\wh{\bm\lambda}_i' \wh{\bm\Phi}_{0} \wh{\bm{\mathcal H}}\wh{\mbf C}_t +\wh{\bm\lambda}_i' \wh{\bm\Phi}_{0}(\wh{\mbf G}_t-\wh{\bm{\mathcal H}}\wh{\mbf C}_t)+\wh{\xi}_{it},\nn
\end{align}
which is the estimated counterpart of the representation given in \eqref{eq:tc2}.

\section{Estimating the cyclical position of the economy\\ and the observation error}\label{sec:emp}
We now use our model to estimate the cyclical position of the US economy and the observation error. In particular, in Section \ref{sec_gdo} we will estimate ``the observation error'' by estimating the non-stationary approximate DFM as explained in Section \ref{sec:ML}. And, in Section \ref{sec_ogap} we will estimate ``the cyclical position of the economy'' by decomposing the common factors into common trends and common cycles using the TC decomposition discussed in Sections \ref{sec:Representation} and \ref{sec:TC2}. 

The following analysis is carried out on a large macroeconomic dataset comprising $n=103$ quarterly series from 1960:Q1 to 2017:Q1 describing the US economy. The complete list of variables and transformations is reported in \ref{sec:data}.

Compared to the papers that use small DFMs to estimate the cyclical position of the economy, which typically estimate the output gap using only high level variables such as GDP, the unemployment rate, and PCE price inflation, we include several other indicators, thus being able to capture information coming from a wider spectrum of the economy. Specifically, our datasets includes national account statistics, industrial production indexes, various price indexes including CPIs, PPIs, and PCE price indexes, various labor market indicators including indicators from both the household survey and the establishment survey as well as labor cost and compensation indexes, monetary aggregates, credit and loans indicators, housing market indicators, interest rates, the oil price, and the S\&P500 index. Broadly speaking, all the variables that are $I(1)$ are not transformed, while all the variables that are $I(2)$ are differenced once. Notice that some variables should from a theoretical economic point of view always be considered as $I(0)$ (e.g. inflation rates, unemployment rate, and interest rates) but since they exhibit a great deal of persistence are here treated as $I(1)$. Finally, a linear trend is estimated where necessary before applying our methodology, thus accounting for the deterministic component in \eqref{eq:trend}.

A thorough empirical analysis requires tackling two main preliminary problems. First, we need to determine the number of  common trends $(q-d)$, of common shocks $q$, and of \textit{static} factors $r$. To determine the number of common trends $(q-d)$ we use the criterion by \citet{BLL2}, which exploits the behaviour of the eigenvalues described in condition B2. This criterion indicates the presence of $(q-d)=1$ common trend, which is in line with many theoretical models assuming a common productivity trend as the sole driver of long-run dynamics \citep[e.g.][]{negro07}. To determine the number of common shocks $q$ we use the test by \citet{onatski09} and the criterion by \citet{hallinliska07}, which exploit the behaviour of the eigenvalues described in condition B1. Both methods indicate the presence of $q=3$ common shocks. Having determined $q$,  as we explained in Section \ref{sec:StaticFM} by virtue of R2  we can set the number of \textit{static} factors $r$ according to their explained variance. By looking at Table \ref{tab_explainedvariance} we can clearly see that $r\simeq2q$, and therefore in our benchmark specification we set $q=3$ and $r=6$.\footnote{An alternative way to select the number of \textit{static} factors $r$ is to resort to one of the many available methods, such as, for example, the criterion of \citet{baing02}, which for our dataset gives results in line with our choice of $r$.}

\begin{table}[t!]\caption{Percentage of explained variance}\label{tab_explainedvariance}									
\centering																					
\begin{tabular*}{.95\textwidth}{@{}@{\extracolsep{\fill}}ccccccccccccc@{}} \hline\hline																					
& 	&	1	&	2	&	3	&	4	&	5	&	6	&	7	&	8	&	9	&	10	&\\\hline
&	$q$ 	&	33.4	&	45.8	&	53.3	&	58.9	&	63.6	&	67.4	&	70.6	&	73.4	&	75.8	&	77.9&	\\
&	$r$ 	&	23.4	&	33.9	&	42.1	&	47.9	&	51.8	&	55.3	&	58.2	&	60.6	&	62.7	&	64.9&	\\\hline
\end{tabular*}				
\begin{tabular}{p{.95\textwidth}} \scriptsize
This table reports the percentage of total variance explained by the $q$ largest eigenvalues of the spectral density matrix of $\Delta \mbf x_t$  and by the $r$ largest eigenvalues of the covariance matrix of $\Delta \mbf x_t$. 
\end{tabular}																	
\end{table}		

Second, we need to choose which idiosyncratic components to model as random walk, and which as white noises. Following the methodology proposed by \citet{baing04}, we can explicitly test the null-hypothesis $H_0$: ${\rho}_i=1$, and if we do not reject $H_0$, we set $\rho_i=1$, while if we reject  $H_0$, we set $\rho_i=0$. This approach is applied to all variables in the dataset except GDP, GDI, unemployment rate, Federal funds rate, and CPI, core CPI, PCE, and core PCE inflation, for which we impose a priori $\rho_i=0$. That is, while for most of the variables in the dataset we let the data determine what is driving their long run dynamics, we impose that the long-run dynamics of GDP, GDI, unemployment rate, Federal funds rate, and CPI, core CPI, PCE, and core PCE inflation are driven exclusively by macroeconomic shocks, with the idiosyncratic shocks accounting only for short-run movement.

\subsection{Measuring Gross Domestic Output} \label{sec_gdo}
A fundamental issue in economics is the measurement of aggregate real output, henceforth Gross Domestic Output (GDO). Historically, GDO has been measured mainly by the Gross Domestic Product (GDP), but GDP, which tracks all expenditures on final goods and services produced, is just an estimate of GDO. An equally acceptable estimate of the concept of GDO is represented by the Gross Domestic Income (GDI), which tracks all income received by those who produced the output. GDP is almost always preferred to GDI, the main reason being that it is released before GDI.\footnote{The first estimate of GDP is released one month after the reference quarter, while GDI is generally released two months after the reference quarter, together with the second release of GDP.} However it has been shown that GDI reflects the business cycle fluctuations in true output growth better than GDP and moreover GDI is better than GDP in recognising the start of a recession (\citealp{jeremy10,jeremy12}).

In recent years, there has been interest in combining GDP and GDI to come up with a better estimate of GDO, where the rationale for doing so is that the difference between GDP and GDI is exclusively the result of measurement error --- using the NIPA table definition ``statistical discrepancy'' --- as these two statistics are in fact measuring the same thing. For example, starting from November 4, 2013, the Philadelphia Fed releases an estimate of GDO, called ``GDPplus'' proposed by \citet{ADNSS}, which is defined as the common component of a bivariate one-factor model built with GDP and GDI growth rates. Similarly, and starting from July 30, 2015, the Bureau of Economic Analysis (BEA) releases ``the average of GDP and GDI'', which the Council of Economic Advisers refers to as GDO \citep*{CEA15}. 

Our approach differs from those mentioned above in that our estimate of GDO is not obtained by combining GDP and GDI, rather it is obtained by using all the 103 variables included in our dataset. In detail, we define GDO as that part of GDP/GDI that is driven by the macroeconomic (common) shocks, i.e., $\textrm{GDO}_t=\chi_t^\textrm{GDP}=\chi_t^\textrm{GDI}$. To estimate GDO in this way, we estimate a constrained version of model \eqref{eq:SS1}-\eqref{eq:SS2}, where we impose the restriction of equal common components: $\chi_t^\textrm{GDP}=\chi_t^\textrm{GDI}$. This restriction is indeed corroborated by the data, as even if we do not impose it, the estimated $\chi_t^\textrm{GDP}$ and $\chi_t^\textrm{GDI}$ are nearly identical. In numbers, the standard deviation of $(\Delta y_t^\textrm{GDP}-\Delta y_t^\textrm{GDI})$ is 1.93, while the standard deviation of $(\Delta \chi_t^\textrm{GDP}-\Delta \chi_t^\textrm{GDI})$ is reduced to 0.28.

\begin{figure}[t!]\caption{Gross Domestic Output}\label{fig_gdo}
\centering \smallskip \noindent
\begin{tabular}{cc}
\it \footnotesize Quarterly annualised percentage change& \it \footnotesize 4-quarter percentage change\\
\includegraphics[width=.475\textwidth,trim=0cm .5cm 0cm 0cm,clip]{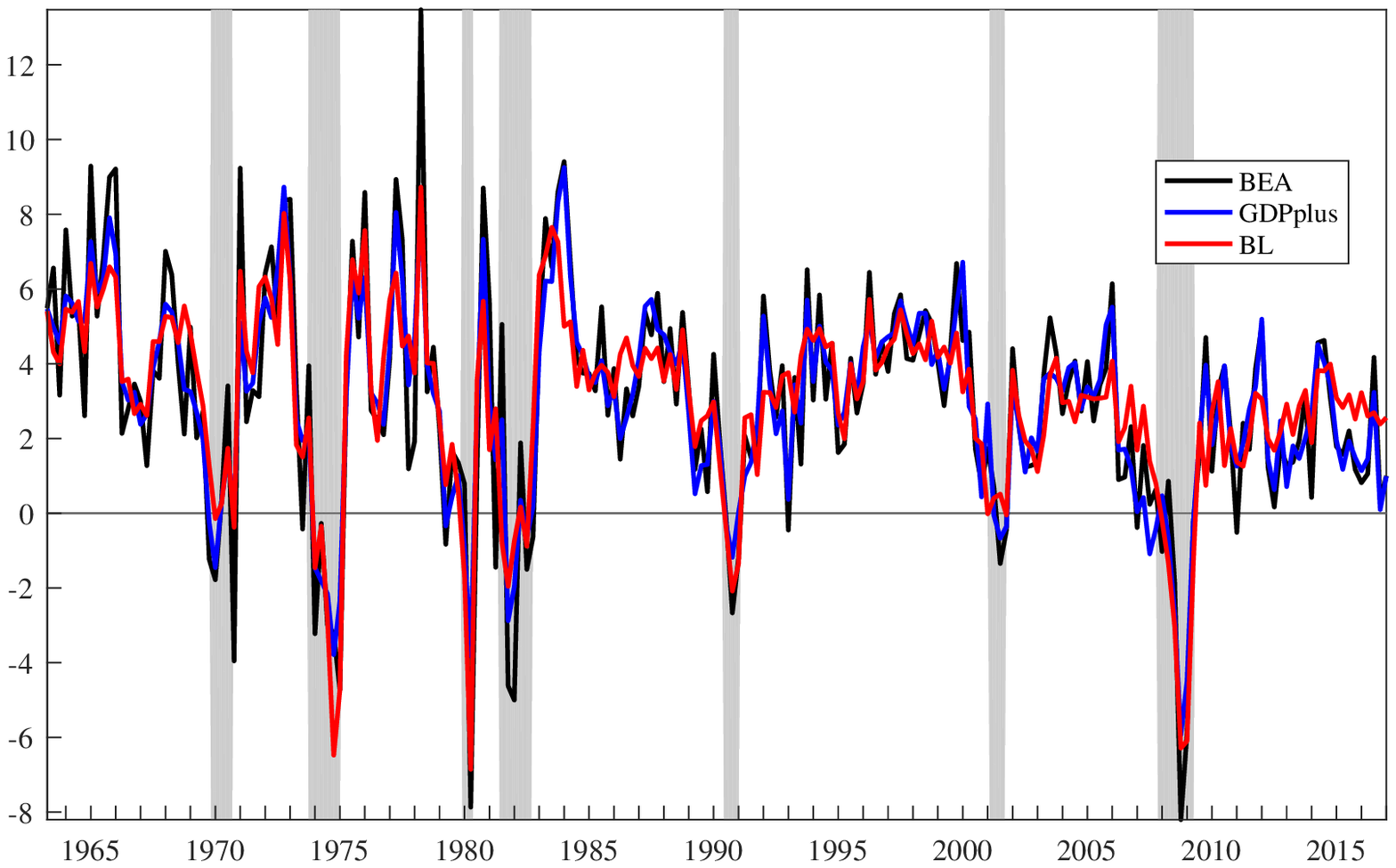}&
\includegraphics[width=.475\textwidth,trim=0cm .5cm 0cm 0cm,clip]{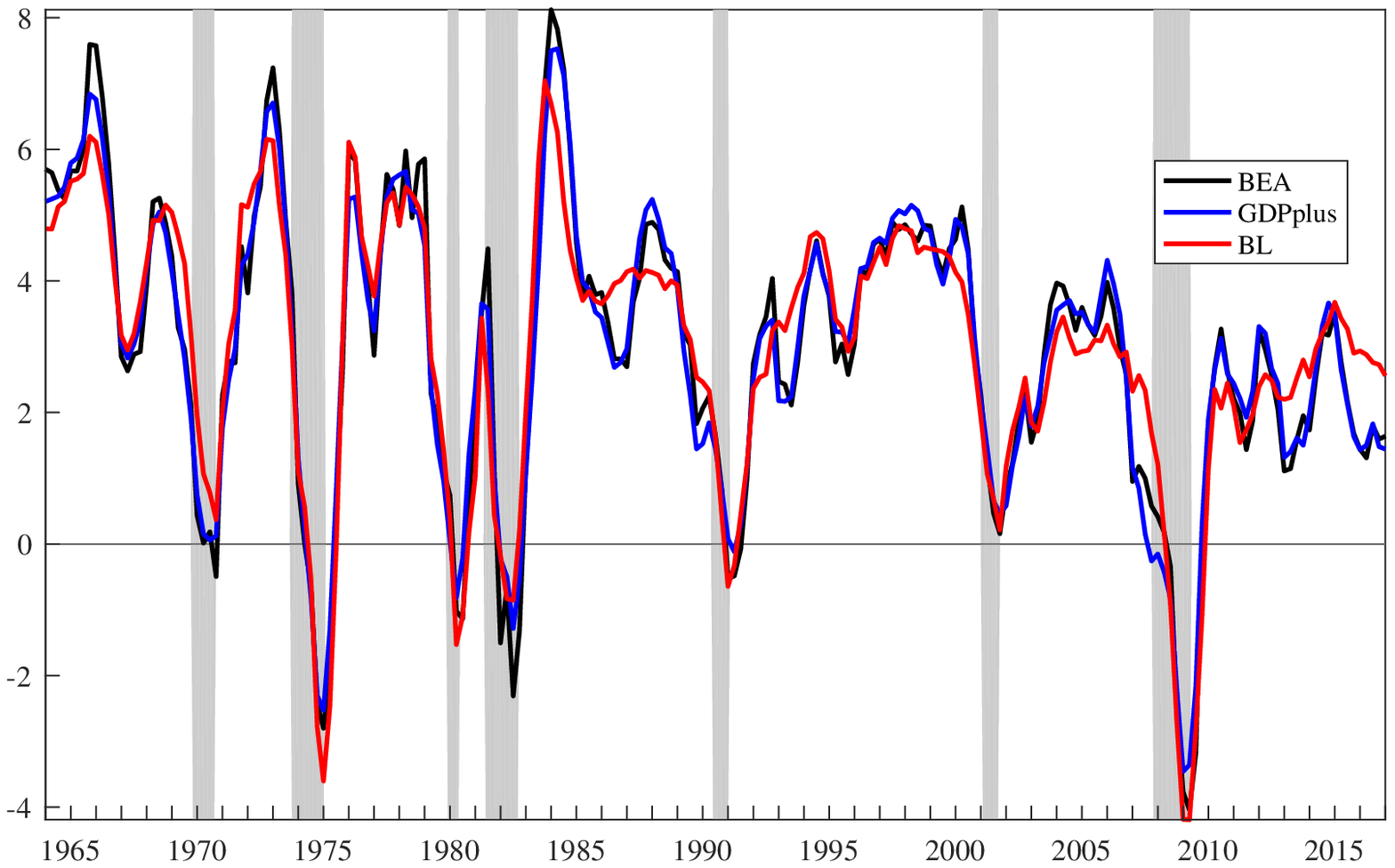} \\
\end{tabular}
\begin{tabular}{p{.95\textwidth}} \scriptsize
This figure reports different estimates of GDO. Black line: ``the average of GDP and GDI'' released by the BEA; blue line: ``GDPplus'' released by the Philadelphia Fed; red line: our estimate.
\end{tabular}
\end{figure}

Figure \ref{fig_gdo} shows our proposed estimate of GDO (red line) together with ``GDPplus'' (blue line) and the ``the average of GDP and GDI'' released by the BEA (black line). Overall, the three measures are very similar, which is not surprising, as they are attempting to estimate the same quantity. However, three important differences emerges. 

First, our estimate of GDO is smoother than the other two. This is not surprising. Compared to  ``GDPplus'' and ``the average of GDP and GDI'', our estimate of GDO is constructed to contain a larger low frequency component, because it is estimated on data in levels rather than on growth rates. Moreover, because it is derived under the assumption that the idiosyncratic components of GDP and GDI are stationary, by construction our estimate of GDO captures all the low frequency movements of GDP and GDI. 

\begin{figure}[t!]\caption{Residual Seasonality}\label{fig_residseas}
\centering \smallskip
\begin{tabular}{cc}
\it \footnotesize GDP& \it \footnotesize BEA\\
\includegraphics[width=.4\textwidth,trim=0cm 0cm 0cm 0cm,clip]{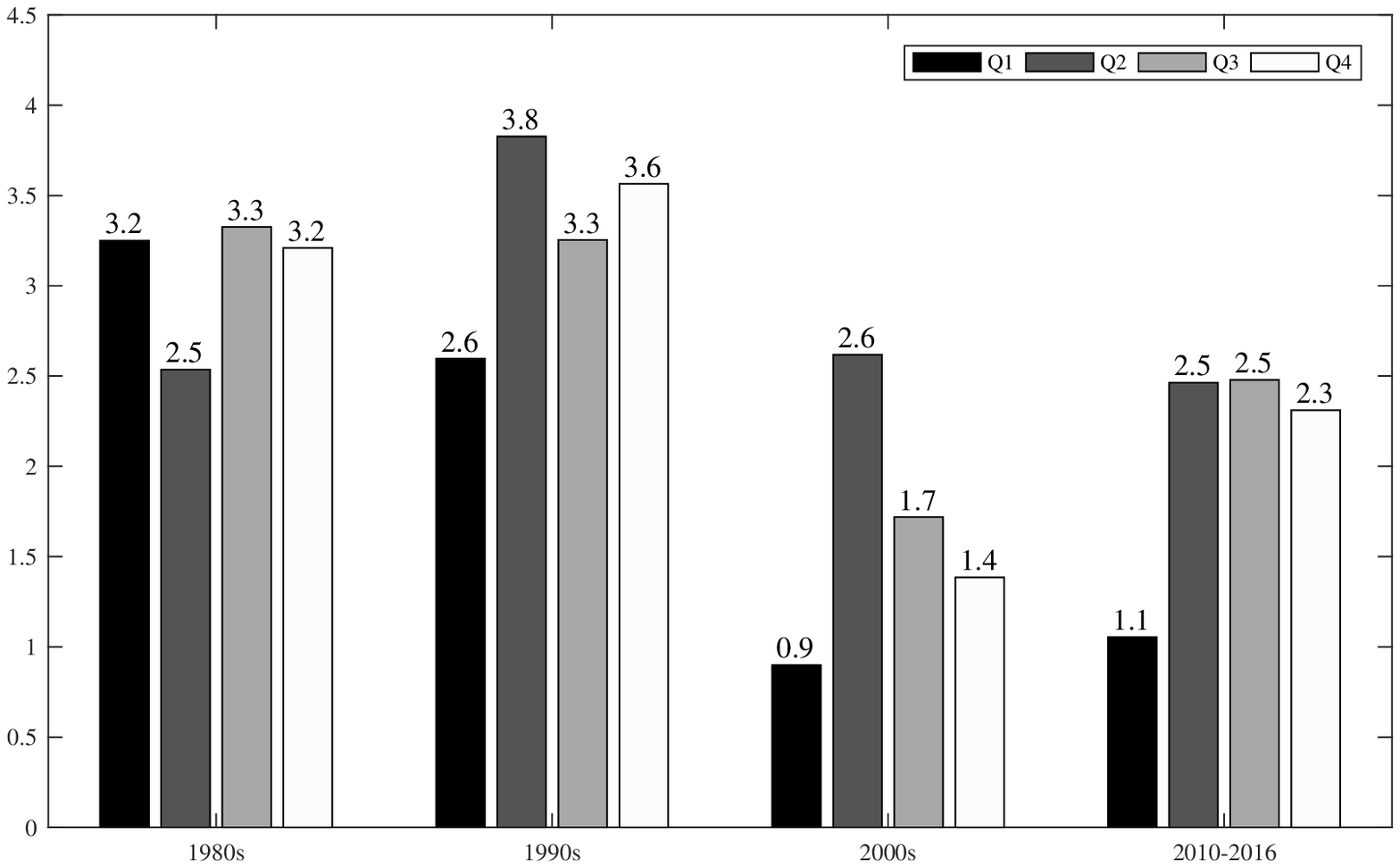}&
\includegraphics[width=.4\textwidth,trim=0cm 0cm 0cm 0cm,clip]{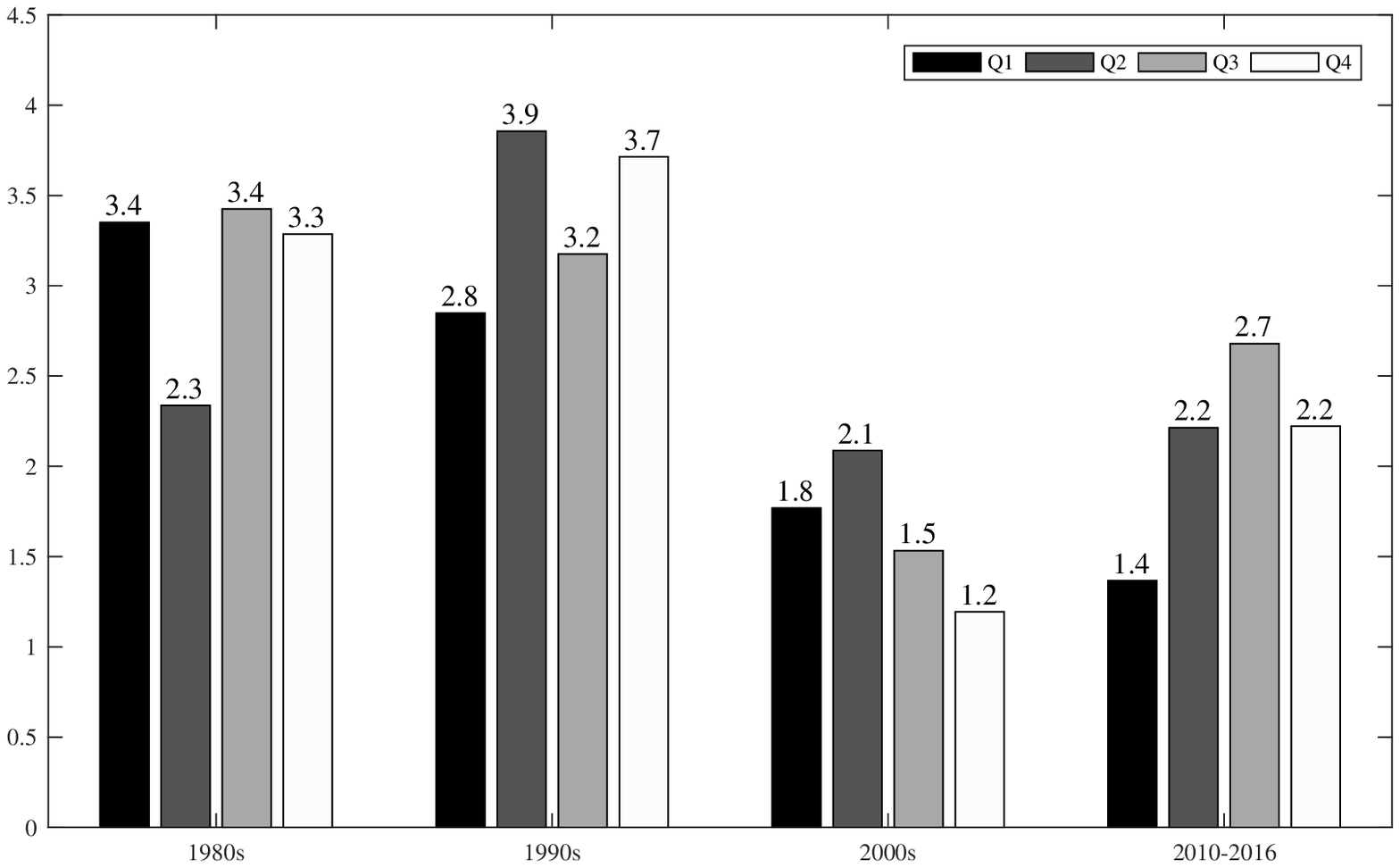} \\
\it \footnotesize GDPplus& \it \footnotesize BL\\
\includegraphics[width=.4\textwidth,trim=0cm 0cm 0cm 0cm,clip]{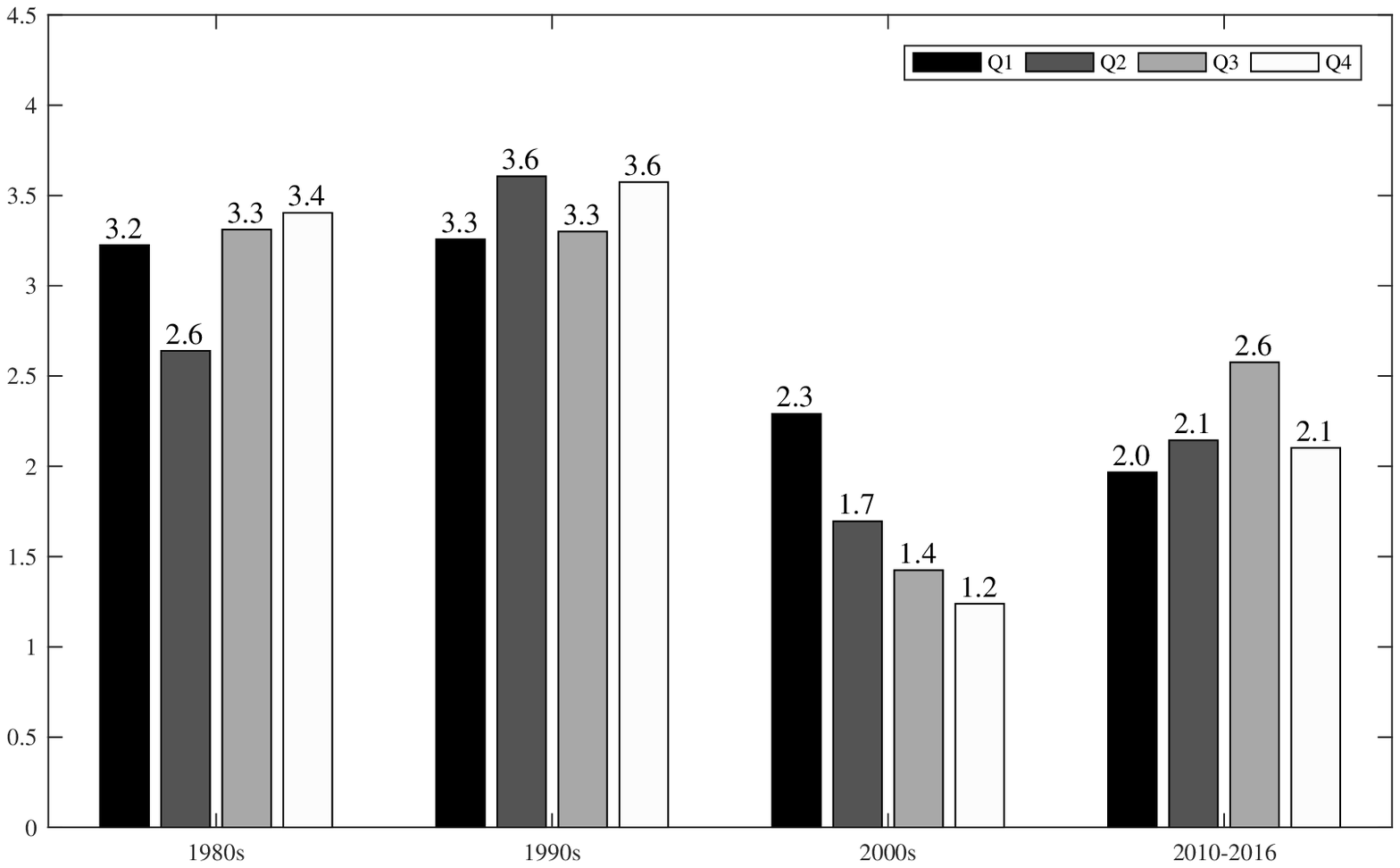}&
\includegraphics[width=.4\textwidth,trim=0cm 0cm 0cm 0cm,clip]{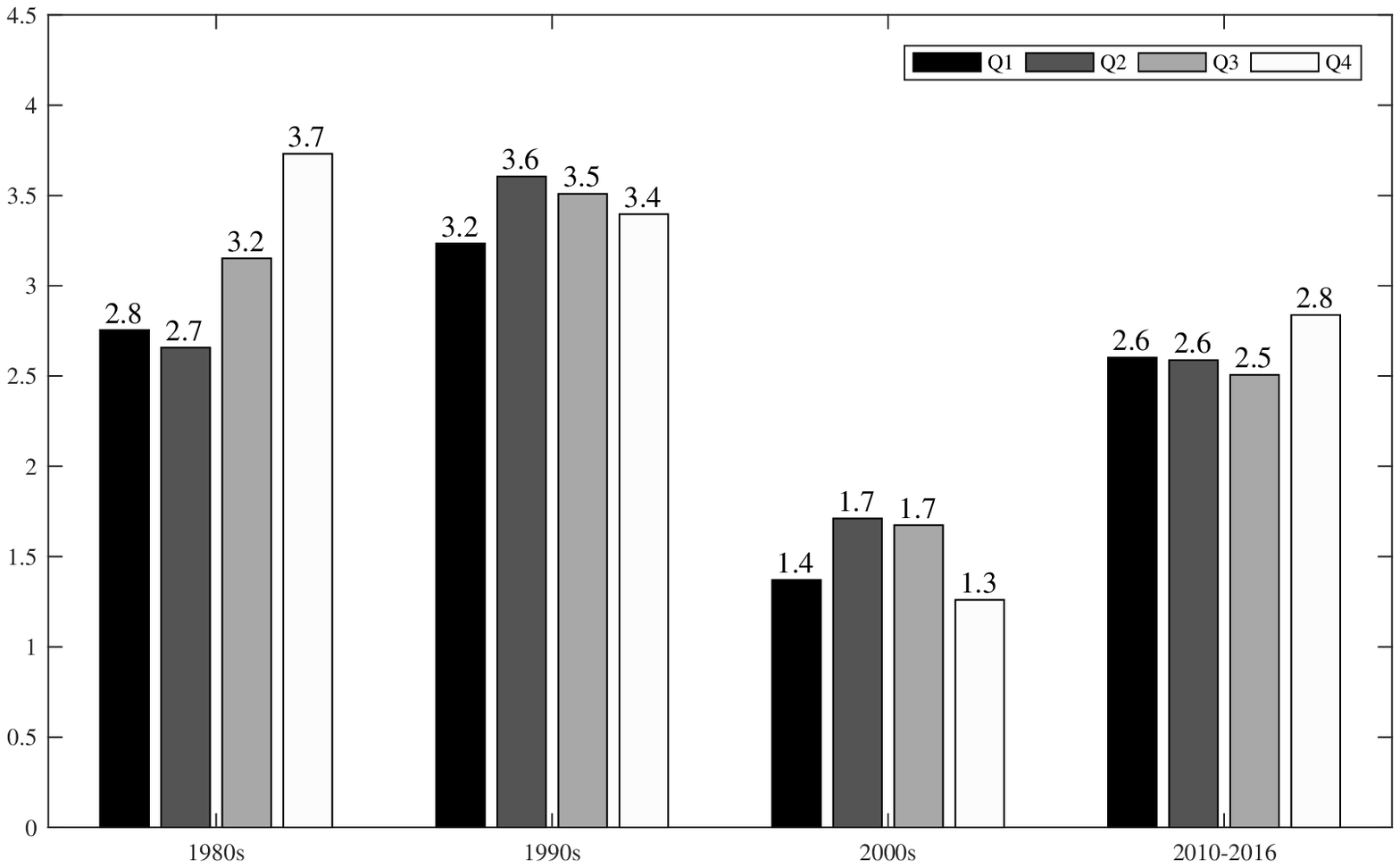} \\
\end{tabular}
\begin{tabular}{p{.82\textwidth}} \scriptsize
This figure reports average growth at an annual rate by quarter for GDP, ``the average of GDP and GDI'' released by the BEA, the Philadelphia Fed estimate of GDO (GDPplus), and our estimate of GDO (BL).
\end{tabular}
\end{figure}

Second, our estimate of GDO does not show any kind of residual seasonality in the last fifteen years, where the term ``residual seasonality'' refers to the presence of ``lingering  seasonal  effects  even  after  seasonal  adjustment processes have been applied to the data'' \citep{Moulton16}. Mainly motivated by the fact that since 2010 GDP growth in Q1 has been on average more than 1 percentage point lower than in the other quarters (NW plot of Figure \ref{fig_residseas}), in recent years there has been lots of discussion on whether US GDP exhibit residual seasonality or not. The profession is not in agreement on this issue, as some authors \citepalias[e.g.][]{Claudia1,Claudia2} conclude that US GDP does not exhibit residual seasonality, while others \citep[e.g.][]{rudebuschetal2017,lunsford2017} find evidence of residual seasonality --- see \citet{Moulton16} for a technical discussion on causes and remedies for residual seasonality in US GDP.  Figure \ref{fig_residseas} shows average real GDO growth by quarter for our estimate of GDO (SE plot), ``GDPplus'' (SW plot), and ``the average of GDP and GDI'' (NE plot). As can be clearly seen, our estimate of GDO exhibits no residual seasonality whatsoever in the last 15 years.

Third, our estimate of GDO in the recent years gives a different signal about the economy than the one given by `GDPplus'' and ``the average of GDP and GDI''. According to our estimate, since 2010 quarterly annualized GDO growth was on average \sfrac{1}{2} of a percentage point higher than estimated by the BEA or the Philadelphia Fed, where this difference comes mainly from our estimate of GDO growth in the first quarter (see Figure \ref{fig_residseas}), and therefore from the fact that our measure do not suffer of residual seasonality. In other words, based on the commonality in the data, the US economy grew at a faster pace than measured by national account statistics.

\subsection{Measuring the output gap}\label{sec_ogap}
Decomposing  aggregate real output into potential output and output gap is a critical task for both monetary and fiscal policy, as the former is a key input for long-term projections, and the latter can be an important gauge of inflationary pressure. There exist many definitions of potential output and of output gap --- see \citealp{outputgaps}, for a survey of different methods and definitions. Here we use the definition implied by the TC decomposition discussed in Sections \ref{sec:Representation} and \ref{sec:TC2}. Among the many existing approaches the most similar to ours are \citet{charlesjohn} and \citet{jarocinskilenza}, who use small dynamic factor models, \citet{aastveittrovik14}, who use a large stationary dynamic factor model combined with the Hodrick Prescott filter, and  \citet{morleywong2017}, who use a large stationary BVAR combined with the Beveridge and Nelson decomposition.

We compare our output gap estimate with the one produced by the Congressional Budget Office (CBO). The CBO estimates potential output and the output gap by using the so-called ``production function approach'' according to which potential output is that level of output consistent with current technologies and normal utilisation of capital and labour, and the output gap is the residual part of output. Specifically, the CBO model is based upon a textbook Solow growth model, with a neoclassical production function. Labour and productivity trends are estimated by using a variant of the Okun's law, so that actual output is above its potential (the output gap is positive), when the unemployment rate is below the natural rate of unemployment, which is in turn defined as the non-accelerating inflation rate of unemployment (NAIRU), i.e., that level of unemployment consistent with a stable inflation --- for further details see \citet*{CBO01}.

\begin{figure}[t!]\caption{Output gap}\label{fig_OG}
\centering
\begin{tabular}{cc}
\it \footnotesize Level& \it \footnotesize 4-quarter percentage change\\
\includegraphics[width=.475\textwidth,trim=0cm .5cm 0cm 0cm,clip]{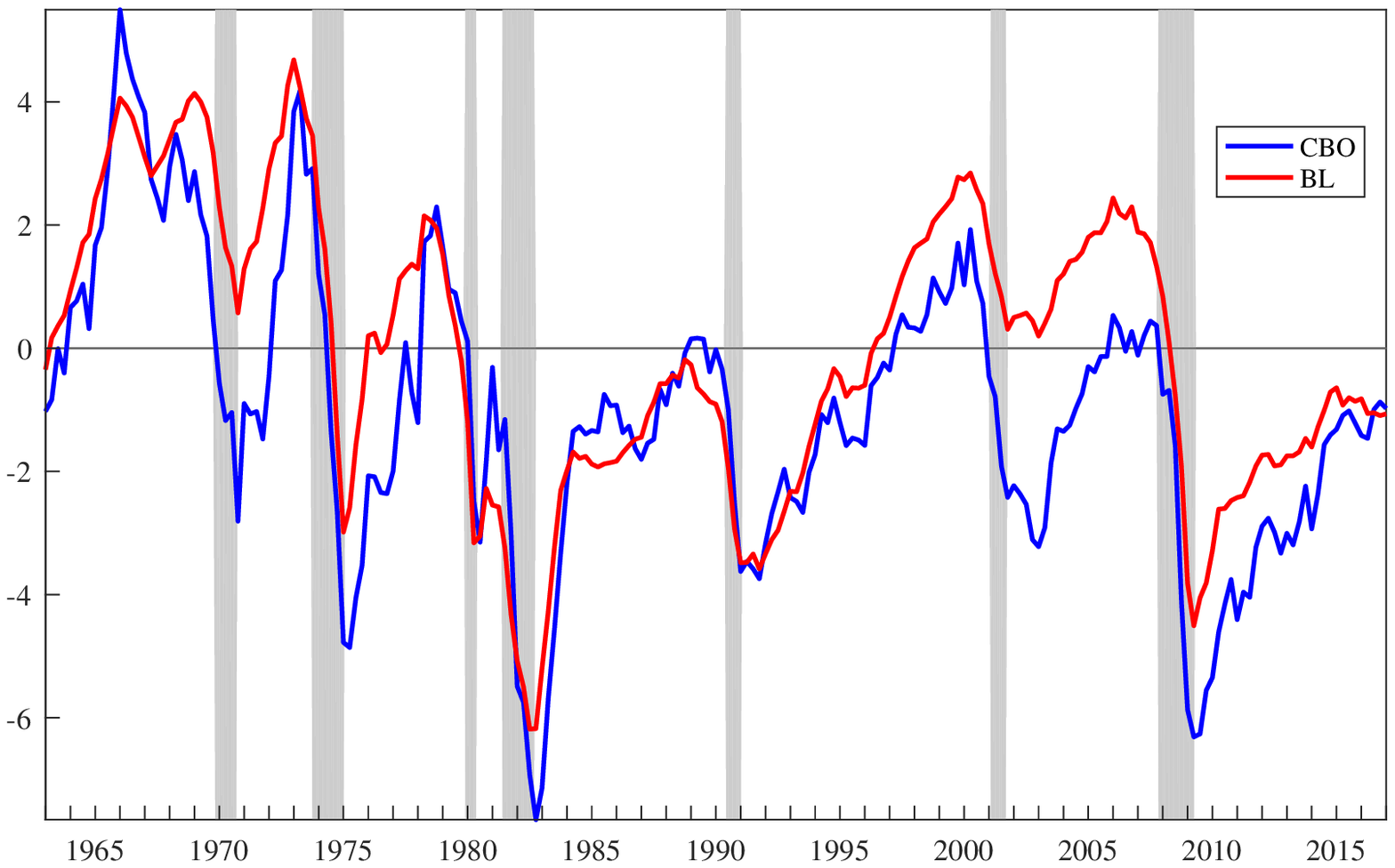} &
\includegraphics[width=.475\textwidth,trim=0cm .5cm 0cm 0cm,clip]{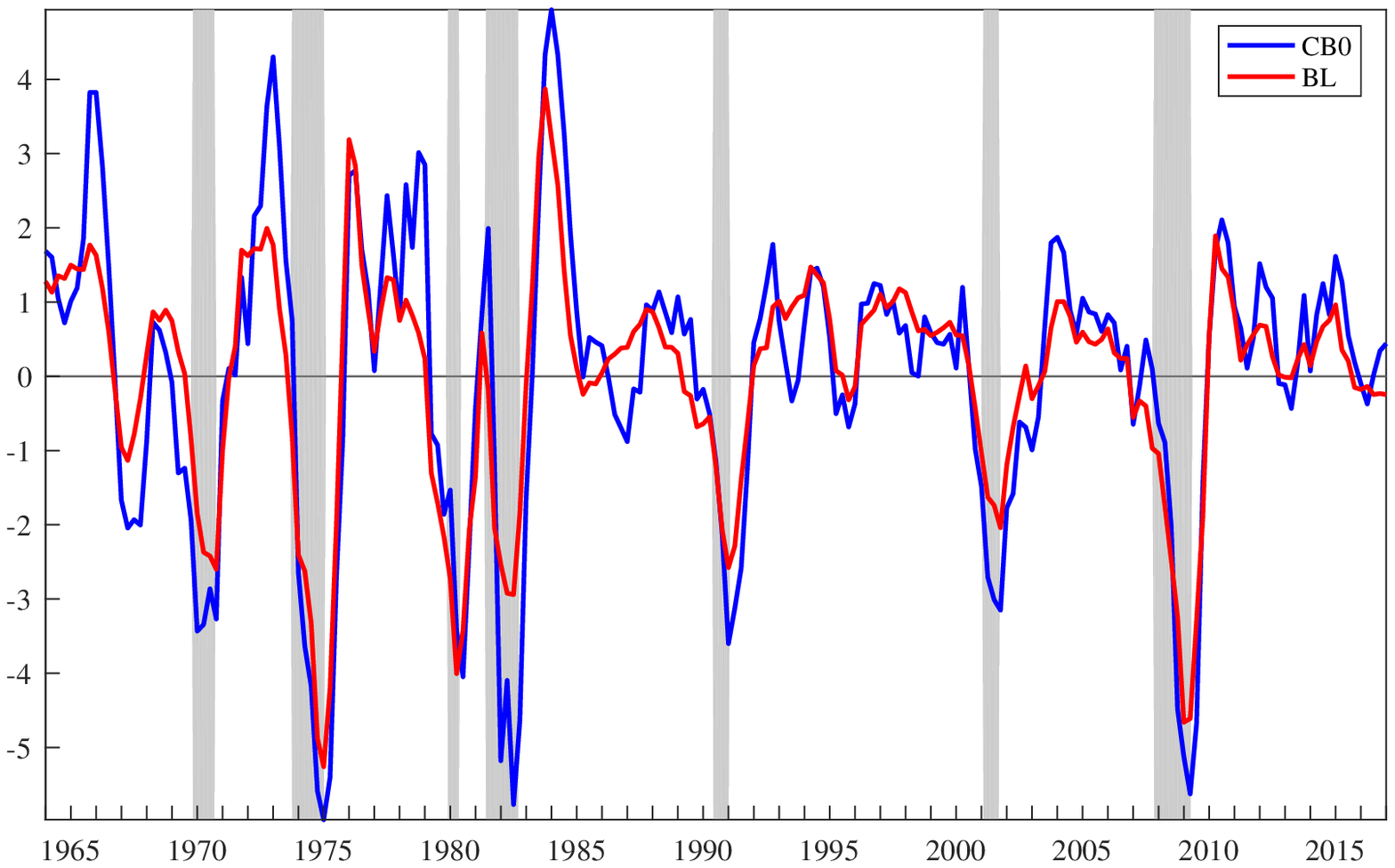}\\
\end{tabular}
\begin{tabular}{p{.98\textwidth}} \scriptsize
The left plot shows the level of the output gap estimated by the CBO (blue line) together with our estimate (red line). The right plot shows the 4-quarter percentage change of the output gap. 
\end{tabular}
\end{figure}

In Figure \ref{fig_OG}, we compare our measure of the output gap (red line) with the one produced by the CBO (blue line), where the left plot  shows the level of the output gap, while the right plot shows the 4-quarter percentage change of the output gap. The main result emerging from Figure \ref{fig_OG} is that our estimate of the output gap is remarkably similar to that of the CBO. However, there are a few periods in which the two estimates diverge, among which the main one is from the late nineties to the financial crisis. In particular, while according to the CBO the level of the output gap was negative between 2001:Q1 and 2005:Q4, according to our estimate in that same period the output gap was positive --- on average 2\sfrac{1}{2} percentage points higher than estimated by the CBO. Therefore, according to our estimate the level of the output gap right before the great financial crisis in 2007:Q4 was 1.3\%, while according to the CBO was -0.7\%, and hence we estimate that the level of slack in the economy at the trough of the crisis in 2009:Q2 was -4.5\%, approximately 1\sfrac{3}{4} percentage points higher than estimated by the CBO. 

To conclude, let us emphasize that the fact that our estimate of the output gap is close to that of the CBO is a remarkable result, particularly so because our estimate of the output gap is very different from that of the CBO from both a technical and an interpretational point of view.~Indeed, while the CBO constructs the output gap so that its level has a specific economic meaning, our measure of the output gap is simply the transitory/stationary part of the common component of output --- i.e., that part of aggregate real output that will disappear in the long-run.\footnote{Notice that also for the CBO the output gap is assumed to revert to zero in the long-run as it imposes in its forecast that in 10 years the output gap will be zero --- see e.g. \citet*{CBO04}.}  Therefore, our output gap estimate provides different and complementary information on the cyclical position of the economy than that contained in the CBO estimate. In particular, our estimate of the output gap seems more suitable to answer the question ``which part of current growth is due to temporary factors?'', while the measure of the CBO is certainly more suitable as a gauge of inflation pressure. This can explain in part the divergence of the two estimates in the 2000s. This period is characterized by stable and low inflation --- on average core CPI inflation between 2001:Q1 and 2007:Q4 was approximately 2.1\%. Accordingly, the CBO estimates that slack is positive (i.e., the output gap negative). By contrast, our measure, which is not specifically affected by inflation, but it is more broadly influenced by the co-movement in the data, estimates that a part of the aggregate real output was transitory. This makes sense given that the years before the crisis were characterized by several factors that proved indeed transitory, such as the housing boom, a historically high share of sub-prime loan origination \citep{Haughwout2009}, and a large amount of equity withdrawal from housing \citep{HouseATM}. And, since our model includes a large number of variables, including housing indicators as well as loan and credit indicators, these transitory factors are captured by our model.

\section{Discussion and conclusions}\label{sec:Conclusions}
In this paper we disentangle long-run co-movements (common trends) from short-run co-movements (common cycles) in large datasets. To this end, we first estimate a non-stationary dynamic factor model by means of a Quasi Maximum Likelihood estimator based on the Expectation Maximisation algorithm, combined with the Kalman Filter and the Kalman Smoother estimators of the factors. We then disentangle common trends from common cycles by applying a non-parametric Trend-Cycle decomposition to the latent common factors and based on eigenanalysis of their long-run covariance. The asymptotic properties of this estimator are derived and discussed in the paper.

We estimate our model on a large panel of US quarterly macroeconomic time series with the goal of estimating the cyclical position of the economy and the observation error. After backing out the observation error, we show that our model naturally produces an estimate of aggregate real output, which we refer to as Gross Domestic Output (GDO). According to our estimate of GDO, since 2010 the US economy grew at a faster pace than measured by national account statistics.

We then use a Trend-Cycle decomposition to estimate the output gap. We compare our estimate of the output gap, which is entirely data-driven, with that produced by the Congressional Budget Office (CBO), which is instead based on theoretical economic models. It turns out that our estimate of the output gap is remarkably similar to that of the CBO except from the late nineties to the financial crisis, when our measure suggests that a greater part of the produced output was driven by transitory factors. 

There are a number of aspects of our model that we have not fully developed in our empirical analysis and that are left for future research. First, due to the use of the Kalman Filter, our factor estimator is in principle able to handle both mixed frequency and missing data \citep[e.g.][]{marianomurasawa03,JKVW2011,banburamodugno14} and, therefore, it can be used for real-time analysis \citep{Nowcasting}. This aspect is well-known to be particularly relevant when estimating the output gap, since as shown by \citet{orphanidesvannorden}, end-of-sample revisions of GDP are of the same order of magnitude as the gap itself. Second, the use of the Kalman Filter makes our model suitable for scenario and counterfactual analysis based on conditional forecasts \citep{banburagiannonelenza15}. Third, as shown in equation \eqref{eq:FTCW}, our model naturally produces a Trend-Cycle decomposition for each variable in the dataset, and therefore it is possible to estimate other policy-relevant indicators, such as the unemployment gap (in our framework, the cycle component of the unemployment rate) or trend inflation (in our framework, the trend component of core CPI or the core PCE price indexes).

Our approach has been so far deliberately entirely data driven, and we have been careful in imposing the least possible amount of restrictions to let the data speak freely. This approach has undeniably some important merits, as estimation of GDO seems to fit naturally in our framework, and the Trend-Cycle decomposition that we obtain for GDO is economically sensible.  However, we believe that imposing the statistical restrictions described in Section \ref{sec:StaticFM}, thus eliminating the miss-specification error when computing the Trend-Cycle decomposition, as well as imposing economically meaningful constraints, seems to be an essential step forward. Our view is that one way to proceed is to consider Bayesian estimation of the model, so that our economic and statistical knowledge of the data can be included by means of suitable priors. All this is the subject of our current research. 


\singlespacing
{\small{
\setlength{\bibsep}{.2cm}
\bibliographystyle{chicago}
\bibliography{BL_biblio}
}}

\setcounter{section}{0}%
\setcounter{subsection}{0}
\setcounter{equation}{0}
\setcounter{table}{0}
\setcounter{figure}{0}
\setcounter{footnote}{0}
\gdef\thesection{Appendix \Alph{section}}
\gdef\thesubsection{\Alph{section}.\arabic{subsection}}
\gdef\thefigure{\Alph{section}\arabic{figure}}
\gdef\theequation{\Alph{section}\arabic{equation}}
\gdef\thetable{\Alph{section}\arabic{table}}
\gdef\thefootnote{\Alph{section}\arabic{footnote}}

\clearpage
\section{Representation results}\label{sec:proofsRepresentation}

Hereafter, and throughout all appendices, we consider restriction R2 when $s=1$ as found empirically in Section \ref{sec:emp}. Therefore, $r=2q$.

\subsection{Proof of restriction R3}
For the dynamic factors consider the VECM(2)
\beq\label{vecmdyn}
\Delta \bm f_t =  - \mbf a\mbf b' \bm f_{t-3} + \bm\Gamma_1 \Delta \bm f_{t-1}+ \bm\Gamma_2 \Delta \bm f_{t-2} +\mbf u_t,
\eeq
where $\mbf a$ and $\mbf b$ are $q\times d$ and for simplicity we consider just the case of two lags since this will imply a VECM(1) and therefore a VAR(2) for the static factors as implemented in \eqref{eq:SS2}. 
 
\medskip
\noindent
First assume that in R1 we have $\mbf K=\mbf I_r$. Our aim is then to find the correct VECM representation for $\mbf F_t=(\bm f_t'\;\bm f_{t-1}')'$ when the VECM in \eqref{vecmdyn}, and restrictions R1 and R2 hold. Since we model $\mbf F_t$ as a VAR(2) we know that we must have a VECM(1) with reduced rank innovations by R1, hence
\beq\label{vecmsta}
\Delta \mbf F_t = -\bm\alpha\bm\beta'\mbf F_t+ \mbf M\Delta\mbf F_{t-1} + \mbf H\mbf u_t,
\eeq
where $\bm\alpha$ and $\bm\beta$ are $r\times c$ with $c<r$ and  $\mbf H$ is $r\times q$. Moreover, from \citet{BLL1} we have  $d\le c\le (r-q+d)$. We are then interested in finding $c$, and the expressions of $\mbf M$, $\bm\alpha$, $\bm\beta$, and $\mbf H$ as functions of the parameters $\mbf a$, $\mbf b$, $\bm\Gamma_1$, and $\bm\Gamma_2$ in \eqref{vecmdyn}. Let us write $\bm\alpha=(\bm\alpha_1'\;\bm\alpha_2')'$ and $\bm\beta=(\bm\beta_1'\;\bm\beta_2')'$ where $\bm\alpha_1$, $\bm\alpha_2$, $\bm\beta_1$, $\bm\beta_2$ are all $q\times c$. We also denote as $\mbf M_{ij}$ for $i,j=1,2$ the four $q\times q$ blocks of $\mbf M$ and as $\mbf H_1$ and $\mbf H_2$ the two $q\times q$ blocks of $\mbf H$.
Following \citet{proietti97}, we define the $(2r+c)$-dimensional vector
\beq
\mbf G_t=\l(\ba{l}
\Delta \mbf F_t\\
\Delta \mbf F_{t-1}\\
\bm \beta'\mbf F_{t-2}
\ea
\r)=\l(\ba{l}
\Delta \bm f_t\\
\Delta \bm f_{t-1}\\
\Delta \bm f_{t-1}\\
\Delta \bm f_{t-2}\\
\bm \beta_1'\bm f_{t-2}+\bm \beta_2'\bm f_{t-3}
\ea
\r).\nn
\eeq
Then, the state-space form of \eqref{vecmsta} is given by
\begin{align}
\Delta \mbf F_t &= {\mbf Z} \mbf G_t,\nn\\
\mbf G_t &={\mbf T}\mbf G_{t-1} +{\mbf Z}'\mbf H\mbf u_t,\label{ss2b}
\end{align} 
with the $r\times (2r+c)$ matrix ${\mbf Z}=(\mbf I_r \;\mbf 0_r\;\mbf 0_{r\times c})$. Then, 
\beq\label{Ztilde}
{\mbf Z}'\mbf H=
\l(\ba{l}
\mbf I_r \\
\mbf 0_r\\
\mbf 0_{c\times r}
\ea
\r)\l(\ba{l}
\mbf H_1\\
\mbf H_2
\ea
\r)
=
\l(\ba{l}
\mbf H_1\\
\mbf H_2\\
\mbf 0_q\\
\mbf 0_q\\
\mbf 0_{c\times q}
\ea
\r).\nn
\eeq
and the $(2r+c)\times (2r+c)$ matrix ${\mbf T}$ is given by
\beq\label{Ttilde}
{\mbf T} = \l(\ba{lll}
\mbf M & -\bm\alpha\bm\beta' & -\bm\alpha\\
\mbf I_r & \mbf 0_r & \mbf 0_{r\times c}\\
\mbf 0_{c\times r} &\bm\beta'&\mbf I_c
\ea
\r)
=\l(\ba{ll | ll | l}
\mbf M_{11} & \mbf M_{12}& -\bm \alpha_1\bm \beta_1'& -\bm \alpha_1\bm \beta_2'& -\bm \alpha_1\\
\mbf M_{21} & \mbf M_{22}& -\bm \alpha_2\bm \beta_1'& -\bm \alpha_2\bm \beta_2'& -\bm \alpha_2\\
\hline
\mbf I_q & \mbf 0_q& \mbf 0_q & \mbf 0_q & \mbf 0_{q\times c}\\
\mbf 0_q & \mbf I_q & \mbf 0_q & \mbf 0_q & \mbf 0_{q\times c}\\
\hline 
\mbf 0_{c\times q} & \mbf 0_{c\times q} & \bm \beta_1'& \bm \beta_2' & \mbf I_{c}\\
\ea
\r).\nn
\eeq
Now using these definitions into \eqref{ss2b} we have five $q$-dimensional equations. The first one is
\begin{align}
\Delta \bm f_t 
&=\mbf M_{11} \Delta \bm f_{t-1}+ \mbf M_{12} \Delta \bm f_{t-2}-\bm\alpha_1\bm\beta_1' \bm f_{t-2}-\bm\alpha_1\bm\beta_2' \bm f_{t-3}+\mbf H_1\mbf u_t,\nn
\end{align}
which is equivalent to \eqref{vecmdyn} when
\beq\label{map2}
\mbf M_{11}=\bm\Gamma_1,\quad \mbf M_{12}=\bm\Gamma_2, \quad \bm\alpha_1=\mbf a,\quad\bm\beta_1 =\mbf 0_{q\times c},\quad\bm\beta_2 =\mbf b,  \quad \mbf H_1=\mbf I_q,\quad c=d.
\eeq
The second equation is
\begin{align}
\Delta\bm f_{t-1} &=\mbf M_{21}\Delta\bm f_{t-1}+\mbf M_{22}\Delta \bm f_{t-2}-\bm\alpha_2\bm\beta_1'\Delta\bm f_{t-2} -\bm\alpha_2\bm\beta_2'\Delta\bm f_{t-3} -\bm\alpha_2\bm \beta_1'\bm f_{t-3}-\bm\alpha_2\bm \beta_2'\bm f_{t-4}+\mbf H_2\mbf u_t,\nn
\end{align}
from which we see that we must also have
\beq\label{map3}
\mbf M_{21}=\mbf I_q,\qquad \mbf M_{22}=\mbf 0_q, \qquad\bm \alpha_2 =\mbf 0_{q\times c}, \qquad \mbf H_2=\mbf 0_q.
\eeq
Under \eqref{map2} and \eqref{map3} the third, fourth and fifth equation in \eqref{ss2b} are just identities.

\medskip
\noindent
By imposing these restrictions we have the mapping between the VECM(1) for $\mbf F_t$ in \eqref{vecmsta} and the VECM(2) for $\bm f_t$ in \eqref{vecmdyn}
\beq\label{map4}
\mbf M = \l(\ba{ll}
\bm\Gamma_1 &\bm\Gamma_2\\
\mbf I_q &\mbf 0_q
\ea
\r),\qquad 
\bm\alpha = \l(\ba{l}
\mbf a\\
\mbf 0_{q\times d}
\ea
\r),\qquad
\bm\beta=\l(\ba{l}
\mbf 0_{q\times d}\\
\mbf b
\ea
\r), \qquad \mbf H=\l(\ba{c}
\mbf I_q\\
\mbf 0_q
\ea
\r).
\eeq
If we now consider a generic $\mbf K$ in R1, then \eqref{vecmsta} holds for $\mbf F_t= \mbf K(\bm f_t'\; \bm f_{t-1}')'$ and 
\eqref{map4} becomes
\beq\label{map5}
\mbf M = \mbf K\l(\ba{ll}
\bm\Gamma_1 &\bm\Gamma_2\\
\mbf I_q &\mbf 0_q
\ea
\r)\mbf K^{-1},\qquad 
\bm\alpha = \mbf K\l(\ba{l}
\mbf a\\
\mbf 0_{q\times d}
\ea
\r),\qquad
\bm\beta={\mbf K^{-1}}'\l(\ba{l}
\mbf 0_{q\times d}\\
\mbf b
\ea
\r), \qquad \mbf H=\mbf K\l(\ba{c}
\mbf I_q\\
\mbf 0_q
\ea
\r).\nn
\eeq
The cointegration rank $c$ of $\mbf F_t$ is given by $\mathsf{rk}(\bm\alpha\bm\beta')=d$.

\subsection{Reduced and structural form of the state-space representation}
Consider \eqref{eq:SS1}-\eqref{eq:SS2} written in matrix notation and using the companion form of the VAR
\begin{align}
\mbf x_{t}&=\bm\Lambda \mbf F_t + \bm\xi_{t},\label{eq:kf1_app}\\
\l(\ba{l}
\mbf F_t\\
\mbf F_{t-1}
\ea
\r)
&=\l(
\ba{cc}
\mbf A_1&\mbf A_2\\
\mbf I_r &\mbf 0_r
\ea
\r)
\l(\ba{c}
\mbf F_{t-1}\\
\mbf F_{t-2}
\ea
\r)
+\l(\ba{c}
\mbf H\\
\mbf 0_{r\times q}
\ea
\r)
\mbf u_t,\label{eq:kf2_app}
\end{align}
with $\bm\Lambda=(\bm\lambda_1\cdots \bm\lambda_n)'$ the $n\times r$ loadings matrix.  We call \eqref{eq:kf1_app}-\eqref{eq:kf2_app} the reduced form of the model. Similarly consider the structural form, where, for convenience, in the VAR we write twice the same equation: 
\begin{align}
\mbf x_{t} &=  \mbf B_0 \bm f_t+\mbf B_1\bm f_{t-1}+ \bm\xi_{t},\label{eq:gdfm1_app}\\
\l(\ba{c}
\bm f_{t}\\
\bm f_{t-1}\\
\bm f_{t-1}\\
\bm f_{t-2}
\ea
\r)&=
\l(\ba{cccc}
\bm\Pi_1&\bm\Pi_2&\mbf 0_q &\bm\Pi_3\\
\mbf I_q &\mbf 0_q&\mbf 0_q&\mbf 0_q\\
\mbf I_q &\mbf 0_q&\mbf 0_q&\mbf 0_q\\
\mbf 0_q &\mbf I_q&\mbf 0_q&\mbf 0_q\\
\ea
\r)
\l(\ba{c}
\bm f_{t-1}\\
\bm f_{t-2}\\
\bm f_{t-2}\\
\bm f_{t-3}
\ea
\r)+\l(\ba{c}
\mbf I_q\\
\mbf 0_q\\
\mbf 0_q\\
\mbf 0_q
\ea
\r)
\mbf u_t,\label{eq:gdfm2_app}
\end{align}
where $\mbf B_0=(\bm b_{01}\cdots \bm b_{0n})'$, $\mbf B_1=(\bm b_{11}\cdots \bm b_{1n})$ are both $n\times q$. Because of R1 there exists an invertible $r\times r$ matrix $\mbf K$ such that
\begin{align}
\mbf F_t = \mbf K (\bm f_t'\,\bm f_{t-1}')', &\qquad  (\bm f_t'\,\bm f_{t-1}')'= \mbf K^{-1}\mbf F_t,\label{eq:R1_app}\\
\bm\Lambda = (\mbf B_0\,\mbf B_1)\mbf K^{-1}, &\qquad  (\mbf B_0\,\mbf B_1)=\bm\Lambda \mbf K.\label{eq:R1_app_2}
\end{align}
By comparing \eqref{eq:kf1_app}-\eqref{eq:kf2_app} with \eqref{eq:gdfm1_app}-\eqref{eq:gdfm2_app}  and using \eqref{eq:R1_app}-\eqref{eq:R1_app_2}, we have the parameters of the reduced form 
\begin{align}
\mbf A_1 =\mbf K\l(\ba{cc}
\bm\Pi_1&\bm\Pi_2\\
\mbf I_q&\mbf 0_q
\ea\r)\mbf K^{-1},\qquad \mbf A_2 =\mbf K\l(\ba{cc}
\mbf 0_q&\bm\Pi_3\\
\mbf 0_q&\mbf 0_q
\ea\r)\mbf K^{-1},\qquad
\mbf H &= \mbf K\l(\ba{c}
\mbf I_q\\
\mbf 0_q
\ea\r).\label{eq:R1_app_3}
\end{align}
The relations \eqref{eq:R1_app}, \eqref{eq:R1_app_2} and \eqref{eq:R1_app_3} are used throughout the following. Moreover, since a VAR(2) of dimension $r$ can always be written as a VAR(1) of dimension $2r$, to avoid introducing further notation hereafter we consider the case of a VAR(1) for $\mbf F_t$, where $\mbf A\equiv \mbf A_1$. 

\subsection{Properties of the structural and reduced form of the linear system}
\begin{lem}\label{lem:control}
The structural model \eqref{eq:gdfm1_app}-\eqref{eq:gdfm2_app} is stabilizable and detectable.
\end{lem}

\noindent{\bf Proof.} Equations \eqref{eq:gdfm1_app}-\eqref{eq:gdfm2_app} define a linear system with $r=2q$ latent states $(\bm f_t'\;\bm f_{t-1}')'$. We say that a linear system is stabilizable if its unstable (non-stationary) states are controllable and all uncontrollable states are stable (see \citealp{AM79}, page 342), where stability is dictated by the eigenvalues of the matrix of VAR coefficients, which we denote as
\begin{align}
\widetilde{\mbf A}={\l(\ba{cc}
\bm\Pi_1&\bm\Pi_2\\
\mbf I_q &\mbf 0_q
\ea
\r)}
\end{align}
Because of cointegration, $\widetilde{\mbf A}$ has $(q-d)$ unit eigenvalues corresponding to $(q-d)$ unstable states. Moreover, $(\mbf I_q-\bm\Pi_1-\bm\Pi_2)=\mbf a\mbf b'$, where $\mbf a$ and $\mbf b$ have full column-rank $q\times d$ matrices, so that $\mathsf{rk}(\mbf a\mbf b')=d$. Define the $q\times (q-d)$ matrices $\mbf a_{\perp}$ and $\mbf b_{\perp}$ such that
$\mbf a_{\perp}'\mbf a=\mbf b_{\perp}'\mbf b=\mbf 0_{(q-d)\times d}$. Then, since $\mathsf{rk}(\mbf a_{\perp}'\mbf I_q)=(q-d)$, the unstable states are controllable because they satisfy the Popov-Belevitch-Hautus rank test (see \citealp{franchi}, Theorem 2.1, and \citealp{AM07}, Corollary 6.11, page 249).

\medskip
\noindent
Now, by looking at \eqref{eq:gdfm2_app}, we see that $\widetilde{\mbf A}$ has also $(r-q+d)=(q+d)$ eigenvalues which are smaller than one in absolute value. Of these $q$ correspond to states which are uncontrollable because they are not driven by any shock, but are also stable since have no dynamics (see the second equation in \eqref{eq:gdfm2_app}). The remaining $d$ states follow a stable VAR, hence are controllable.

\medskip
\noindent
Similarly, we say that a linear system is detectable if its unstable states are observable and all unobservable states are stable (see \citealp{AM79}, page 342). First, notice that $\mathsf{rk}(\mbf B_0)=q$ and $\mathsf{rk}(\mbf B_1)=q$ because of C2 and \eqref{eq:R1_app_2}, therefore $\mathsf{rk}(\mbf B_0 \mbf b_{\perp})=(q-d)$ and $\mathsf{rk}(\mbf B_1 \mbf b_{\perp})=(q-d)$, which implies that the unstable states are observable because they satisfy the Popov-Belevitch-Hautus rank test (see \citealp{franchi}, Theorem 2.1, and \citealp{AM07}, Corollary 6.11, page 249). Since $\mbf B_0$ and $\mbf B_1$ have full column-rank there are no unstable unobservable states. This completes the proof. $\Box$
 

\setcounter{footnote}{0}
\setcounter{equation}{0}
\setcounter{subsection}{0}

\section{Details of estimation}\label{sec:proofsDetails}
This appendix provides details on estimation of factors and parameters which are necessary to introduce the notation required in the  proofs in \ref{sec:proofsProposition}. The model considered is \eqref{eq:SS1}-\eqref{eq:SS3}, where for simplicity of exposition we consider a VAR(1) for the factors. As explained in the text without loss of generality we can assume $\bm\xi_t\sim I(0)$, thus considering as latent states only the $r$ static factors. 

\medskip
\noindent
Adding $I(1)$ idiosyncratic components as latent states does not increase the dimension of the parameter space but it increases the dimension of the latent states vector. However, since the idiosyncratic components are assumed to be orthogonal (see D2), and moreover they are orthogonal to the static factors (see A2), the results in \ref{sec:proofsProposition} can be generalized to this case by treating each new state separately.

\subsection{Expectation Maximization algorithm}
In what follows we denote the whole sample of observed data as $\bm{\mathcal X}_T:= (\mbf x_1\cdots \mbf x_T)'$ and the whole history of the unknown factors as $\bm{\mathcal F}_T:= (\mbf F_1\cdots\mbf F_T)'$. Recall that the vector of parameters  is given by
${\bm\Theta}:=(\mbox{vec}({\bm\Lambda})' \;\mbox{vec}({\mbf A})'\; \mbox{vec}({\mbf H})'\; \;\mbox{diag}({\mbf R}))'$. 
To avoid heavier notation we use $\bm\Theta$ to indicate both a generic value of the parameters and the true value, whether we refer  to one or the other is either clearly implied by the context or explicitly stated.

\medskip
\noindent
The joint pdf of data and factors is denoted as $f(\bm{\mathcal X}_T,\bm{\mathcal F}_T;\bm\Theta)$ and the corresponding joint log-likelihood is denoted as $\ell(\bm{\mathcal X}_T,\bm{\mathcal F}_T;\bm\Theta):= \log f(\bm{\mathcal X}_T,\bm{\mathcal F}_T;\bm\Theta)$ and it is such that 
\beq\label{eq:LL1}
\ell(\bm{\mathcal X}_T,\bm{\mathcal F}_T;\bm\Theta)=\ell(\bm{\mathcal X}_T|\bm{\mathcal F}_T;\bm\Theta)+\ell(\bm{\mathcal F}_T;\bm\Theta), 
\eeq
where $\ell(\bm{\mathcal X}_T|\bm{\mathcal F}_T;\bm\Theta)$ is the log-likelihood of the data conditional on the factors and $\ell(\bm{\mathcal F}_T;\bm\Theta)$ is the marginal log-likelihood of the factors. Because of D2 and D3 all log-likelihoods are Gaussian and in particular
\begin{align}
\ell(\bm{\mathcal X}_T|\bm{\mathcal F}_T;\bm\Theta) &=-\frac{nT}2\log(2\pi)-\frac T2\log\det (\mbf R)-\frac 12\mathsf{tr}\l[(\bm{\mathcal X}_T-\bm{\mathcal F}_T\bm\Lambda')\mbf R^{-1}(\bm{\mathcal X}_T-\bm{\mathcal F}_T\bm\Lambda')'\r].\label{eq:loglikXcondF}
\end{align}
We first briefly review the steps of the EM algorithm, while in Section \ref{sec:EMML} we prove that the values of the parameters obtained at convergence of the EM algorithm converge to the QML estimator.

\subsubsection{Initialization}
The EM algorithm is initialised with estimated parameters 
\beq\label{initparam}
\wh{\bm\Theta}_0:=(\mbox{vec}(\wh{\bm\Lambda}_0)' \;\mbox{vec}(\wh{\mbf A}_0)'\; \mbox{vec}(\wh{\mbf H}_0)'\; \;\mbox{diag}(\wh{\mbf R}_0))'.
\eeq
These are obtained as follows. From the integration of the first $r$ principal components of $\Delta\mbf x_t$  we have an estimator of the factors, $\wt{\mbf F}_t$, and then of the loadings, $\wh{\bm\Lambda}_0$. 
The VAR parameters, $\wh{\mbf A}_0$ are obtained by fitting a VAR on the estimated factors $\widetilde{\mbf F}_t$  
and the columns of $\wh{\mbf H}_0$ are given by the $q$ leading eigenvectors of the covariance matrix of VAR residuals. Finally, the diagonal entries of $\wh{\mbf R}_0$, are obtained  as sample variances of $\wh{\xi}_{it,0}=x_{it}-\wh{\bm\Lambda}_0\wt{\mbf F}_t$. Consistency of these estimators is discussed in Section \ref{app:KFBLL}.

\subsubsection{E-step}
At iteration $k\ge 0$, given $\bm{\mathcal X}_T$ and an estimate of the parameters $\wh{\bm\Theta}_k$, we compute the expected log-likelihood as function of a generic value of the parameters $\bm\Theta$, where the expectation is computed with respect to the conditional distribution of $\bm{\mathcal F}_T$ given $\bm{\mathcal X}_T$ and when using $\wh{\bm\Theta}_k$:
\beq\label{eq:Estep}
\mathcal Q(\bm\Theta;\wh{\bm\Theta}_k) := \int_{\mathbb R^{r\times T}} \ell(\bm{\mathcal X}_T, \bm{\mathcal F}_T;\bm\Theta)
f(\bm{\mathcal F}_T|\bm{\mathcal X}_T;\wh{\bm\Theta}_k) \mathrm d \bm{\mathcal F}_T=\E_{\wh{\bm\Theta}_k} [\ell(\bm{\mathcal X}_T,\bm{\mathcal F}_T;\bm\Theta)|\bm{\mathcal X}_T].
\eeq
In the Gaussian case \eqref{eq:Estep} depends on the conditional mean of the factors and their conditional second moments, which are obtained with the KS when using the parameters $\wh{\bm\Theta}_k$ and are given by (see Section \ref{sec:KFKS} for details)
\begin{align}\label{eq:FKS_PKS}
{\mbf F}_{t|T,k} := \E_{\wh{\bm\Theta}_k}[\mbf F_t |\bm{\mathcal X}_T],\qquad{\mbf P}_{t|T,k} := \E_{\wh{\bm\Theta}_k}[(\mbf F_t-{\mbf F}_{t|T,k})(\mbf F_t-{\mbf F}_{t|T,k})' |\bm{\mathcal X}_T].
\end{align}

\subsubsection{M-step}
A new estimator of the parameters is obtained by maximising the expected log-likelihood over all possible values of the parameters:
\beq\label{eq:Mstep}
\wh{\bm\Theta}_{k+1} = \arg\!\!\!\!\!\max_{\bm\Theta\in\Omega\subseteq \mathbb R^Q}\; \mathcal Q(\bm\Theta;\wh{\bm\Theta}_k).
\eeq
Thus, maximizing the conditional expectation of \eqref{eq:loglikXcondF} and using \eqref{eq:FKS_PKS}, we have the loadings estimator
\begin{align}\label{eq:loadEM}
\wh{\bm\Lambda}_{k+1}&= \l(\sum_{t=1}^T \E_{\wh{\bm\Theta}_k}[ \mbf x_t\mbf F_t' |\bm{\mathcal X}_T]\r)\l(\sum_{t=1}^T \E_{\wh{\bm\Theta}_k}[ \mbf F_t\mbf F_t' |\bm{\mathcal X}_T]\r)^{-1}\nn\\
&=\l(\sum_{t=1}^T \mbf x_t {\mbf F}_{t|T,k}'\r)\l(\sum_{t=1}^T \l({\mbf F}_{t|T,k}{\mbf F}_{t|T,k}'+{\mbf P}_{t|T,k}\r)\r)^{-1}.\nn
\end{align}
Similarly we can obtain estimates of the other parameters $\wh{\mbf A}_{k+1}$ and $\wh{\mbf R}_{k+1}$ (see e.g. \citealp{banburamodugno14}, for their expressions). The columns of $\wh{\mbf H}_{k+1}$ are obtained as the $q$ leading eigenvectors of the matrix
\beq\label{eq:sigmaEM}
\wh{\bm\Sigma}_{k+1} = \frac 1 T\l(\sum_{t=1}^T \E_{\wh{\bm\Theta}_k}[ \mbf F_t\mbf F_t' |\bm{\mathcal X}_T] -\wh{\mbf A}_{k+1}\sum_{t=1}^T\E_{\wh{\bm\Theta}_k}[ \mbf F_t\mbf F_{t-1}' |\bm{\mathcal X}_T] \r),\nn
\eeq
which is an estimator of the covariance of the VAR residuals, and where the second expectation can also be computed from the output of the KS.

\subsubsection{Convergence}\label{sec:EMML}
Denote the QML estimator of the parameters as 
\beq\label{MLparam}
\wh{\bm\Theta}^*:=(\mbox{vec}(\wh{\bm\Lambda}^*)' \;\mbox{vec}(\wh{\mbf A}^*)'\; \mbox{vec}(\wh{\mbf H}^*)'\; \;\mbox{diag}(\wh{\mbf R}^*))',
\eeq
then by definition we have
\beq
\wh{\bm\Theta}^*= \arg\!\!\!\!\!\max_{\bm\Theta\in\Omega\subseteq \mathbb R^Q}\;\ell(\bm{\mathcal X}_T;\bm\Theta).
\eeq
where $\ell(\bm{\mathcal X}_T;\bm\Theta)$ is the log-likelihood of the data such that 
\beq\label{eq:LLX}
\ell(\bm{\mathcal X}_T;\bm\Theta) = \ell(\bm{\mathcal X}_T, \bm{\mathcal F}_T;\bm\Theta)-\ell(\bm{\mathcal F}_T|\bm{\mathcal X}_T;\bm\Theta),
\eeq 
where the first term on the rhs is given by \eqref{eq:LL1} and the second can be computed using the output of the KS for a given value of $\bm\Theta$.
Define the expectation
\begin{align}
\mathcal H(\bm\Theta;\wh{\bm\Theta}_k) &:=  \int_{\mathbb R^{r\times T}}\ell(\bm{\mathcal F}_T|\bm{\mathcal X}_T;\bm\Theta) f(\bm{\mathcal F}_T|\bm{\mathcal X}_T;\wh{\bm\Theta}_k) \mathrm d \bm{\mathcal F}_T=\E_{\wh{\bm\Theta}_k}[\ell(\bm{\mathcal F}_T|\bm{\mathcal X}_T;\bm\Theta)|\bm{\mathcal X}_T],\label{eq:H} 
\end{align}
and recall the definition of $\mathcal Q(\bm\Theta;\wh{\bm\Theta}_k)$ in the E-step in \eqref{eq:Estep}.
Since the lhs of \eqref{eq:LLX} does not depend on $\bm{\mathcal F}_T$, by taking its expectation with respect to the conditional distribution of $\bm{\mathcal F}_T$ given $\bm{\mathcal X}_T$ and when using $\wh{\bm\Theta}_k$, for any $\bm\Theta\in \Omega$, we have
\begin{align}
\ell(\bm{\mathcal X}_T;\bm\Theta) 
&=\mathcal Q(\bm\Theta;\wh{\bm\Theta}_k)-\mathcal H(\bm\Theta;\wh{\bm\Theta}_k).\label{eq:LLX2}
\end{align}
Now, by definition of Kullback-Leibler divergence, we have (see also Lemma 1 in \citealp{DLR77})
\begin{align}
\mathcal H(\wh{\bm\Theta}_{k+1};\wh{\bm\Theta}_k)&\leq \mathcal H(\wh{\bm\Theta}_k;\wh{\bm\Theta}_k).\label{eq:HHH}
\end{align}
Hence, from \eqref{eq:LLX2} and \eqref{eq:HHH}, for any $k$,
\[
\ell(\bm{\mathcal X}_T;\wh{\bm\Theta}_{k+1})- \ell(\bm{\mathcal X}_T;\wh{\bm\Theta}_{k})\geq \mathcal Q(\wh{\bm\Theta}_{k+1};\wh{\bm\Theta}_{k})- \mathcal Q(\wh{\bm\Theta}_k;\wh{\bm\Theta}_k)\geq 0
\]
where the last inequality is a consequence of the M-step in \eqref{eq:Mstep}.
This shows that the log-likelihood increases monotonically as $k$ increases. Moreover, since due to Gaussianity $\mathcal Q(\bm{\Theta};\bm{\Theta}')$ is continuous in $\bm{\Theta}$ and $\bm{\Theta}'$ and its gradient $\nabla_{\bm\Theta}\mathcal Q(\bm{\Theta};\bm{\Theta}')$ is continuous in $\bm{\Theta}$, then conditions for Theorems 1 and 2 and Corollary 1 in \citet{wu83} are satisfied and we have convergence of the log-likelihood to its unique maximum and of the parameters to the corresponding QML estimators
\begin{align}\label{eq:EMtoML}
&\lim_{k\to\infty}\ell(\bm{\mathcal X}_T;\wh{\bm\Theta}_{k})=\ell(\bm{\mathcal X}_T;\wh{\bm\Theta}^*),\qquad \lim_{k\to\infty}\wh{\bm\Theta}_{k} = \wh{\bm\Theta}^*.
\end{align}
The previous result holds in the limit $k\to\infty$, but in practice we can run the EM algorithm only for a finite number of iterations $k_{\max}$. Define, for any $k$,
\[
\Delta\ell_k = \frac{\vert \ell(\bm{\mathcal X}_T,{\bm{\mathcal F}}_{T,k+1};\wh{\bm\Theta}_{k+1})
-\ell(\bm{\mathcal X}_T,{\bm{\mathcal F}}_{T,k};\wh{\bm\Theta}_{k})\vert}
{\vert \ell(\bm{\mathcal X}_T,{\bm{\mathcal F}}_{T,k+1};\wh{\bm\Theta}_{k+1})
\vert+\vert\ell(\bm{\mathcal X}_T,{\bm{\mathcal F}}_{T,k};\wh{\bm\Theta}_{k})\vert},
\]
where ${\bm{\mathcal F}}_{T,k}:=({\mbf F}_{1|T,k} \cdots {\mbf F}_{T|T,k})'$. We say that the algorithm has converged at iteration $k^*<k_{\max}$ according to the following rule, which is defined for a given threshold $\eta$,
\beq
\Delta\ell_{k^*}<\eta, \;\mbox{ but }\; \Delta\ell_{k^*-1}\ge \eta.\nn
\eeq
Once we find $k^*$, our estimator of the parameters is defined as $\wh{\bm\Theta}:= \wh{\bm\Theta}_{k^*}$.
The corresponding estimator of the factors is then defined as $\wh{\mbf F}_{t}:= {\mbf F}_{t|T,k^*}$, thus running the KS once last time using $\wh{\bm\Theta}_{k^*}$. The rate of convergence of $\wh{\bm\Theta}$ to $\wh{\bm\Theta}^*$ in \eqref{eq:EMtoML} is studied in Lemma \ref{lem:convEM} below.

\subsection{Kalman filter and Kalman smoother}\label{sec:KFKS}
For ease of notation assume to know the true parameter collected in the vector $\bm\Theta$. When using the KF-KS in the EM algorithm at a given iteration $k$, the factors' estimators given below are obtained by replacing $\bm\Theta$ with  $\wh{\bm\Theta}_k$ throughout this section. We denote the conditional expectation and covariance of the factors as
\beq\label{eq:condmean}
\mbf F_{t|s} := \E_{\bm\Theta}[\mbf F_t|\bm{\mathcal X}_s], \qquad \mbf P_{t|s} := \E_{\bm\Theta}[(\mbf F_t-\mbf F_{t|s})(\mbf F_t-\mbf F_{t|s})'|\bm{\mathcal X}_s],
\eeq
where $\bm{\mathcal X}_s:=(\mbf x_1\cdots \mbf x_s)'$. Under Gaussianity (D2 and D3) these can be computed with the KF-KS. Specifically, when $s=t-1$ we have the optimal one-step-ahead prediction, when $s=t$ we have the optimal in-sample estimator, when $s=T$ we have the optimal smoother. The KF gives the first two cases while the KS gives the latter. 
In particular, we denote the KF-KS estimators respectively as: $\mbf F_{t|t}$ and $\mbf F_{t|T}$ when using the true value $\bm\Theta$ and as ${\mbf F}_{t|t,k}$ and ${\mbf F}_{t|T,k}$ when using $\wh{\bm\Theta}_k$ (see also \eqref{eq:FKS_PKS}).

\subsubsection{Forward iterations - Filtering}
For given initial conditions $\mbf F_{0|0}$ and $\mbf P_{0|0}$, the KF is based on the forward iterations for $t=1,\ldots, T$:
\begin{align}
&\mbf F_{t|t-1} = \mbf A \mbf F_{t-1|t-1},\label{eq:pred1}\\
&\mbf P_{t|t-1} = \mbf A\mbf P_{t-1|t-1} \mbf A' + \mbf H\mbf H',\label{eq:pred2}\\
&\mbf F_{t|t} =\mbf F_{t|t-1}+\mbf P_{t|t-1}\bm\Lambda'(\bm\Lambda\mbf P_{t|t-1}\bm\Lambda'+\mbf R)^{-1}(\mbf x_t-\bm\Lambda\mbf F_{t|t-1}),\label{eq:up1}\\
&\mbf P_{t|t} =\mbf P_{t|t-1}-\mbf P_{t|t-1}\bm\Lambda'(\bm\Lambda\mbf P_{t|t-1}\bm\Lambda'+\mbf R)^{-1}\bm\Lambda\mbf P_{t|t-1}.\label{eq:up2}
\end{align}
Moreover, by combining \eqref{eq:pred2} and \eqref{eq:up2}, we obtain  the Riccati difference equation
\beq\label{eq:riccati}
\mbf P_{t+1|t}-\mbf A\mbf P_{t|t-1}\mbf A'+\mbf A\mbf P_{t|t-1}\bm\Lambda'(\bm\Lambda\mbf P_{t|t-1}\bm\Lambda'+\mbf R)^{-1}\bm\Lambda\mbf P_{t|t-1}\mbf A'=\mbf H\mbf H'.
\eeq
The KF is started with given values of ${\mbf F}_{0|0}$ and ${\mbf P}_{0|0}$. The latter can be obtained with a diffuse prior run for $t<0$ (see \citealp{koopman97}, and \citealp{KD00}, for details). 

\subsubsection{Backward iterations - Smoothing}
The KS is then based on the backward iterations for $t=T,\ldots, 1$:
\begin{align}
\mbf F_{t|T} &=\mbf F_{t|t}+\mbf P_{t|t}\mbf A'\mbf P_{t+1|t}^{-1}(\mbf F_{t+1|T}-\mbf F_{t+1|t}),\label{eq:KS1}\\
\mbf P_{t|T}&=\mbf P_{t|t} + \mbf P_{t|t} \mbf A' \mbf P_{t+1|t}^{-1}
(\mbf P_{t+1|T}-\mbf P_{t+1|t})\mbf P_{t+1|t}^{-1} \mbf A \mbf P_{t|t}.\label{eq:KS2}
\end{align}
The KS iterations in \eqref{eq:KS1} require $T$ inversions of $\mbf P_{t|t-1}$ and in the singular case $r>q$ these matrices are likely to be singular (see also Lemma \ref{lem:steady3}). There are two possible solutions to this problem. \citet{KA83} suggest to use a generalized inverse of $\mbf P_{t|t-1}$, like the Moore-Penrose one. Alternatively, it can be proved that \eqref{eq:KS1} can be written in an equivalent way, which does not require matrix inversion, and which is defined by the backward iterations for $t=T,\ldots, 1$:
\begin{align}
&\mbf F_{t|T}=\mbf F_{t|t-1}+\mbf P_{t|t-1}\mbf r_{t-1},\label{eq:KS3}\\
&\mbf r_{t-1}=\bm\Lambda'(\bm\Lambda\mbf P_{t|t-1}\bm\Lambda'+\mbf R)^{-1}(\mbf x_t-\bm\Lambda\mbf F_{t|t-1})+\mbf L'_t\mbf r_t,\label{eq:KS4}\\
&\mbf P_{t|T}=\mbf P_{t|t-1}-\mbf P_{t|t-1}\mbf N_{t-1}\mbf P_{t|t-1},\label{eq:KS5}\\
&\mbf N_{t-1}=\bm\Lambda'(\bm\Lambda\mbf P_{t|t-1}\bm\Lambda'+\mbf R)^{-1}\bm\Lambda+\mbf L_t'\mbf N_t\mbf L_t,\label{eq:KS6}\\
&\mbf L_t= \mbf A-\mbf A \mbf P_{t|t-1} \bm\Lambda' (\bm\Lambda\mbf P_{t|t-1}\bm\Lambda'+\mbf R)^{-1} \bm\Lambda,\label{eq:KS7}
\end{align}
where $\mbf r_T=\mbf 0_{r\times 1}$, $\mbf N_T=\mbf 0_{r}$ and by consturction $\mbf A\mbf P_{t|t}=\mbf L_t\mbf P_{t|t-1}$ (see also \citealp{DK01}, pp.70-73). Although numerically no appreciable differences emerge with respect to the chosen method, \eqref{eq:KS3}-\eqref{eq:KS7} are particularly useful for our proofs.

\section{Consistency of the EM algorithm}\label{sec:proofsProposition}
\subsection{Preliminary results}
\begin{lem}\label{lem:wood}
For $m< n$, and given symmetric positive definite matrices $\bm A$ of dimension $m\times m$, $\bm B$ of dimension $n\times n$, and for $\bm C$ of dimension $n\times m$ with full column-rank, the following holds
\beq\label{statement}
\bm A \bm C' (\bm C\bm A\bm C'+\bm B)^{-1} = (\bm A^{-1}+\bm C'\bm B^{-1}\bm C)^{-1}\bm C'\bm B^{-1}.
\eeq
\end{lem}

\noindent{\bf Proof.} 
Recall the Woodbury forumla
\begin{align}\label{woodbury}
(\bm C\bm A\bm C'+\bm B)^{-1}=\bm B^{-1}-\bm B^{-1}\bm C(\bm A^{-1}+\bm C'\bm B^{-1}\bm C)^{-1}\bm C'\bm B^{-1}.
\end{align}
Denote $\bm D=(\bm A^{-1}+\bm C'\bm B^{-1}\bm C)^{-1}$ then 
from \eqref{woodbury} the lhs of \eqref{statement} is equivalent to
\begin{align}
\bm A\bm C'\l[\bm B^{-1}-\bm B^{-1}\bm C\bm D\bm C'\bm B^{-1}\r] = \bm A\l[\bm C'\bm B^{-1}-\bm C'\bm B^{-1}\bm C\bm D\bm C'\bm B^{-1}\r]=\bm A\l[\bm I-\bm C'\bm B^{-1}\bm C\bm D\r]\bm C'\bm B^{-1}.\nn
\end{align}
Then, \eqref{statement} becomes
\[
\bm A\l[\bm I-\bm C'\bm B^{-1}\bm C\bm D\r]\bm C'\bm B^{-1}=\bm D\bm C'\bm B^{-1},
\]
or equivalently multiplying both sides on the right by $\bm B\bm C(\bm C'\bm C)^{-1}$ 
\beq\label{statement2}
\bm A\l[\bm I-\bm C'\bm B^{-1}\bm C\bm D\r]=\bm D.
\eeq
Now notice that
\begin{align}
\bm D = (\bm A^{-1}+\bm C'\bm B^{-1}\bm C)^{-1}&= \bm A(\bm I+\bm A\bm C'\bm B^{-1}\bm C)^{-1}.\label{D1}
\end{align}
Substituting \eqref{D1} in \eqref{statement2} and multiplying both sides on the left by $\bm A^{-1}$
\[
\l[\bm I-\bm C'\bm B^{-1}\bm C\bm A(\bm I+\bm A\bm C'\bm B^{-1}\bm C)^{-1}\r]=(\bm I+\bm A\bm C'\bm B^{-1}\bm C)^{-1}.
\]
Multiplying both sides on the right by $(\bm I+\bm A\bm C'\bm B^{-1}\bm C)$ we have that \eqref{statement} is equivalent to
\beq\label{proof1}
\bm I+\bm A\bm C'\bm B^{-1}\bm C-\bm C'\bm B^{-1}\bm C\bm A = \bm I
\eeq
Therefore \eqref{statement} is correct provided that $\bm A\bm C'\bm B^{-1}\bm C=\bm C'\bm B^{-1}\bm C\bm A$ which is always true since both $\bm A$ and $\bm C'\bm B^{-1}\bm C$ are symmetric. $\Box$

\begin{lem}\label{lem:denom}
For $m<n$ with $m$ independent of $n$ and given
\ben
\item[(a)] an $m\times m$ matrix $\bm A$ symmetric and positive definite with $\mu_j^{A}\le M$ for $j=1,\ldots,m$; 
\item[(b)] an $n\times n$ matrix $\bm B$ symmetric and positive definite with $\mu_j^{B}\le M$ for $j=1,\ldots,n$;
\item[(c)] an $n\times m$ matrix $\bm C$ such that $\bm C'\bm C$ is positive definite with $\mu_j^{C'C}=M_j n$ for $j=1,\ldots ,m$;
\een
then the following holds
\[
(\bm A^{-1}+\bm C'\bm B^{-1}\bm C)^{-1}\bm C'\bm B^{-1}\bm C = \mbf I_m+ O(n^{-1}).
\]
\end{lem}

\noindent{\bf Proof.} First notice that for two matrices $\bm K$ and $\bm H$ we have
\begin{align}
(\bm H+ \bm K)^{-1}&= (\bm H + \bm K)^{-1}- \bm K^{-1} + \bm K^{-1}= (\bm H+ \bm K)^{-1}(\bm K - (\bm H + \bm K))\bm K^{-1}+ \bm K^{-1}\nn\\
&= (\bm H + \bm K)^{-1}(-\bm H)\bm K^{-1} + \bm K^{-1}= \bm K^{-1}- (\bm H + \bm K)^{-1}\bm H\bm K^{-1}.\label{eq:inverse}
\end{align}
Then setting $\bm K=\bm C'\bm B^{-1}\bm C$ and $\bm H=\bm A^{-1}$ from \eqref{eq:inverse} we have
\begin{align}\nn
(\bm A^{-1}+\bm C'\bm B^{-1}\bm C)^{-1}=(\bm C'\bm B^{-1}\bm C)^{-1}-(\bm A^{-1}+\bm C'\bm B^{-1}\bm C)^{-1}\bm A^{-1}(\bm C'\bm B^{-1}\bm C)^{-1}.
\end{align}
which implies
\beq\label{eq:abcinv}
(\bm A^{-1}+\bm C'\bm B^{-1}\bm C)^{-1}\bm C'\bm B^{-1}\bm C = \mbf I_m - (\bm A^{-1}+\bm C'\bm B^{-1}\bm C)^{-1}\bm A^{-1}.
\eeq
Now consider the second term on the rhs of \eqref{eq:abcinv}
\begin{align}
\Vert (\bm A^{-1}+\bm C'\bm B^{-1}\bm C)^{-1}\bm A^{-1}\Vert^2&\le\Vert (\bm A^{-1}+\bm C'\bm B^{-1}\bm C)^{-1}\Vert^2\,\Vert\bm A^{-1}\Vert^2\nn\\
&\le\Vert (\bm C'\bm B^{-1}\bm C)^{-1}\Vert^2\,(\mu_n^A)^{-1}\le \Vert (\bm C'\bm B^{-1}\bm C)^{-1}\Vert^2\, M_1^{-1},\label{eq:abcinv2}
\end{align}
where we use norm sub-additivity and the fact that by condition (a) $\bm A$ and $\bm A^{-1}$ are positive definite and therefore  $\mu_n^{A}\geq M_1>0$ and moreover $\mu_n^{A^{-1}}\geq M_2>0$ thus by Weyl's inequality
\[
\mu_n^{A^{-1}+ C' B^{-1} C} \ge \mu_n^{A^{-1}}+\mu_n^{ C' B^{-1} C}\ge \mu_n^{ C' B^{-1} C},
\]
therefore,
\[
\Vert (\bm A^{-1}+\bm C'\bm B^{-1}\bm C)^{-1}\Vert = (\mu_n^{A^{-1}+ C' B^{-1} C})^{-1}\le( \mu_n^{ C' B^{-1} C})^{-1} = \Vert (\bm C'\bm B^{-1}\bm C)^{-1}\Vert.
\]
Then, the first term on the rhs of \eqref{eq:abcinv2} is
\begin{align}
\Vert (\bm C'\bm B^{-1}\bm C)^{-1}\Vert^2\le \mathsf{tr}\l[(\bm C'\bm B^{-1}\bm C)^{-2}\r]=\sum_{j=1}^m \frac 1{{(\mu_j^{C'B^{-1}C})}^2}=O(n^{-2}).\label{eq:abcinv3}
\end{align} 
Indeed, the $m$ eigenvalues of $\bm C'\bm B^{-1}\bm C$ are also the $m$ non-zero eigenvalues of $\bm B^{-1/2}\bm C\bm C' \bm B^{-1/2}$, which are all $O(n)$ by conditions (b) and (c). 
By using  \eqref{eq:abcinv2} and \eqref{eq:abcinv3} in \eqref{eq:abcinv} we prove the Lemma. $\Box$ 

\subsection{Consistency of KF and KS using the true value of the parameters}
\begin{lem}\label{lem:steady1}
For the conditional covariance $\mbf P_{t|t-1}$ of the static factors given $\bm{\mathcal X}_{t-1}$, there exists a steady state for the reduced form denoted as ${\mbf P}$ solving the algebraic Riccati equation (ARE) and such that 
\[
{\mbf P}_{t|t-1} = {\mbf P}+ O(e^{-t}).
\]
Moreover, as $n\to\infty$,
\[
{\mbf P} = \mbf K\l(\ba{cc}
\mbf I_q &\mbf 0_q\\
\mbf 0_q&\mbf 0_q
\ea
\r)\mbf K' + O(n^{-1}) = \mbf H\mbf H'+ O(n^{-1}).
\]
\end{lem}

\noindent{\bf Proof.} Define $\widetilde{\mbf P}_{t|t-1}$ as the conditional covariance matrix for the vector $(\bm f_t'\,\bm f_{t-1}')'$ given $\bm{\mathcal X}_{t-1}$. Then, due to stabilizability and detectability proved in Lemma \ref{lem:control}, there exists a steady state for the structural model denoted as $\widetilde{\mbf P}$ solving the algebraic Riccati equation (ARE) and such that (see \citealp{AM79}, pp.76-77, and \citealp{harvey90}, pp.118-119)
\[
\widetilde{\mbf P}_{t|t-1} = \widetilde{\mbf P}+ O(e^{-t}).
\]
In presence of a diffuse prior its effect is limited to the first few periods, say $t_0$ (see \citealp{koopman97}), then the result above holds for $t>t_0$. The ARE for the structural model is then (see also \eqref{eq:riccati})
\beq\label{eq:AREstruct}
\widetilde{\mbf P}-\widetilde{\mbf A}\widetilde{\mbf P}\widetilde{\mbf A}'+\widetilde{\mbf A}\widetilde{ \mbf P}\widetilde{\mbf B}'(\widetilde{\mbf B}\widetilde{ \mbf P}\widetilde{\mbf B}'+\mbf R)^{-1}\widetilde{\mbf B}\widetilde{ \mbf P}\widetilde{\mbf A}'=\l(\ba{cc}
\mbf I_q&\mbf 0_q\\
\mbf 0_q&\mbf 0_q
\ea\r),
\eeq
where
\beq
\widetilde{\mbf A}=\l(\ba{cc}
\bm\Pi_1&\bm\Pi_2\\
\mbf I_q&\mbf 0_q
\ea
\r), \qquad \widetilde{\mbf B} =(\mbf B_0\,\mbf B_1).\label{AtildeBtilde}
\eeq
Now since the structural model has only $q$ controllable and observable states (see Lemma \ref{lem:control}) and $\widetilde{\mbf P}$ is the steady state covariance of those states, $\mathrm{rk}(\widetilde{\mbf P})=q$. Define as $\mbf V$ the $r\times r$ matrix of eigenvectors of $\widetilde{\mbf P}$ and as $\mbf D$ the $q\times q$ diagonal matrix of its non zero eigenvalues, then
\beq\label{eq:PWtilde}
\widetilde{\mbf P}=\mbf V\l(\ba{cc}
\mbf D&\mbf 0_q\\
\mbf 0_q&\mbf 0_q
\ea\r)\mbf V'=\mbf V\l(\ba{cc}
\mbf D^{1/2}&\mbf 0_q\\
\mbf 0_q&\mbf 0_q
\ea\r)\l(\ba{cc}
\mbf D^{1/2}&\mbf 0_q\\
\mbf 0_q&\mbf 0_q
\ea\r)\mbf V'=
\mbf W\l(\ba{cc}
\mbf I_q&\mbf 0_q\\
\mbf 0_q&\mbf 0_q
\ea\r)
\mbf W',
\eeq
with 
\[
\mbf W = \mbf V\l(\ba{cc}
\mbf D^{1/2}&\mbf 0_q\\
\mbf 0_q &\mbf I_q
\ea
\r).
\] 
Define $\mbf B_0^*$ and $\mbf B_1^*$ as the $n\times q$ matrices such that $\widetilde{\mbf B}\mbf W=(\mbf B_0^*\,\mbf B_1^*)$. Then, from \eqref{eq:PWtilde}
\beq\label{eq:PWtilde1}
\widetilde{\mbf B}\widetilde{\mbf P}\widetilde{\mbf B}'=\widetilde{\mbf B}\mbf W
\l(\ba{cc}
\mbf I_q&\mbf 0_q\\
\mbf 0_q&\mbf 0_q
\ea\r)\mbf W'\widetilde{\mbf B}' =
(\mbf B_0^*\,\mbf B_1^*)
\l(\ba{cc}
\mbf I_q&\mbf 0_q\\
\mbf 0_q&\mbf 0_q
\ea\r)
\l(\ba{c}
{\mbf B_0^*}'\\
{\mbf B_1^*}'
\ea\r)
=\mbf B_0^*\mbf {B_0^*}',
\eeq
and
\beq\label{eq:PWtilde2}
\l(\ba{cc}
\mbf I_q&\mbf 0_q\\
\mbf 0_q&\mbf 0_q
\ea\r)\mbf W'\widetilde{\mbf B}' =
\l(\ba{cc}
\mbf I_q&\mbf 0_q\\
\mbf 0_q&\mbf 0_q
\ea\r)
\l(\ba{c}
{\mbf B_0^*}'\\
{\mbf B_1^*}'
\ea\r)=
\l(\ba{c}
{\mbf B_0^*}'\\
\mbf 0_{q\times n}
\ea\r).
\eeq
From \eqref{eq:PWtilde}, \eqref{eq:PWtilde1}, \eqref{eq:PWtilde2}, Lemmas \ref{lem:wood} and \ref{lem:denom}, we have
\begin{align}
\l(\ba{cc}
\mbf I_q&\mbf 0_q\\
\mbf 0_q&\mbf 0_q
\ea\r)\mbf W'\widetilde{\mbf B}'(\widetilde{\mbf B}\widetilde{\mbf P}\widetilde{\mbf B}'+\mbf R)^{-1}\widetilde{\mbf B}\mbf W\l(\ba{cc}
\mbf I_q&\mbf 0_q\\
\mbf 0_q&\mbf 0_q
\ea\r)&=
\l(\ba{c}
{\mbf B_0^*}'({\mbf B_0^*}{\mbf B_0^*}'+\mbf R)^{-1}\\
\mbf 0_{q\times n}
\ea\r)(\mbf B_0^*\,\mbf B_1^*)\l(\ba{cc}
\mbf I_q&\mbf 0_q\\
\mbf 0_q&\mbf 0_q
\ea\r)\nn\\
&=
\l(\ba{cc}
({\mbf B_0^*}'\mbf R^{-1}{\mbf B_0^*}+\mbf I_q)^{-1}{\mbf B_0^*}'\mbf R^{-1}{\mbf B_0^*}&\mbf 0_{q}\\
\mbf 0_{q} & \mbf 0_{q}
\ea\r)\nn\\
&=\l(\ba{cc}
\mbf I_q +O(n^{-1})&\mbf 0_q\\
\mbf 0_q &\mbf 0_q
\ea
\r).
\label{eq:PWtilde4}
\end{align}
Notice that we can apply Lemma \ref{lem:denom} to the top left $q\times q$ block of \eqref{eq:PWtilde4} since: $\mbf I_q$ trivially satisfies condition (a), $\mbf R^{-1}$ satisfies condition (b) because of D2 and 
${\mbf B_0^*}'\mbf B_0^*$ satisfies condition (c). Indeed, from definition \eqref{eq:R1_app_2} we have
\[
\frac1 n \l(\ba{cc}
{\mbf B_0^*}'\\
{\mbf B_1^*}'
\ea
\r)(\mbf B_0^*\,\mbf B_1^*) =\mbf W'\mbf K'\frac{\bm\Lambda'\bm\Lambda}n\mbf K\mbf W,
\]
and because of assumption C2 the top left $q\times q$ block of this matrix which is $n^{-1}{\mbf B_0^*}'\mbf B_0^*$ has full column-rank for any $n$. By substituting \eqref{eq:PWtilde} and \eqref{eq:PWtilde4} into \eqref{eq:AREstruct} we have
\beq\label{Ptildesteady}
\widetilde{\mbf P} = \l(\ba{cc}
\mbf I_q &\mbf 0_q\\
\mbf 0_q&\mbf 0_q
\ea
\r) + O(n^{-1}).
\eeq
Now, notice that by construction $\mbf P_{t|t-1} =\mbf K\widetilde {\mbf P}_{t|t-1}\mbf K'$ and since $\mbf K$ is full-rank then also the reduced form system is stabilizable and detectable, thus it has a steady state $\mbf P$ such that 
\beq\label{Ptt1}
\mbf P_{t|t-1} = \mbf P + O(e^{-t}).
\eeq
Moreover, since $\mbf K$ does not depend on $t$ nor $n$, we have $\mbf P=\mbf K\widetilde {\mbf P}\mbf K'$ and the result follows directly from \eqref{Ptildesteady}. Last, from the definition of $\mbf H$ in \eqref{eq:R1_app_3} we have also $\mbf P=\mbf H\mbf H'$. $\Box$ 
 
\begin{lem}\label{lem:steady3}
For the static factors estimated via KF and KS when using the true value of the parameters $\bm\Theta$, under condition \eqref{eq:exprate} in the text, the following hold, for all $\bar t\le t \le T$ and as $n\to\infty$,
\begin{align}
&\sqrt n\, \Vert\mbf F_{t|t}-\mbf F_t\Vert = O_p(1),\nn\\
&\sqrt n\,\Vert\mbf F_{t|T}-\mbf F_t\Vert = O_p(1).\nn
\end{align}
\end{lem}

\noindent{\bf Proof.} By Lemma \ref{lem:steady1}, the conditional covariance $\mbf P_{t|t}$ of the static factors given $\bm{\mathcal X}_{t}$ has a steady state $\mbf S$ such that (see \eqref{eq:up2}) 
\beq\label{Pkf1}
\mbf S = \mbf P-\mbf P\bm\Lambda'(\bm\Lambda\mbf P\bm\Lambda'+\mbf R)^{-1}\bm\Lambda\mbf P.
\eeq
Then, notice that by Lemma \ref{lem:steady1} and \eqref{eq:R1_app_2}
\begin{align}
\mbf P\bm\Lambda' &= \mbf K\l(\ba{cc}
\mbf I_q &\mbf 0_q\\
\mbf 0_q&\mbf 0_q
\ea
\r)\mbf K'(\mbf K')^{-1}\l(\ba{c}\mbf B_0'\\\mbf B_1'\ea\r)+O(n^{-1})=\mbf K\l(\ba{c}
\mbf B_0'\\
\mbf 0_q
\ea
\r)+O(n^{-1}),\label{Pkf2}\\
\bm\Lambda\mbf P\bm\Lambda' &= (\mbf B_0\,\mbf B_1)\mbf K^{-1}\mbf K\l(\ba{cc}
\mbf I_q &\mbf 0_q\\
\mbf 0_q&\mbf 0_q
\ea
\r)\mbf K'(\mbf K')^{-1}\l(\ba{c}\mbf B_0'\\\mbf B_1'\ea\r)+O(n^{-1})=\mbf B_0\mbf B_0'+O(n^{-1}).\label{Pkf3}
\end{align}
Using \eqref{Pkf2} and \eqref{Pkf3} and by applying Lemmas \ref{lem:wood} and \ref{lem:denom} we have
\begin{align}
\mbf P\bm\Lambda'(\bm\Lambda\mbf P\bm\Lambda'+\mbf R)^{-1}\bm\Lambda\mbf P
&=
\mbf K\l(\ba{cc}
\mbf B_0'(\mbf B_0\mbf B_0'+\mbf R)^{-1}\mbf B_0&\mbf 0_q\\
\mbf 0_q&\mbf 0_q
\ea\r)
\mbf K'+O(n^{-1})\nn\\
&=\mbf K\l(\ba{cc}
(\mbf B_0'\mbf R^{-1}\mbf B_0+\mbf I_q)^{-1}\mbf B_0'\mbf R^{-1}\mbf B_0&\mbf 0_q\\
\mbf 0_q&\mbf 0_q
\ea\r)
\mbf K'+O(n^{-1})\nn\\
&=\mbf K\l(\ba{cc}
\mbf I_q+O(n^{-1})&\mbf 0_q\\
\mbf 0_q&\mbf 0_q
\ea\r)
\mbf K'+O(n^{-1}).\label{Pkf4}
\end{align}
Notice that we can apply Lemma \ref{lem:denom} to the top $q\times q$ block of \eqref{Pkf4} since: $\mbf I_q$ trivially satisfies condition (a), $\mbf R^{-1}$ satisfies condition (b) because of D2 and 
${\mbf B_0}'\mbf B_0$ satisfies condition (c) because of assumption C2 and definition \eqref{eq:R1_app_2} (see also \eqref{eq:PWtilde4} in the proof of Lemma \ref{lem:steady1}). By substituting \eqref{Pkf4} into \eqref{Pkf1} and because of Lemma \ref{lem:steady1}, we have
\beq\label{Pkf5}
\mbf S = \mbf K\l(\ba{cc}
\mbf I_q &\mbf 0_q\\
\mbf 0_q&\mbf 0_q
\ea
\r)\mbf K' + O(n^{-1}) - \mbf K\l(\ba{cc}
\mbf I_q&\mbf 0_q\\
\mbf 0_q&\mbf 0_q
\ea\r)
\mbf K'+O(n^{-1})=O(n^{-1}).
\eeq
By substituting \eqref{Ptt1} in \eqref{eq:up2}, from \eqref{Pkf1} we have
\beq\label{Pkf0}
\mbf P_{t|t} = \mbf S +O(e^{-t}).
\eeq
Therefore, by substituting \eqref{Pkf5} into \eqref{Pkf0} and letting $n=T^{\gamma}$ for $\gamma>0$ and $\bar t\equiv \bar t (T)$, because of \eqref{eq:exprate}  for $\bar t\le t \le T$ we have
\beq\label{Ptt}
\mbf P_{t|t} = O(n^{-1})+O(e^{-t})=O(n^{-1}).
\eeq
Now, let us consider $\mbf P_{t|T}$ defined in \eqref{eq:KS5}. From \eqref{eq:up2}
\beq\label{eq:B6}
\mbf P_{t|t-1} = \mbf P_{t|t}+\mbf P_{t|t-1}\bm\Lambda'(\bm\Lambda\mbf P_{t|t-1}\bm\Lambda'+\mbf R)^{-1}\bm\Lambda\mbf P_{t|t-1}.
\eeq
By substituting  \eqref{eq:B6} and \eqref{eq:KS6} in \eqref{eq:KS5} we have
\begin{align}
\mbf P_{t|T} &=\mbf P_{t|t}+\mbf P_{t|t-1}\mbf L_t'\mbf N_{t}\mbf L_t\mbf P_{t|t-1}.\label{eq:PtTeasy}
\end{align}
Since $\mbf N_t$ is function of $\mbf P_{t|t-1}$, because of Lemma \ref{lem:steady1}, it has a steady state $\mbf N$ such that $\Vert\mbf N\Vert=O(1)$ and
\beq\label{steadyN}
\mbf N_t = \mbf N + O(e^{-t}).
\eeq
Now, since $\mbf A\mbf P_{t|t}=\mbf L_t\mbf P_{t|t-1}$, using \eqref{Ptt} and \eqref{steadyN}, because of \eqref{eq:exprate} for $\bar t\le t \le T$ we have
\beq\label{eq:L}
\mbf P_{t|t-1}\mbf L_t'\mbf N_{t}\mbf L_t\mbf P_{t|t-1}=\mbf P_{t|t}\mbf A'\mbf N_t\mbf A\mbf P_{t|t}=O(n^{-2}).
\eeq
By using  \eqref{Ptt} and \eqref{eq:L} into \eqref{eq:PtTeasy}, for $\bar t\le t \le T$ we have
\beq\label{PtT}
\mbf P_{t|T} = O(n^{-1}).
\eeq
By the law of iterated expectations, for $\bar t\le t \le T$ we have (see also the definitions in \eqref{eq:condmean})
\begin{align}
&\E_{\bm\Theta}[(\mbf F_t-\mbf F_{t|t})(\mbf F_t-\mbf F_{t|t})'] = \E_{\bm\Theta}[\E_{\bm\Theta}[(\mbf F_t-\mbf F_{t|t})(\mbf F_t-\mbf F_{t|t})'|\bm {\mathcal X}_t] ]= \E_{\bm\Theta}[\mbf P_{t|t}]=O(n^{-1}),\nn\\
&\E_{\bm\Theta}[(\mbf F_t-\mbf F_{t|T})(\mbf F_t-\mbf F_{t|T})'] = \E_{\bm\Theta}[\E_{\bm\Theta}[(\mbf F_t-\mbf F_{t|T})(\mbf F_t-\mbf F_{t|T})'|\bm {\mathcal X}_T] ]= \E_{\bm\Theta}[\mbf P_{t|T}]=O(n^{-1}),\nn
\end{align}
which imply mean-square convergence of the KF and KS when the parameters are known and for all $\bar t\le t \le T$:
\begin{align}
&\E_{\bm\Theta}[\Vert\mbf F_t-\mbf F_{t|t}\Vert^2] = \sum_{j=1}^r\E_{\bm\Theta}[(F_{j,t}-F_{j,t|t})^2] = \mathsf{tr}\l\{\E_{\bm\Theta}[\mbf P_{t|t}]\r\} = O(n^{-1}),\nn\\
&\E_{\bm\Theta}[\Vert\mbf F_t-\mbf F_{t|T}\Vert^2] = \sum_{j=1}^r\E_{\bm\Theta}[(F_{j,t}-F_{j,t|T})^2] = \mathsf{tr}\l\{\E_{\bm\Theta}[\mbf P_{t|T}]\r\} = O(n^{-1}).\nn
\end{align}
The result follows from Chebychev's inequality. $\Box$

\subsection{Consistency of KF and KS using estimated parameters}\label{app:KFBLL}
\begin{lem}\label{lem:eststar}
Consider the QML estimator of the parameters $\wh{\bm\Theta}^*$ defined in \eqref{MLparam} and obtained using the true values of the static factors $\mbf F_t$,  then, as $T\to\infty$:
\begin{align}
&\sqrt T\,\Vert\wh{\bm\lambda}_{i}^{*}-\bm\lambda_i \Vert = O_p(1), \quad i=1,\ldots, n,\nn\\
&\sqrt T\,\Vert\wh{\mbf A}^*-\mbf A\Vert = O_p(1),\nn\\
&\sqrt T\,\Vert\wh{\mbf H}^*-\mbf H\Vert = O_p(1),\nn\\
&\sqrt T\,\vert[\wh{\mbf R}]_{ii}^*-[\mbf R]_{ii}\vert = O_p(1), \quad i=1,\ldots, n.\nn
\end{align}
\end{lem}

\noindent{\bf Proof.} 
The QML estimator of the loadings, for any $i=1,\ldots,n$, is given by
\begin{align}
\wh{\bm\lambda}_i^{*'}=\l(\sum_{t=1}^T  x_{it} {\mbf F}_{t}'\r)
\l(\sum_{t=1}^T {\mbf F}_{t}{\mbf F}_{t}'\r)^{-1}.
\end{align}
We know that $\mbf F_t$ is driven by $(q-d)$ common trends (see C2), therefore we can find an orthonormal linear basis of dimension $(q-d)$ such that the projection of $\mbf F_t$ onto this basis span the same space as the common trends. Collect the elements of this basis in the $r\times (q-d)$ matrix $\bm\gamma$, and denote as $\bm\gamma_\perp$ the $r\times(r-q+ d)$ matrix such that $\bm\gamma_\perp'\bm\gamma=\mbf 0_{(r-q+d)\times (q-d)}$. Then, consider the $r\times r$ linear  transformation 
\beq\label{DFZ}
\bm{\mathcal D}\mbf F_t = \l(\ba{c}
\bm\gamma'\mbf F_t\\
\bm\gamma_\perp'\mbf F_t\\
\ea
\r)=\l(\ba{c}
\mbf Z_{1t}\\
\mbf Z_{0t}
\ea\r),
\eeq
where $\mbf Z_{1t}$ has all $(q-d)$ components which are $I(1)$ while $\mbf Z_{0t}\sim I(0)$ and is of dimension $(r-q+d)$. Moreover, for $\mbf Z_{1t}$ we have the MA representation  
\beq\label{z1Q}
\Delta \mbf Z_{1t} = \mbf Q(L) \bm\zeta_t,
\eeq
$\bm\zeta_t\stackrel{w.n.}{\sim} (\mbf 0_{q-d},\bm\Sigma_{\zeta})$ with $\bm\Sigma_{\zeta}$ positive definite and ${\mbf Q}(L)$ is a $(q-d)\times (q-d)$ one-sided, infinite matrix polynomial with square-summable coefficients and $\mathsf{rk}({\mbf Q}(1))=(q-d)$. \medskip
 
\noindent 
Because of orthonormality $\bm{\mathcal D}'\bm{\mathcal D}=\mbf I_r$. Then, the corresponding transformation of the loadings gives $\bm\lambda_i'\bm{\mathcal D}'=(\bm\lambda_{i1}'\; \bm\lambda_{i0}')$ such that $x_{it}=\bm\lambda_{i1}'\mbf Z_{1t}+\bm\lambda_{i0}'\mbf Z_{0t}+\xi_{it}$ and we also have $\wh{\bm\lambda}_i^{*'}\bm{\mathcal D}'=(\wh{\bm\lambda}_{i1}^{*'}\;\wh{\bm\lambda}_{i0}^{*'})$. Recall that $\mbf Z_{1t}$ and $\mbf Z_{0t}$ are orthogonal by construction, then
we have
\begin{align}
&\l(\ba{c}
\wh{\bm\lambda}_{i1}^{*'}-\bm\lambda_{i1}'\\
\wh{\bm\lambda}_{i0}^{*'}-\bm\lambda_{i0}'
\ea\r)\label{lambdaDD}\\
&=\l(\ba{cc}
\l(\frac 1{T^2} \sum_{t=1}^T  \xi_{it} {\mbf Z}_{1t}'\r)
\l(\frac 1{T^2}\sum_{t=1}^T {\mbf Z}_{1t}{\mbf Z}_{1t}'\r)^{-1}&\mbf 0_{(q-d)\times (r-q+d)}\\
\mbf 0_{(r-q+d)\times (q-d)}&
\l(\frac 1{T} \sum_{t=1}^T  \xi_{it} {\mbf Z}_{0t}'\r)
\l(\frac 1{T}\sum_{t=1}^T {\mbf Z}_{0t}{\mbf Z}_{0t}'\r)^{-1}
\ea
\r).\nn
\end{align}
By Theorem 1 in \citet{penaponcela97,penaponcela04}, under C1 and C3, and from \eqref{z1Q}, as $T\to\infty$,
\beq
\frac 1{T^2}\sum_{t=1}^T {\mbf Z}_{1t}{\mbf Z}_{1t}'\Rightarrow \mbf Q(1)\bm\Sigma_{\zeta}^{1/2}\l(\int_0^1\bm{\mathcal W}(u)\bm{\mathcal W}(u)'\mathrm d u\right)\bm\Sigma_{\zeta}^{1/2}\mbf Q(1)',\label{PPAPP}
\eeq
where $\bm{\mathcal W}(\cdot)$ is a $(q-d)$-dimensional standard Wiener process. Thus this term is $O_p(1)$ and positive definite therefore invertible. Last, from \eqref{z1Q} we see that each component of $\Delta \mbf Z_{1t}$ has an MA representation with square summable coefficients ($\Delta \mbf Z_{1t}\sim I(0)$ by construction), therefore $\Var(t^{-1}Z_{1jt})= O(1)$ for any $j=1,\ldots, (q-d)$ and any $t=1,\ldots, T$. Thus, by using Gaussianity (see D2 and D3) and by A2 also independence of factors and idiosyncratic components, we can prove that, as $T\to\infty$,
\beq\label{casino}
\frac 1 {T^2}\sum_{t=1}^T\xi_{it} {\mbf Z}_{1t}' = O_p\l({T}^{-1}\r).
\eeq
Moreover, from C1 it is easy to see that ${\mbf Z}_{0t}$ has an MA representation with square summable coefficients (it is stationary) and because of A2 and C3 we have, as $T\to\infty$,
\begin{align}
&\frac 1{T} \sum_{t=1}^T  \xi_{it} {\mbf Z}_{0t}' = O_p(T^{-1/2}),\qquad \frac 1{T} \sum_{t=1}^T  \mbf Z_{0t} {\mbf Z}_{0t}' = \E[\mbf Z_{0t} {\mbf Z}_{0t}']+O_p(T^{-1/2}) = O_p(1).\label{lambda0cons1}
\end{align}
From \eqref{lambdaDD}, \eqref{PPAPP}, \eqref{casino}, and \eqref{lambda0cons1}, and since $\bm{\mathcal D}$ does not depend on $T$, we obtain the result.

\medskip
\noindent
Consider the VAR 
\beq
\bm{\mathcal D}\mbf F_t = (\bm{\mathcal D}\mbf A\bm{\mathcal D}')\bm{\mathcal D}\mbf F_{t-1}+\bm{\mathcal D}\mbf H\mbf u_t.
\eeq
such that $\bm{\mathcal D}\mbf H\mbf u_t=(\mbf e_{1t}'\;\mbf {e}_{0t}')'$ where $\mbf e_{1t}$ and $\mbf e_{0t}$ are white noise processes of dimensions $(q-d)$ and $(r-q+d)$, respectively. Then, similarly to \eqref{lambdaDD} we have
\begin{align}
&\bm{\mathcal D}(\wh{\mbf A}^*-\mbf A)\bm{\mathcal D}'=\label{DAD}\\
&=\l(\ba{cc}
\l(\frac 1{T^2} \sum_{t=1}^T  \mbf e_{1t} {\mbf Z}_{1t-1}'\r)
\l(\frac 1{T^2}\sum_{t=1}^T {\mbf Z}_{1t-1}{\mbf Z}_{1t-1}'\r)^{-1}&
\l(\frac 1{T} \sum_{t=1}^T  \mbf e_{0t} {\mbf Z}_{1t-1}'\r)
\l(\frac 1{T}\sum_{t=1}^T {\mbf Z}_{0t-1}{\mbf Z}_{0t-1}'\r)^{-1}\\
\l(\frac 1{T^2} \sum_{t=1}^T  \mbf e_{1t} {\mbf Z}_{0t-1}'\r)
\l(\frac 1{T^2}\sum_{t=1}^T {\mbf Z}_{1t-1}{\mbf Z}_{1t-1}'\r)^{-1}&
\l(\frac 1{T} \sum_{t=1}^T  \mbf e_{0t} {\mbf Z}_{0t-1}'\r)
\l(\frac 1{T}\sum_{t=1}^T {\mbf Z}_{0t-1}{\mbf Z}_{0t-1}'\r)^{-1}
\ea
\r)\nn
\end{align}
Then, using the fact that $\mbf e_{1t}$ and $\mbf e_{0t}$ are white noise, it can be shown that
\begin{align}
&\frac 1{T^2} \sum_{t=1}^T  \mbf e_{1t} {\mbf Z}_{1t-1}' = O_p(T^{-1}), && \frac 1{T^2} \sum_{t=1}^T  \mbf e_{1t} {\mbf Z}_{0t-1}' = O_p(T^{-1/2}),\label{altriOP}\\
&\frac 1{T} \sum_{t=1}^T  \mbf e_{1t} {\mbf Z}_{0t-1}' = O_p(T^{-1}),&&\frac 1{T} \sum_{t=1}^T  \mbf e_{0t} {\mbf Z}_{0t-1}' = O_p(T^{-1/2}).\nn
\end{align}
Substituting \eqref{PPAPP}, \eqref{lambda0cons1}, and \eqref{altriOP} into \eqref{DAD} and since $\bm{\mathcal D}$ does not depend on $T$, we have the result for the VAR parameters. Similar results can be proved for all other parameters. $\Box$

\begin{lem}\label{lem:steady_est1} 
Define the KF and KS estimators of the static factors and their conditional covariances when using the 
QML estimator of the parameters $\wh{\bm\Theta}^*$ as
\begin{align}
&\mbf F_{t|t}^* = \E_{\wh{\bm\Theta}^*}[\mbf F_t|\bm{\mathcal X}_t], \qquad
\mbf P_{t|t}^* = \E_{\wh{\bm\Theta}^*}[(\mbf F_t-\mbf F_{t|t}^*)(\mbf F_t-\mbf F_{t|t}^*)'|\bm{\mathcal X}_t],\nn\\
&\mbf F_{t|T}^* = \E_{\wh{\bm\Theta}^*}[\mbf F_t|\bm{\mathcal X}_T],\qquad \mbf P_{t|T}^* = \E_{\wh{\bm\Theta}^*}[(\mbf F_t-\mbf F_{t|T}^*)(\mbf F_t-\mbf F_{t|T}^*)'|\bm{\mathcal X}_T].\nn
\end{align}
Then, under condition \eqref{eq:exprate} in the text, the following hold, for all $\bar t\le t \le T$ and as $n,T\to\infty$,
\begin{align}
&\min(\sqrt n,\sqrt T)\, \Vert\mbf F_{t|t}^*-\mbf F_t\Vert = O_p(1),\nn\\
&\min(\sqrt n,\sqrt T)\,\Vert\mbf F_{t|T}^*-\mbf F_t\Vert = O_p(1),\nn\\
&\min( n,\sqrt T)\,\Vert\mbf P_{t|t}^*\Vert = O_p(1),\nn\\
&\min( n,\sqrt T)\,\Vert\mbf P_{t|T}^*\Vert = O_p(1).\nn
\end{align}
\end{lem}

\noindent{\bf Proof.} We start with three preliminary results. First, from \eqref{Ptt} in the proof of Lemma \ref{lem:steady3} we have
\beq\label{res1}
\mbf F_{t-1|t-1}-\mbf F_{t-1}= O(e^{-(t-1)/2})+O(n^{-1/2}).
\eeq
Second, because of Lemma \ref{lem:steady3} and \eqref{res1}, we have
\begin{align}\label{eq:OP1}
\l\Vert\frac{\mbf x_t}{\sqrt n}-\frac{{\bm\Lambda}\mbf F_{t|t-1}}{\sqrt n}\r\Vert&=\l\Vert\frac{\bm\Lambda\mbf F_t+\bm \xi_t}{\sqrt n}-\frac{{\bm\Lambda}\mbf F_{t|t-1}}{\sqrt n}\r\Vert\le \l\Vert\frac{{\bm\Lambda}}{\sqrt n}\r\Vert\l(\Vert \mbf A\Vert\;\Vert \mbf F_{t-1}-\mbf F_{t-1|t-1}\Vert+\Vert\mbf H\mbf u_t\Vert\r) +\l\Vert \frac{\bm\xi_t}{\sqrt n}\r\Vert\nn\\
&=\l\Vert\frac{{\bm\Lambda}}{\sqrt n}\r\Vert\;\l[\Vert \mbf A\Vert\;  \l(O(e^{-(t-1)/2})+O(n^{-1/2})\r)+\Vert\mbf H\mbf u_t\Vert\r] +\l\Vert \frac{\bm\xi_t}{\sqrt n}\r\Vert= O_p(1),
\end{align}
since $\mbf u_t\stackrel{w.n.}{\sim}(\mbf 0_q,\mbf I_q)$ and $\bm\xi_t\sim I(0)$. 

\medskip
\noindent
Third, from transformation \eqref{DFZ}, defined in the proof of Lemma \ref{lem:eststar}, $\bm{\mathcal D}\mbf F_t=(\mbf Z_{1t}'\;\mbf Z_{0t}')'$, we have $\Vert\mbf Z_{1t}\Vert=O_p(\sqrt T)$ and $\Vert\mbf Z_{0t}\Vert=O_p(1)$. Then, since $\bm{\mathcal D}'\bm{\mathcal D}=\mbf I_r$, as a consequence of Lemma \ref{lem:eststar} (see in particular \eqref{lambdaDD} and \eqref{DAD}), the following hold
\begin{align}
&(\wh{\mbf A}^*-\mbf A)\mbf F_t=\bm{\mathcal D}'\bm{\mathcal D}(\wh{\mbf A}^*-\mbf A)\bm{\mathcal D}'\bm{\mathcal D}\mbf F_t=O_p(T^{-1/2}),\label{DF1}\\
&(\wh{\bm{\lambda}}_i^{*'}-{\bm{\lambda}}_i')\mbf A\mbf F_t=(\wh{\bm{\lambda}}_i^{*'}-{\bm{\lambda}}_i')\bm{\mathcal D}'\bm{\mathcal D}\mbf A\bm{\mathcal D}'\bm{\mathcal D}\mbf F_t=O_p(T^{-1/2}), \quad i=1,\ldots, n.\label{DF2}
\end{align}
Now, we compare the KF iterations, \eqref{eq:pred1}-\eqref{eq:up2}, with those obtained when using $\wh{\bm\Theta}^*$:
\begin{align}
&\mbf F_{t|t-1}^* = \wh{\mbf A}^* \mbf F_{t-1|t-1}^*,\label{eq:pred1hat}\\
&\mbf P_{t|t-1}^* = \wh{\mbf A}^* \mbf P_{t-1|t-1}^* \wh{\mbf A}^{*'} + \wh{\mbf H}^*\wh{\mbf H}^{*'},\label{eq:pred2hat}\\
&\mbf F_{t|t}^* =\mbf F_{t|t-1}^*+\mbf P_{t|t-1}^*\wh{\bm\Lambda}^{*'}(\wh{\bm\Lambda}^*\mbf P_{t|t-1}^*\wh{\bm\Lambda}^{*'}+\wh{\mbf R}^*)^{-1}(\mbf x_t-\wh{\bm\Lambda}^*\mbf F_{t|t-1}^*),\label{eq:up1hat}\\
&\mbf P_{t|t}^* =\mbf P_{t|t-1}^*-\mbf P_{t|t-1}^*\wh{\bm\Lambda}^{*'}(\wh{\bm\Lambda}^*\mbf P_{t|t-1}^*\wh{\bm\Lambda}^{*'}+\wh{\mbf R}^*)^{-1}\wh{\bm\Lambda}^*\mbf P_{t|t-1}^*.\label{eq:up2hat}
\end{align}
From \eqref{eq:pred1hat} we have
\begin{align}
\mbf F_{t|t-1}^*-\mbf F_{t|t-1} &=\mbf A(\mbf F_{t-1|t-1}^*-\mbf F_{t-1|t-1})+(\wh{\mbf A}^*-\mbf A)(\mbf F_{t-1|t-1}^*-\mbf F_{t-1|t-1})+(\wh{\mbf A}^*-\mbf A)\mbf F_{t-1|t-1}\nn\\
&=\mbf A(\mbf F_{t-1|t-1}^*-\mbf F_{t-1|t-1})+(\wh{\mbf A}^*-\mbf A)(\mbf F_{t-1|t-1}^*-\mbf F_{t-1|t-1})+O_p(T^{-1/2}),\label{predFerror}
\end{align}
since
\begin{align}
(\wh{\mbf A}^*-\mbf A)\mbf F_{t-1|t-1}&=(\wh{\mbf A}^*-\mbf A)(\mbf F_{t-1|t-1}-\mbf F_{t-1})+(\wh{\mbf A}^*-\mbf A)\mbf F_{t-1}\nn\\
&=O_p(T^{-1/2})O(e^{-(t-1)/2})+O_p(T^{-1/2})O(n^{-1/2})+O_p(T^{-1/2}),\nn
\end{align}
because of Lemma \ref{lem:eststar}, \eqref{res1} and \eqref{DF1}. Similarly, from \eqref{eq:pred2hat} we have
\begin{align}
\mbf P_{t|t-1}^* &-\mbf P_{t|t-1} =\mbf A(\mbf P_{t-1|t-1}^*-\mbf P_{t-1|t-1})\mbf A'+(\wh{\mbf A}^*-\mbf A)(\mbf P_{t-1|t-1}^*-\mbf P_{t-1|t-1})\mbf A'\label{predPerror}\\
&\;+(\wh{\mbf A}^*-\mbf A)(\mbf P_{t-1|t-1}^*-\mbf P_{t-1|t-1})(\wh{\mbf A}^*-\mbf A)'+\mbf A(\mbf P_{t-1|t-1}^*-\mbf P_{t-1|t-1})(\wh{\mbf A}^*-\mbf A)'\nn\\
&\;+(\wh{\mbf A}^*-\mbf A)\mbf P_{t-1|t-1}\mbf A'+(\wh{\mbf A}^*-\mbf A)\mbf P_{t-1|t-1}(\wh{\mbf A}^*-\mbf A)'+\mbf A\mbf P_{t-1|t-1}(\wh{\mbf A}^*-\mbf A)'\nn\\
&\;+(\wh{\mbf H}^*-{\mbf H})\mbf H'+(\wh{\mbf H}^*-{\mbf H})(\wh{\mbf H}^*-{\mbf H})'+\mbf H(\wh{\mbf H}^*-{\mbf H})'\nn\\
&=\mbf A(\mbf P_{t-1|t-1}^*-\mbf P_{t-1|t-1})\mbf A'+(\wh{\mbf A}^*-\mbf A)(\mbf P_{t-1|t-1}^*-\mbf P_{t-1|t-1})\mbf A'\nn\\
&\;+(\wh{\mbf A}^*-\mbf A)(\mbf P_{t-1|t-1}^*-\mbf P_{t-1|t-1})(\wh{\mbf A}^*-\mbf A)'+\mbf A(\mbf P_{t-1|t-1}^*-\mbf P_{t-1|t-1})(\wh{\mbf A}^*-\mbf A)'+O_p(T^{-1/2}),\nn
\end{align}
since
\begin{align}
&(\wh{\mbf A}^*-\mbf A)\mbf P_{t-1|t-1}\mbf A'=\mbf A\mbf P_{t-1|t-1}(\wh{\mbf A}^*-\mbf A)'=O_p(T^{-1/2})O(e^{-(t-1)})+O_p(T^{-1/2})O(n^{-1}),\nn\\
&(\wh{\mbf A}^*-\mbf A)\mbf P_{t-1|t-1}(\wh{\mbf A}^*-\mbf A)'=O_p(T^{-1})O(e^{-(t-1)})+O_p(T^{-1})O(n^{-1}).\nn
\end{align}
because of Lemma \ref{lem:eststar} and \eqref{Ptt} in the proof of Lemma \ref{lem:steady3} and 
\beq
\wh{\mbf H}^*\wh{\mbf H}^{*'}-{\mbf H}{\mbf H}'=(\wh{\mbf H}^*-{\mbf H})\mbf H'+\mbf H(\wh{\mbf H}^*-{\mbf H})'+(\wh{\mbf H}^*-{\mbf H})(\wh{\mbf H}^*-{\mbf H})'=O_p(T^{-1/2}),\nn
\eeq
because of Lemma \ref{lem:eststar}. 

\medskip
\noindent
Define 
\[
{\bm{\mathcal K}}_t = \mbf P_{t|t-1}{\bm\Lambda}'({\bm\Lambda}\mbf P_{t|t-1}{\bm\Lambda}'+{\mbf R})^{-1}, 
\]
and analogously define $\wh{\bm{\mathcal K}}^*_t$ when using $ \mbf P_{t|t-1}^*$, $\wh{\bm\Lambda}^*$ and $\wh{\mbf R}^*$. From \eqref{eq:up1hat} we have
\begin{align}
\mbf F_{t|t}^*-\mbf F_{t|t}&=\mbf F_{t|t-1}^*-\mbf F_{t|t-1}+(\wh{\bm{\mathcal K}}^*_t-{\bm{\mathcal K}}_t)(\mbf x_t-\bm\Lambda \mbf F_{t|t-1})\nn\\
&\;+(\wh{\bm{\mathcal K}}^*_t-{\bm{\mathcal K}}_t)(\bm\Lambda \mbf F_{t|t-1}-\wh{\bm\Lambda}^* \mbf F_{t|t-1}^*)+{\bm{\mathcal K}}_t(\bm\Lambda \mbf F_{t|t-1}-\wh{\bm\Lambda}^* \mbf F_{t|t-1}^*).\label{upFerror}
\end{align}
Moreover, because of \eqref{DF2}
\begin{align}
\frac{\wh{\bm\Lambda}^* \mbf F_{t|t-1}^*-{\bm\Lambda} \mbf F_{t|t-1}}{\sqrt n}&=\frac{\bm\Lambda}{\sqrt n}(\mbf F_{t|t-1}^*-\mbf F_{t|t-1})+\l(\frac{\wh{\bm\Lambda}^*-\bm\Lambda}{\sqrt n}\r)(\mbf F_{t|t-1}^*-\mbf F_{t|t-1})+\l(\frac{\wh{\bm\Lambda}^*-\bm\Lambda}{\sqrt n}\r)\mbf F_{t|t-1}\nn\\
&=\frac{\bm\Lambda}{\sqrt n}(\mbf F_{t|t-1}^*-\mbf F_{t|t-1})+\l(\frac{\wh{\bm\Lambda}^*-\bm\Lambda}{\sqrt n}\r)(\mbf F_{t|t-1}^*-\mbf F_{t|t-1})+O_p(T^{-1/2}),\label{upFerror2}
\end{align}
since
\begin{align}
\l(\frac{\wh{\bm\Lambda}^*-\bm\Lambda}{\sqrt n}\r)\mbf F_{t|t-1}&=\l(\frac{\wh{\bm\Lambda}^*-\bm\Lambda}{\sqrt n}\r)\mbf A\mbf F_{t-1|t-1}=\l(\frac{\wh{\bm\Lambda}^*-\bm\Lambda}{\sqrt n}\r)\mbf A(\mbf F_{t-1|t-1}-\mbf F_{t-1})+\l(\frac{\wh{\bm\Lambda}^*-\bm\Lambda}{\sqrt n}\r)\mbf A\mbf F_{t-1}\nn\\
&=O_p(T^{-1/2})O(e^{-(t-1)/2})+O_p(T^{-1/2})O(n^{-1/2})+O_p(T^{-1/2}),\nn
\end{align}
because of Lemma \ref{lem:eststar}, \eqref{res1}, \eqref{DF2} and since $\Vert\mbf A\Vert=O(1)$. Similarly, from \eqref{eq:up2hat}
\begin{align}
\mbf P_{t|t}^*-\mbf P_{t|t} &=\mbf P_{t|t-1}^*-\mbf P_{t|t-1}^*-\l[(\wh{\bm{\mathcal K}}_t^*-{\bm{\mathcal K}}_t)\bm\Lambda\mbf P_{t|t-1}\r.\nn\\
&+\;\l.(\wh{\bm{\mathcal K}}_t^*-{\bm{\mathcal K}}_t)(\wh{\bm\Lambda}^*\mbf P_{t|t-1}^*-\bm\Lambda\mbf P_{t|t-1}) + {\bm{\mathcal K}}_t(\wh{\bm\Lambda}^*\mbf P_{t|t-1}^*-\bm\Lambda\mbf P_{t|t-1})\r].\label{upPerror}
\end{align}
Moreover,
\begin{align}
\frac{\wh{\bm\Lambda}^*\mbf P_{t|t-1}^*-\bm\Lambda\mbf P_{t|t-1}}{\sqrt n}&=\frac{\bm\Lambda}{\sqrt n}(\mbf P_{t|t-1}^*-\mbf P_{t|t-1})+\l(\frac{\wh{\bm\Lambda}^*-\bm\Lambda}{\sqrt n}\r)(\mbf P_{t|t-1}^*-\mbf P_{t|t-1})+\l(\frac{\wh{\bm\Lambda}^*-\bm\Lambda}{\sqrt n}\r)\mbf P_{t|t-1}\nn\\
&=\frac{\bm\Lambda}{\sqrt n}(\mbf P_{t|t-1}^*-\mbf P_{t|t-1})+\l(\frac{\wh{\bm\Lambda}^*-\bm\Lambda}{\sqrt n}\r)(\mbf P_{t|t-1}^*-\mbf P_{t|t-1})+O_p(T^{-1/2}),\label{upPerror2}
\end{align}
since
\begin{align}
\l(\frac{\wh{\bm\Lambda}^*-\bm\Lambda}{\sqrt n}\r)\mbf P_{t|t-1}&=\l(\frac{\wh{\bm\Lambda}^*-\bm\Lambda}{\sqrt n}\r)\l(\mbf A\mbf P_{t-1|t-1}\mbf A'+\mbf H\mbf H'\r)\nn\\
&=O_p(T^{-1/2})O(e^{-(t-1)})+O_p(T^{-1/2})O(n^{-1})+O_p(T^{-1/2}),\nn
\end{align}
because of Lemma \ref{lem:eststar} and \eqref{Ptt} in the proof of Lemma \ref{lem:steady3} and since $\Vert\mbf A\Vert=O(1)$ and 
$\Vert\mbf H\Vert=O(1)$. Following the same reasoning we also have
\begin{align}
\frac{\bm\Lambda\mbf P_{t|t-1}}{\sqrt n}&=\frac{\bm\Lambda}{\sqrt n}\l(\mbf A\mbf P_{t-1|t-1}\mbf A'+\mbf H\mbf H'\r)=\frac{\bm\Lambda}{\sqrt n}\l(O(e^{-(t-1)})+O(n^{-1})+\mbf H\mbf H'\r).
\label{upPerror3}
\end{align}

\medskip
\noindent
Now, set $t=1$. Then, by noticing that $\mbf F_{0|0}^*=\mbf F_{0|0}$ and $\mbf P_{0|0}^*=\mbf P_{0|0}$, from \eqref{predFerror} and \eqref{predPerror} at $t=1$ we have
\begin{align}
\mbf F_{1|0}^*-\mbf F_{1|0} = O_p(T^{-1/2}), \qquad \mbf P_{1|0}^*-\mbf P_{1|0} = O_p(T^{-1/2}).\label{F10P10}
\end{align}
Then, because of \eqref{eq:OP1}, \eqref{F10P10} and Lemma \ref{lem:eststar}, at $t=1$ we have
\begin{align}
&\sqrt n(\wh{\bm{\mathcal K}}^*_1-{\bm{\mathcal K}}_1)\l(\frac{\mbf x_1}{\sqrt n}-\frac{\bm\Lambda \mbf F_{1|0}}{\sqrt n}\r)= \label{res2}\\
&=\l[\mbf P_{1|0}^*\frac{\wh{\bm\Lambda}^{*'}}{\sqrt n}\l(\frac{\wh{\bm\Lambda}}{\sqrt n}\mbf P_{1|0}^*\frac{\wh{\bm\Lambda}^{*'}}{\sqrt n}+\frac{\wh{\mbf R}^*}n\r)^{-1}-\mbf P_{1|0}\frac{{\bm\Lambda}'}{\sqrt n}\l(\frac{\bm\Lambda}{\sqrt n}\mbf P_{1|0}\frac{\bm\Lambda'}{\sqrt n}+\frac{\mbf R}n\r)^{-1}\r]\l(\frac{\mbf x_1}{\sqrt n}-\frac{\bm\Lambda \mbf F_{1|0}}{\sqrt n}\r)\nn\\
&=O_p(T^{-1/2}).\nn
\end{align}
Moreover, from \eqref{upPerror3}
\begin{align}
\sqrt n(\wh{\bm{\mathcal K}}_1^*-{\bm{\mathcal K}}_1)\frac{\bm\Lambda\mbf P_{1|0}}{\sqrt n}=O_p(T^{-1/2}).\label{res3}
\end{align}
From \eqref{upFerror} and using \eqref{predFerror}, \eqref{upFerror2}, \eqref{F10P10}, and \eqref{res2} at $t=1$ we have
\begin{align}
\mbf F_{1|1}^*-\mbf F_{1|1}&=\mbf F_{1|0}^*-\mbf F_{1|0}+(\wh{\bm{\mathcal K}}^*_1-{\bm{\mathcal K}}_1)(\mbf x_1-\bm\Lambda \mbf F_{1|0})\nn\\
&\;+(\wh{\bm{\mathcal K}}^*_1-{\bm{\mathcal K}}_1)(\bm\Lambda \mbf F_{1|0}-\wh{\bm\Lambda} \mbf F_{1|0}^*)+{\bm{\mathcal K}}_1(\bm\Lambda \mbf F_{1|0}-\wh{\bm\Lambda} \mbf F_{1|0}^*)= O_p(T^{-1/2}).\label{upFerror11}
\end{align}
Similarly, from \eqref{upPerror} and using \eqref{predPerror}, \eqref{upPerror2}, \eqref{F10P10}, and \eqref{res3} at $t=1$ we have
\begin{align}
\mbf P_{1|1}^*-\mbf P_{1|1} &=\mbf P_{1|0}^*-\mbf P_{1|0}^*-\l[(\wh{\bm{\mathcal K}}_1^*-{\bm{\mathcal K}}_1)\bm\Lambda\mbf P_{1|0}\r.\nn\\
&+\;\l.(\wh{\bm{\mathcal K}}_1^*-{\bm{\mathcal K}}_1)(\wh{\bm\Lambda}^*\mbf P_{1|0}^*-\bm\Lambda\mbf P_{1|0}) + {\bm{\mathcal K}}_1(\wh{\bm\Lambda}^*\mbf P_{1|0}^*-\bm\Lambda\mbf P_{1|0})\r]=O_p(T^{-1/2}).\label{upPerror11}
\end{align}
Then substituting \eqref{upFerror11} into \eqref{upFerror} and  \eqref{upPerror11} into \eqref{upPerror}  we have
\beq
\mbf F_{2|1}^*-\mbf F_{2|1}=O_p(T^{-1/2}), \qquad \mbf P_{2|1}^*-\mbf P_{2|1}=O_p(T^{-1/2}).\label{F21P21}
\eeq
Then, because of \eqref{eq:OP1}, \eqref{F21P21} and Lemma \ref{lem:eststar}, at $t=2$ we have
\begin{align}
&\sqrt n(\wh{\bm{\mathcal K}}^*_2-{\bm{\mathcal K}}_2)\l(\frac{\mbf x_2}{\sqrt n}-\frac{\bm\Lambda \mbf F_{2|1}}{\sqrt n}\r)= O_p(T^{-1/2}),\label{res4}
\end{align}
and from \eqref{upPerror3}
\begin{align}
\sqrt n(\wh{\bm{\mathcal K}}_2^*-{\bm{\mathcal K}}_2)\frac{\bm\Lambda\mbf P_{2|1}}{\sqrt n}=O_p(T^{-1/2}).\label{res5}
\end{align}
From, \eqref{upFerror} and using \eqref{predFerror}, \eqref{upFerror2}, \eqref{F21P21}, and \eqref{res4}, at $t=2$ we have
\[
\mbf F_{2|2}^*-\mbf F_{2|2}= O_p(T^{-1/2}),
\]
and from \eqref{upPerror} and using \eqref{predPerror}, \eqref{upPerror2}, \eqref{F21P21}, and \eqref{res5}, at $t=2$ we have
\[
\mbf P_{2|2}^*-\mbf P_{2|2}= O_p(T^{-1/2}).
\]
By repeating the same reasoning for $t=3,\ldots ,T$ we have
\begin{align}
&\Vert\mbf F_{t|t}^*-\mbf F_{t|t} \Vert = O_p(T^{-1/2}),\qquad \Vert\mbf P_{t|t}^*-\mbf P_{t|t} \Vert = O_p(T^{-1/2}),\label{F110_6}
\end{align}
and also
\begin{align}
&\Vert\mbf F_{t|t-1}^*-\mbf F_{t|t-1} \Vert = O_p(T^{-1/2}),\qquad \Vert\mbf P_{t|t-1}^*-\mbf P_{t|t-1} \Vert = O_p(T^{-1/2}).\label{F110_8}
\end{align}
Because of Lemma \ref{lem:steady3} and \eqref{F110_6}, we have for $\bar t\le t \le T$,
\begin{align}
&\Vert\mbf F_{t|t}^*-\mbf F_{t} \Vert \le \Vert\mbf F_{t|t}^*-\mbf F_{t|t} \Vert + \Vert\mbf F_{t|t}-\mbf F_{t} \Vert = O_p(T^{-1/2}) + O(n^{-1/2}),\nn\\
&\Vert\mbf P_{t|t}^* \Vert \le \Vert\mbf P_{t|t}^*-\mbf P_{t|t} \Vert + \Vert\mbf P_{t|t} \Vert = O_p(T^{-1/2}) + O(n^{-1}).\nn
\end{align}
Now compare the KS iterations, \eqref{eq:KS3}-\eqref{eq:KS7}, with those obtained when using $\wh{\bm\Theta}^*$:
\begin{align}
&\mbf F_{t|T}^*=\mbf F_{t|t-1}^*+\mbf P_{t|t-1}^*\mbf r_{t-1}^*,\label{eq:KS3hat}\\
&\mbf r_{t-1}^*=\wh{\bm\Lambda}^{*'}(\wh{\bm\Lambda}^*\mbf P_{t|t-1}^*\wh{\bm\Lambda}^{*'}+\wh{\mbf R}^*)^{-1}(\mbf x_t-\wh{\bm\Lambda}^*\mbf F_{t|t-1}^*)+\mbf L^{*'}_{t}\mbf r_{t}^*,\label{eq:KS4hat}\\
&\mbf P_{t|T}^*=\mbf P_{t|t-1}^*-\mbf P_{t|t-1}^*\mbf N_{t-1}^*\mbf P_{t|t-1}^*,\label{eq:KS5hat}\\
&\mbf N_{t-1}^*=\wh{\bm\Lambda}^{*'}(\wh{\bm\Lambda}^*\mbf P_{t|t-1}^*\wh{\bm\Lambda}^{*'}+\wh{\mbf R}^*)^{-1}\wh{\bm\Lambda}^*+\mbf L_{t}^{*'}\mbf N_{t}^*\mbf L_{t}^*,\label{eq:KS6hat}\\
&\mbf L_{t}^*= \wh{\mbf A}^*-\wh{\mbf A}^* \mbf P_{t|t-1}^* \wh{\bm\Lambda}^{*'} (\wh{\bm\Lambda}^*\mbf P_{t|t-1}^*\wh{\bm\Lambda}^{*'}+\wh{\mbf R}^*)^{-1} \wh{\bm\Lambda}^*,\label{eq:KS7hat}
\end{align}
where $\mbf r_{T}^*=\mbf 0_{r\times 1}$, $\mbf N_{T}^*=\mbf 0_{r}$. First notice that obviously at $t=T$ both KF and KS give the same result hence \eqref{F110_6} applies also in this case, and because of Lemma \ref{lem:eststar}, \eqref{eq:OP1}, \eqref{upFerror2}, and \eqref{F110_8}, we have
\beq\label{rT1}
\mbf r_{T-1}^*-\mbf r_{T-1}=O_p(T^{-1/2}),\qquad \mbf N_{T-1}^*-\mbf N_{T-1}=O_p(T^{-1/2}).
\eeq
Moreover, from \eqref{eq:KS7hat}, because of Lemma \ref{lem:eststar} and \eqref{F110_8}, we have
\begin{align}
\mbf L_{t}^*-\mbf L_t&=\wh{\mbf A}^*-\mbf A-\sqrt n\l[\wh{\mbf A}^* \wh{\bm{\mathcal K}}_t^* \frac{\wh{\bm\Lambda}^*}{\sqrt n}-{\mbf A} {\bm{\mathcal K}}_t \frac{{\bm\Lambda}}{\sqrt n}\r]= O_p(T^{-1/2}).\label{rNL_2}
\end{align}
Then, from \eqref{eq:KS4hat}, because of \eqref{eq:OP1}, \eqref{upFerror2}, \eqref{F110_8}, \eqref{rT1} and \eqref{rNL_2}, at $t=T-1$ we have
\begin{align}\label{rT2}
\mbf r_{T-2}^*-\mbf r_{T-2}=O_p(T^{-1/2}),\qquad \mbf N_{T-2}^*-\mbf N_{T-2}=O_p(T^{-1/2}).
\end{align}
Therefore, from \eqref{eq:KS3hat} and \eqref{eq:KS5hat}, because of \eqref{F110_8} and \eqref{rT2}, we have
\begin{align}
\mbf F_{T-1|T}^*-\mbf F_{T-1|T}= O_p(T^{-1/2}), \qquad \mbf P_{T-1|T}^*-\mbf P_{T-1|T}= O_p(T^{-1/2}).
\end{align}
By repeating the same reasoning for  $t=(T-2),\ldots ,1$, we have
\beq\label{FtT00cons}
\Vert\mbf F_{t|T}^*-\mbf F_{t|T}\Vert=O_p(T^{-1/2}), \qquad \Vert\mbf P_{t|T}^*-\mbf P_{t|T}\Vert=O_p(T^{-1/2}).
\eeq
Because of Lemma \ref{lem:steady3} and \eqref{FtT00cons}, we have for $\bar t\le t \le T$
\begin{align}
&\Vert\mbf F_{t|T}^*-\mbf F_{t} \Vert \le \Vert\mbf F_{t|T}^*-\mbf F_{t|T} \Vert + \Vert\mbf F_{t|T}-\mbf F_{t} \Vert = O_p(T^{-1/2}) + O(n^{-1/2}),\nn\\
&\Vert\mbf P_{t|T}^* \Vert \le \Vert\mbf P_{t|T}^*-\mbf P_{t|T} \Vert + \Vert\mbf P_{t|T} \Vert = O_p(T^{-1/2}) + O(n^{-1}),\nn
\end{align}
which completes the proof. $\Box$

\begin{lem}\label{lem:est0}
Consider the initial estimator of the parameters $\wh{\bm\Theta}_0$ defined in \eqref{initparam}, then there exists an invertible $r\times r$ matrix $\mbf J$ such that, as $n,T\to\infty$:
\begin{align}
&\min(\sqrt n,\sqrt T)\,\Vert\wh{\bm\lambda}_{i0}'-\bm\lambda_i'\mbf J^{-1}\Vert = O_p(1), \quad i=1,\ldots, n,\nn\\
&\min(\sqrt n,\sqrt T)\,\Vert\wh{\mbf A}_0-\mbf J\mbf A\mbf J^{-1}\Vert = O_p(1),\nn\\
&\min(\sqrt n,\sqrt T)\,\Vert\wh{\mbf H}_0-\mbf J\mbf H\Vert = O_p(1),\nn\\
&\min(\sqrt n,\sqrt T)\,\vert[\wh{\mbf R}]_{ii,0}-[\mbf R]_{ii}\vert = O_p(1), \quad i=1,\ldots, n.\nn
\end{align}
Moreover, under E1 and E2 we have $\mbf J=\mbf I_r$.
\end{lem}

\noindent{\bf Proof.} When E2 holds the proof Lemmas 3 and 5 in \citealp{BLL2} where is shown that $\mbf J$ is a diagonal matrix with entries $\pm 1$. If we impose also E1 the sign indeterminacy is fixed and $\mbf J=\mbf I_r$. $\Box$ 

\begin{lem}\label{lem:convEM} Consider the estimator of the parameters obtained at convergence of the EM algorithm $\wh{\bm\Theta}:= \wh{\bm\Theta}_{k^*}$,  then, as $n,T\to\infty$:
\begin{align}
&\sqrt T\,\Vert\wh{\bm\lambda}_{ik^*}-\bm\lambda_i \Vert = O_p(1), \quad i=1,\ldots, n,\nn\\
&\sqrt T\,\Vert\wh{\mbf A}_{k^*}-\mbf A\Vert = O_p(1),\nn\\
&\sqrt T\,\Vert\wh{\mbf H}_{k^*}-\mbf H\Vert = O_p(1),\nn\\
&\sqrt T\,\vert[\wh{\mbf R}]_{ii,k^*}-[\mbf R]_{ii}\vert = O_p(1), \quad i=1,\ldots, n.\nn
\end{align}
\end{lem}

\noindent{\bf Proof.} Define the $Q\times Q$ matrices 
\begin{align}
&\bm{\mathcal I}(\bm\Theta)=-\nabla^2_{\bm\Theta\bm\Theta'}\ \ell(\bm{\mathcal X}_T;\bm\Theta),\nn\\
&\bm{\mathcal I}_0(\bm\Theta) = -\int_{\mathbb R^{r\times T}}\nabla^2_{\bm\Theta\bm\Theta'}\ \ell(\bm{\mathcal X}_T,\bm{\mathcal F}_T;\bm\Theta) f(\bm{\mathcal F}_T|\bm{\mathcal X}_T;\bm\Theta)\mathrm d\bm{\mathcal F}_T = -\E_{\bm\Theta}[\nabla^2_{\bm\Theta\bm\Theta'}\ \ell(\bm{\mathcal X}_T,\bm{\mathcal F}_T;\bm\Theta)|\bm{\mathcal X}_T],\nn\\
&\bm{\mathcal I}_1(\bm\Theta) = -\int_{\mathbb R^{r\times T}}\nabla^2_{\bm\Theta\bm\Theta'}\ \ell(\bm{\mathcal F}_T|\bm{\mathcal X}_T,\bm{\mathcal F}_T;\bm\Theta) f(\bm{\mathcal F}_T|\bm{\mathcal X}_T;\bm\Theta)\mathrm d\bm{\mathcal F}_T= -\E_{\bm\Theta}[\nabla^2_{\bm\Theta\bm\Theta'}\ \ell(\bm{\mathcal F}_T|\bm{\mathcal X}_T;\bm\Theta)|\bm{\mathcal X}_T],\nn
\end{align}
then, since $\bm{\mathcal I}(\bm\Theta)$ does not depend on $\bm{\mathcal F}_T$ from \eqref{eq:LL1} and \eqref{eq:LLX2}
\beq\label{III}
\bm{\mathcal I}(\bm\Theta)= \bm{\mathcal I}_0(\bm\Theta)-\bm{\mathcal I}_1(\bm\Theta).
\eeq
Moreover, at convergence of the EM algorithm (iteration $k^*$) we have the Taylor approximation
\begin{align}\label{taylorIII}
(\wh{\bm\Theta}_{k^*}-\wh{\bm\Theta}^*) &= (\wh{\bm\Theta}_{k^*-1}-\wh{\bm\Theta}^*)\bm{\mathcal R}(\wh{\bm\Theta}^*) + O\l(\Vert\wh{\bm\Theta}_{k^*}-\wh{\bm\Theta}^*\Vert^2\r),
\end{align}
where following \citet{MR94} we have (see also \eqref{III})
\[
\bm{\mathcal R}(\wh{\bm\Theta}^*) = \bm{\mathcal I}_1(\wh{\bm\Theta}^*)\l(\bm{\mathcal I}_0(\wh{\bm\Theta}^*)\r)^{-1} =  \mbf I_Q-\bm{\mathcal I}(\wh{\bm\Theta}^*)\l(\bm{\mathcal I}_0(\wh{\bm\Theta}^*)\r)^{-1}.
\]
Hence, by iterating \eqref{taylorIII} $k^*$ times and neglecting the second term on the rhs which at convergence is always smaller the the first term, we obtain
\begin{align}
\Vert\wh{\bm\Theta}_{k^*}-\wh{\bm\Theta}^*\Vert& \le \Vert \wh{\bm\Theta}_{0}-\wh{\bm\Theta}^*\Vert \;\Vert\bm{\mathcal R}(\wh{\bm\Theta}^*)\Vert^{k^*}.\label{taylorIII2}
\end{align}
Hereafter, denote $\zeta_{nT}:=\max(n^{-1/2},T^{-1/2})$. Consider \eqref{taylorIII2} for the estimated loadings, for any $i=1,\ldots,n$, and using Gaussianity (see D2 and D3), we have
\begin{align}
\Vert\wh{\bm\lambda}_{ik^*}-\wh{\bm\lambda}_i^*\Vert
&\le \Vert \wh{\bm\lambda}_{i0}-\wh{\bm\lambda}_i^*\Vert \;\Big\Vert\mbf I_r- \Big(\sum_{t=1}^T\mbf F_t\mbf F_t'\Big)\Big(\sum_{t=1}^T\E_{\wh{\bm\Theta}^*}[\mbf F_t\mbf F_t'|\bm{\mathcal X}_T]\Big)^{-1}\Big\Vert^{k^*}\nn\\
&=\Vert \wh{\bm\lambda}_{i0}-\wh{\bm\lambda}_i^*\Vert \;\Big\Vert\mbf I_r- \Big(\frac 1 {T^2}\sum_{t=1}^T\mbf F_t\mbf F_t'\Big)\Big(\frac 1 {T^2}\sum_{t=1}^T\l(\mbf F_{t|T}^*\mbf F_{t|T}^{*'}+\mbf P^*_{t|T}\r)\Big)^{-1}\Big\Vert^{k^*}.\label{convEMML}
\end{align}
Therefore, by condition \eqref{eq:exprate} in the text ${\bar t}=O(\log T)$, because of Lemma \ref{lem:steady_est1}, we have
\begin{align}\label{convEMML1}
\Big(\frac 1 {T^2} \sum_{t=1}^T\l(\mbf F_{t|T}^*\mbf F_{t|T}^{*'}+\mbf P^*_{t|T}\r)\Big)^{-1}&=\Big(\frac 1 {T^2} \sum_{t=1}^{\bar t-1}\l(\mbf F_{t|T}^*\mbf F_{t|T}^{*'}+\mbf P^*_{t|T}\r)+\frac 1 {T^2}\sum_{t=\bar t}^{T}\l(\mbf F_{t|T}^*\mbf F_{t|T}^{*'}+\mbf P^*_{t|T}\r)\Big)^{-1}\nn\\
&=\bigg(\frac 1 {T^2}\sum_{t=1}^{T}\mbf F_{t}\mbf F_{t}'+O_p\bigg(\frac{\zeta_{nT}}{\sqrt T}\bigg)+O_p\bigg(\frac {\log T}{T}\bigg)\bigg)^{-1}\nn\\
&=\Big(\frac 1 {T^2}\sum_{t=1}^{T}\mbf F_{t}\mbf F_{t}'\Big)^{-1}+o_p(T^{-1/2}),
\end{align}
since $\Vert\mbf F_{t|T}^*\mbf F_{t|T}^{*'}+\mbf P^*_{t|T}\Vert = O_p(T)$ and $\Vert\mbf F_t\mbf F_t'\Vert = O_p(T)$, because $\mbf F_t\sim I(1)$ and $\mbf F_{t|T}^*\sim I(1)$, and
\begin{align}
&\frac 1 {T^2} \sum_{t=1}^{\bar t-1}\l(\mbf F_{t|T}^*\mbf F_{t|T}^{*'}+\mbf P^*_{t|T}\r)=O_p\l(\frac {\log T}{T}\r),\qquad \frac 1{T^2}\sum_{t=1}^{T}\mbf F_{t}\mbf F_{t}'- \frac 1 {T^2}\sum_{t=\bar t}^{T}\mbf F_{t}\mbf F_{t}'= O_p\l(\frac {\log T}{T}\r).\nn
\end{align}
Moreover, because of Lemmas \ref{lem:est0} and \ref{lem:eststar}, we have
\begin{align}
\Vert \wh{\bm\lambda}_{i0}-\wh{\bm\lambda}_i^*\Vert&\le \Vert \wh{\bm\lambda}_{i0}-{\bm\lambda}_i\Vert +\Vert \wh{\bm\lambda}_i^*-{\bm\lambda}_i\Vert = O_p(\zeta_{nT}) + O_p(T^{-1/2}). \label{convEMML2}
\end{align}
By substituting \eqref{convEMML1} and \eqref{convEMML2} into \eqref{convEMML}, we have
\begin{align}\label{convEMML3}
\Vert\wh{\bm\lambda}_{ik^*}-\wh{\bm\lambda}_i^*\Vert=o_p(\zeta_{nT}\,T^{-k^*/2}).
\end{align}
Finally, because of Lemma \ref{lem:eststar} and \eqref{convEMML3}, we have
\begin{align}
\Vert \wh{\bm\lambda}_{ik^*}-{\bm\lambda}_i\Vert&\le \Vert \wh{\bm\lambda}_{ik^*}-\wh{\bm\lambda}_i^*\Vert + \Vert \wh{\bm\lambda}_{i}^*-{\bm\lambda}_i\Vert =o_p(T^{-k^*/2})+O_p(T^{-1/2}).\nn
\end{align}
The proof for the other parameters follows the same steps by taking the appropriate second derivatives and applying the results in Lemma \ref{lem:steady_est1}. $\Box$

\subsection{Proof of Proposition \ref{prop:EMcons}}\label{app:KFKSEM}
Consistency of the estimated loadings is proved in Lemma \ref{lem:convEM}. Recalling that $\wh{\bm\Theta}:= \wh{\bm\Theta}_{k^*}$ and also $\bm\lambda_i = \mbf K^{-1'} (\bm b_{i0}'\,\bm b_{i1}')$, because of \eqref{eq:R1_app_2} we prove  \eqref{eq:consEM2}. 

\medskip
\noindent
Consistency of the estimated static factors is then proved as in Lemma \ref{lem:steady_est1} but using the results of Lemma \ref{lem:convEM} for the estimated parameters. In particular, for all $\bar t\le t \le T$
\begin{align}
\Vert{\mbf F}_{t|T,k^*}-\mbf F_{t} \Vert \le \Vert{\mbf F}_{t|T,k^*}-\mbf F_{t|t} \Vert + \Vert\mbf F_{t|t}-\mbf F_{t} \Vert = O_p(T^{-1/2}) + O(n^{-1/2}).\nn
\end{align}
Recalling that $\wh{\mbf F}_t:= \mbf F_{t|T,k^*}$ and also $\mbf F_t=\mbf K (\bm f_t'\, \bm f_{t-1}')'$ because of \eqref{eq:R1_app} we prove \eqref{eq:consEM3}. A similar result can be proved for the KF estimator of the static factors, $\mbf F_{t|t,k^*}$ using again Lemmas \ref{lem:steady_est1} and \ref{lem:convEM}.

\medskip
\noindent
Denote $\zeta_{nT}:=\max(n^{-1/2},T^{-1/2})$. Then, recalling that $\wh{\bm\lambda}_i'\wh{\mbf F}_t:= \wh{\bm\lambda}_{ik^*}'{\mbf F}_{t|T,k^*}$, because of \eqref{eq:consEM2}, \eqref{eq:consEM3} and Lemma \ref{lem:steady3}, we have
\begin{align}
\vert \wh{\chi}_{it} - \chi_{it}\vert &= \vert \wh{\bm\lambda}_{i}'\wh{\mbf F}_{t}-\bm\lambda_i'\mbf F_t\vert \le 
\Vert(\wh{\bm\lambda}_i-\bm\lambda_i)\mbf F_t\Vert + 
\Vert\bm\lambda_i\Vert\;\Vert\wh{\mbf F}_t-\mbf F_t\Vert+
\Vert\wh{\bm\lambda}_i-\bm\lambda_i\Vert\;\Vert\wh{\mbf F}_t-\mbf F_t\Vert\nn\\
&=\Vert(\wh{\bm\lambda}_i-\bm\lambda_i)\mbf F_t\Vert + O_p(\zeta_{nT})\nn\\
&=\Vert\wh{\bm\lambda}_i-\wh{\bm\lambda}_i^*\Vert \;\Vert\mbf F_t\Vert + \Vert(\wh{\bm\lambda}_i^*-\bm\lambda_i)\mbf F_t\Vert + O_p(\zeta_{nT}).\label{ultimo}
\end{align}
Now, from Lemma \ref{lem:convEM} (see in particular \eqref{convEMML3}) and since $\Vert\mbf F_t\Vert=O_p(T^{-1/2})$, we have
\beq
\Vert\wh{\bm\lambda}_i-\wh{\bm\lambda}_i^*\Vert \;\Vert\mbf F_t\Vert= o_p(\zeta_{nT}\,T^{-(k^*-1)/2}).\label{ultimo2}
\eeq
Moreover, from transformation \eqref{DFZ}, defined in the proof of Lemma \ref{lem:eststar}, $\bm{\mathcal D}\mbf F_t=(\mbf Z_{1t}'\;\mbf Z_{0t}')'$, we have $\Vert\mbf Z_{1t}\Vert=O_p(\sqrt T)$ and $\Vert\mbf Z_{0t}\Vert=O_p(1)$. Then, since $\bm{\mathcal D}'\bm{\mathcal D}=\mbf I_r$, as a consequence of Lemma \ref{lem:eststar} (see in particular \eqref{lambdaDD}), we have
\beq
(\wh{\bm\lambda}_{i}^*-\bm\lambda_i)'\bm{\mathcal D}'\bm{\mathcal D}\mbf F_t= O_p(T^{-1/2}).\label{ultimo3}
\eeq
By substituting \eqref{ultimo2} and \eqref{ultimo3} into \eqref{ultimo} and since $k^*\ge 1$ because we run the EM algorithm at least once after initialization, we prove  \eqref{eq:consEM4}. $\Box$

%
\clearpage
\setcounter{footnote}{0}
\section{Data Description and Data Treatment }\label{sec:data}

This Appendix present the dataset used in the analysis.~All variables where downloaded from Haver on June 16$^{th}$ 2017. None of the variables where adjusted for outliers but variables 57, 83, 87, and 94. All variables are from the USECON database but variable 103 that is from the DAILY database. All monthly and daily series are transformed into quarterly observation by simple averages.

In order to choose whether or not to de-trend a variable we apply the following procedure: let $m_i$ be the sample mean of $\Delta y_{it}$, $\gamma_{i}(j)$ be the auto-covariance of order $j$ of $\Delta y_{it}$, and $\bar{\gamma}_i=\sqrt{\frac{1}{T}\sum_{j=1}^J \gamma_i(j)}$, then if  $\frac{|m_i|}{\bar{\gamma}_i}\ge 1.96$ we estimate $a_i$ and $b_i$ from an OLS regression of $y_{it}$ on a constant and a time trend, whereas if $\frac{|m_i|}{\bar{\gamma}_i}< 1.96$ we set $\wh{a}_i=m_i$ and $\wh{b}_i=0$.

\begin{center}

\begin{tabular}{
p{.21\textwidth}
p{.21\textwidth}
p{.1\textwidth}
p{.1\textwidth}
} 
\multicolumn{4}{c}{\textsc{List of Abbreviations}}\\ \hline\hline
\multicolumn{4}{l}{Source:}\\\hline
\multicolumn{4}{l}{BLS=U.S. Department of Labor: Bureau of Labor Statistics}\\
\multicolumn{4}{l}{BEA=U.S. Department of Commerce: Bureau of Economic Analysis}	\\
\multicolumn{4}{l}{ISM = Institute for Supply Management}	\\
\multicolumn{4}{l}{CB=U.S. Department of Commerce: Census Bureau}	\\
\multicolumn{4}{l}{FRB=Board of Governors of the Federal Reserve System}	\\
\multicolumn{4}{l}{EIA=Energy Information Administration}	\\
\multicolumn{4}{l}{WSJ=Wall Street Journal}	\\
\multicolumn{4}{l}{CBO=Congressional Budget Office}	\\
\multicolumn{4}{l}{FRBPHIL=Federal Reserve Bank of Philadelphia}		\\\hline
\\ \hline\hline
F = Frequency 	&	 T=Transformation 	&	 SA 	&	$\xi$=Idiosyncratic	\\	\hline
Q = Quarterly 	&	0 = None	&	 0 = no 	&	0=$I(0)$	\\	
M = Monthly 	&	 1 = $\log$ 	&	 1 = yes 	&	1=$I(1)$	\\	
D = Daily 	&	 2 = $\Delta \log$ 	&	 	&		\\	\hline
\\ \hline\hline
\multicolumn{2}{l}{D = Deterministic Component}	&	 \multicolumn{2}{l}{U=Units}\\ \hline
\multicolumn{2}{l}{0 = $\wh{a}_i=\frac{1}{T}\sum_{t=1}^T\Delta y_{it}$, $\wh{b}_i=0$}	&	\multicolumn{2}{l}{1000--P = Thousands of Persons}\\
\multicolumn{2}{l}{1 = OLS Detrending }	&	\multicolumn{2}{l}{1000--U = Thousands of Units}\\
&	&	\multicolumn{2}{l}{BoC = Billions of Chained}\\
&	&	\multicolumn{2}{l}{\$--B = Dollars per Barrel}\\\hline

\end{tabular}
\end{center}

\setlength{\tabcolsep}{.01\textwidth}

\centering \footnotesize

\begin{tabular}{
p{.015\textwidth}
p{.1\textwidth}
p{.45\textwidth}
p{.1\textwidth}
p{.02\textwidth}
p{.04\textwidth}
P{.025\textwidth}
p{.01\textwidth}
p{.01\textwidth}
p{.01\textwidth}
}\hline\hline
\bf N	&	\bf Series ID	&	\bf Definition	&	\bf Unit	&	\bf F	&	\bf S	&	\bf SA	&	\bf T	&	\bf D	&	$\bm \xi$	\\\hline
1	&	GDPH	&	Real Gross Domestic Product	&	BoC 2009\$	&	Q	&	BEA	&	1	&	1	&	1	&	0	\\
2	&	GDYH	&	Real gross domestic income	&	BoC 2009\$	&	Q	&	BEA	&	1	&	1	&	1	&	0	\\
3	&	FSH	&	Real Final Sales of Domestic Product	&	BoC 2009\$	&	Q	&	BEA	&	1	&	1	&	1	&	1	\\
4	&	IH	&	Real Gross Private Domestic Investment	&	BoC 2009\$	&	Q	&	BEA	&	1	&	1	&	1	&	1	\\
5	&	GSH	&	Real State \& Local$^\ast$	&	BoC 2009\$	&	Q	&	BEA	&	1	&	1	&	1	&	1	\\
6	&	FRH	&	Real Private Residential Fixed Investment	&	BoC 2009\$	&	Q	&	BEA	&	1	&	1	&	1	&	0	\\
7	&	FNH	&	Real Private Nonresidential Fixed Investment	&	BoC 2009\$	&	Q	&	BEA	&	1	&	1	&	1	&	1	\\
8	&	MH	&	Real Imports of Goods \& Services	&	BoC 2009\$	&	Q	&	BEA	&	1	&	1	&	1	&	0	\\
9	&	GH	&	Real Government$^\ast$	&	BoC 2009\$	&	Q	&	BEA	&	1	&	1	&	1	&	1	\\
10	&	XH	&	Real Exports of Goods \& Services	&	BoC 2009\$	&	Q	&	BEA	&	1	&	1	&	1	&	0	\\
14	&	CH	&	Real Personal Consumption Expenditures	(PCE)&	BoC 2009\$	&	Q	&	BEA	&	1	&	1	&	1	&	0	\\
11	&	CNH	&	Real PCE: Nondurable Goods	&	BoC 2009\$	&	Q	&	BEA	&	1	&	1	&	1	&	1	\\
12	&	CSH	&	Real PCE: Services	&	BoC 2009\$	&	Q	&	BEA	&	1	&	1	&	1	&	0	\\
13	&	CDH	&	Real PCE: Durable Goods	&	BoC 2009\$	&	Q	&	BEA	&	1	&	1	&	1	&	0	\\
15	&	GFDIH	&	Real National Defense Gross Investment	&	BoC 2009\$	&	Q	&	BEA	&	1	&	1	&	1	&	0	\\
16	&	GFNIH	&	Real Federal Nondefense Gross Investment	&	BoC 2009\$	&	Q	&	BEA	&	1	&	1	&	1	&	0	\\
17	&	YPDH	&	Real Disposable Personal Income	&	BoC 2009\$	&	Q	&	BEA	&	1	&	1	&	1	&	0	\\ \hline
18	&	JI	&	Gross Private Domestic Investment:$^\star$ 	&	 2009=100	&	Q	&	BEA	&	1	&	2	&	0	&	0	\\
19	&	JGDP	&	Gross Domestic Product:$^\star$ 	&	 2009=100	&	Q	&	BEA	&	1	&	2	&	0	&	1	\\\hline
20	&	LXNFU	&	Unit Labor Cost$^\dag$	&	 2009=100	&	Q	&	BLS	&	1	&	1	&	1	&	1	\\
21	&	LXNFR	&	Real Compensation Per Hour$^\dag$	&	2009=100	&	Q	&	BLS	&	1	&	1	&	1	&	1	\\
22	&	LXNFC	&	Compensation Per Hour$^\dag$	&	 2009=100	&	Q	&	BLS	&	1	&	1	&	1	&	1	\\
23	&	LXNFH	&	Hours of All Persons$^\dag$	&	 2009=100	&	Q	&	BLS	&	1	&	1	&	1	&	0	\\
24	&	LXNFA	&	Output Per Hour of All Persons$^\dag$	&	 2009=100	&	Q	&	BLS	&	1	&	1	&	1	&	0	\\
25	&	LXMU	&	Unit Labor Cost$^\ddag$	&	 2009=100	&	Q	&	BLS	&	1	&	1	&	1	&	1	\\
26	&	LXMR	&	Real Compensation Per Hour$^\ddag$	&	 2009=100	&	Q	&	BLS	&	1	&	1	&	1	&	1	\\
27	&	LXMC	&	Compensation Per Hour$^\ddag$	&	 2009=100	&	Q	&	BLS	&	1	&	1	&	1	&	0	\\
28	&	LXMH	&	Hours of All Persons$^\ddag$	&	 2009=100	&	Q	&	BLS	&	1	&	1	&	1	&	0	\\
29	&	LXMA	&	Output Per Hour of All Persons$^\ddag$	&	 2009=100	&	Q	&	BLS	&	1	&	1	&	1	&	1	\\ \hline
30	&	IP	&	Industrial Production (IP) Index	&	 2012=100	&	M	&	FRB	&	1	&	1	&	1	&	0	\\
31	&	IP521	&	IP: Business Equipment	&	 2012=100	&	M	&	FRB	&	1	&	1	&	1	&	1	\\
32	&	IP511	&	IP: Durable Consumer Goods	&	 2012=100	&	M	&	FRB	&	1	&	1	&	1	&	0	\\
33	&	IP531	&	IP: Durable Materials	&	 2012=100	&	M	&	FRB	&	1	&	1	&	1	&	1	\\
34	&	IP512	&	IP: Nondurable Consumer Goods	&	 2012=100	&	M	&	FRB	&	1	&	1	&	1	&	0	\\
35	&	IP532	&	IP: nondurable Materials	&	 2012=100	&	M	&	FRB	&	1	&	1	&	1	&	0	\\\hline
\end{tabular}																			
																			
\begin{tabular}{p{\textwidth}}																			
\scriptsize $^\ast$ Consumption Expenditures \& Gross Investment\\																			
\scriptsize $^\star$ Chain-type Price Index\\																			
\scriptsize $^\dag$ Nonfarm Business Sector\\  																			
\scriptsize $^\ddag$ Manufacturing Sector  																			
\end{tabular}																			
																			
\newpage																			
\begin{tabular}{
p{.015\textwidth}
p{.1\textwidth}
p{.45\textwidth}
p{.1\textwidth}
p{.02\textwidth}
p{.04\textwidth}
P{.025\textwidth}
p{.01\textwidth}
p{.01\textwidth}
p{.01\textwidth}
}\hline\hline
\bf N	&	\bf Series ID	&	\bf Definition	&	\bf Unit	&	\bf F	&	\bf S	&	\bf SA	&	\bf T	&	\bf D	&	$\bm \xi$	\\\hline
36	&	PCU	&	CPI-U: All Items 	&	 82-84=100	&	M	&	BLS	&	1	&	2	&	0	&	0	\\
37	&	PCUSE	&	CPI-U: Energy 	&	 82-84=100	&	M	&	BLS	&	1	&	2	&	0	&	0	\\
38	&	PCUSLFE	&	CPI-U: All Items Less Food and Energy 	&	 82-84=100	&	M	&	BLS	&	1	&	2	&	0	&	0	\\
39	&	PCUFO	&	CPI-U: Food 	&	 82-84=100	&	M	&	BLS	&	1	&	2	&	0	&	0	\\ \hline
40	&	JCBM	&	PCE: Chain Price Index 	&	 2009=100	&	M	&	BEA	&	1	&	2	&	0	&	0	\\
41	&	JCEBM	&	PCE: Energy Goods \& Services--price index 	&	 2009=100	&	M	&	BEA	&	1	&	2	&	0	&	0	\\
42	&	JCNFOM	&	PCE: Food \& Beverages--price index$^\ast$	&	 2009=100	&	M	&	BEA	&	1	&	2	&	0	&	0	\\
43	&	JCXFEBM	&	PCE less Food \& Energy--price index	&	 2009=100	&	M	&	BEA	&	1	&	2	&	0	&	0	\\
44	&	JCSBM	&	PCE: Services--price index 	&	 2009=100	&	M	&	BEA	&	1	&	2	&	0	&	0	\\
45	&	JCDBM	&	PCE: Durable Goods--price index 	&	 2009=100	&	M	&	BEA	&	1	&	2	&	0	&	0	\\
46	&	JCNBM	&	PCE: Nondurable Goods--price index	&	 2009=100	&	M	&	BEA	&	1	&	2	&	0	&	0	\\\hline
47	&	PC1	&	PPI: Intermediate Demand Processed Goods 	&	 1982=100	&	M	&	BLS	&	1	&	2	&	0	&	0	\\
48	&	P05	&	PPI: Fuels and Related Products and Power 	&	 1982=100	&	M	&	BLS	&	0	&	2	&	0	&	0	\\
49	&	SP3000	&	PPI: Finished Goods 	&	1982=100	&	M	&	BLS	&	1	&	2	&	0	&	0	\\
50	&	PIN	&	PPI: Industrial Commodities	&	1982=100	&	M	&	BLS	&	0	&	2	&	0	&	0	\\
51	&	PA	&	PPI: All Commodities 	&	1982=100	&	M	&	BLS	&	0	&	2	&	0	&	0	\\
52	&	PC1	&	PPI: Intermediate Demand Processed Goods 	&	1982=100	&	M	&	BLS	&	1	&	2	&	0	&	0	\\ \hline
53	&	FMC	&	Money Stock: Currency 	&	Bil. of \$	&	M	&	FRB	&	1	&	2	&	0	&	0	\\
54	&	FM1	&	Money Stock: M1 	&	Bil. of \$	&	M	&	FRB	&	1	&	2	&	0	&	1	\\
55	&	FM2	&	Money Stock: M2 	&	Bil. of \$	&	M	&	FRB	&	1	&	2	&	0	&	0	\\ \hline
56	&	FABWC	&	C \& I Loans in Bank Credit:$^\dag$ 	&	Bil. of \$	&	M	&	FRB	&	1	&	1	&	1	&	1	\\
57	&	FABWQ	&	Consumer Loans in Bank Credit:$^\dag$	&	Bil. of \$	&	M	&	FRB	&	1	&	1	&	1	&	1	\\
58	&	FAB	&	Bank Credit:$^\dag$ 	&	Bil. of \$	&	M	&	FRB	&	1	&	1	&	1	&	1	\\
59	&	FABW	&	Loans \& Leases in Bank Credit:$^\dag$ 	&	Bil. of \$	&	M	&	FRB	&	1	&	1	&	1	&	1	\\
60	&	FABYO	&	Other Securities in Bank Credit:$^\dag$ 	&	Bil. of \$	&	M	&	FRB	&	1	&	1	&	1	&	1	\\
61	&	FABWR	&	Real Estate Loans in Bank Credit:$^\dag$ 	&	Bil. of \$	&	M	&	FRB	&	1	&	1	&	1	&	0	\\
62	&	FOT	&	Consumer Credit Outstanding 	&	Bil. of \$	&	M	&	FRB	&	1	&	1	&	1	&	0	\\\hline
63	&	HSTMW	&	Housing Starts: Midwest	&	1000--U	&	M	&	CB	&	1	&	1	&	0	&	0	\\
64	&	HSTNE	&	Housing Starts: Northeast	&	1000--U	&	M	&	CB	&	1	&	1	&	0	&	0	\\
65	&	HSTS	&	Housing Starts: South	&	1000--U	&	M	&	CB	&	1	&	1	&	0	&	0	\\
66	&	HSTGW	&	Housing Starts: West 	&	1000--U	&	M	&	CB	&	1	&	1	&	0	&	0	\\
67	&	HPT	&	Building Permit\scriptsize $^\star$	&	1000--U	&	M	&	CB	&	1	&	1	&	0	&	0	\\\hline
68	&	FBPR	&	Bank Prime Loan Rate 	&	Percent	&	M	&	FRB	&	0	&	0	&	0	&	0	\\
69	&	FFED	&	Federal Funds [effective] Rate 	&	Percent	&	M	&	FRB	&	0	&	0	&	0	&	0	\\
70	&	FCM1	&	1-Year Treasury Bill Yield$^\ddag$ 	&	Percent	&	M	&	FRB	&	0	&	0	&	0	&	0	\\
71	&	FCM10	&	10-Year Treasury Note Yield$^\ddag$ 	&	Percent	&	M	&	FRB	&	0	&	0	&	0	&	0	\\\hline
\end{tabular}																			
\begin{tabular}{p{\textwidth}}																			
\scriptsize $^\ast$ Purchased for Off-Premises Consumption\\																			
\scriptsize $^\dag$  All Commercial Banks\\																			
\scriptsize $^\star$ New Private Housing Units Authorized by 																			
\scriptsize $^\ddag$ at Constant Maturity																			
\end{tabular}																			
																			
\newpage																			

\begin{tabular}{
p{.015\textwidth}
p{.1\textwidth}
p{.45\textwidth}
p{.1\textwidth}
p{.02\textwidth}
p{.04\textwidth}
P{.025\textwidth}
p{.01\textwidth}
p{.01\textwidth}
p{.01\textwidth}
}\hline\hline
\bf N	&	\bf Series ID	&	\bf Definition	&	\bf Unit	&	\bf F	&	\bf S	&	\bf SA	&	\bf T	&	\bf D	&	$\bm \xi$	\\\hline
72	&	LP	&	Civilian Participation Rate: 16 yr +	&	Percent	&	M	&	BLS	&	0	&	0	&	0	&	1	\\
73	&	LQ	&	Civilian Employment/Population Ratio: 16 yr +	&	Percent	&	M	&	BLS	&	0	&	0	&	0	&	1	\\
74	&	LE	&	Civilian Employment: Sixteen Years \& Over 	&	1000--P	&	M	&	BLS	&	0	&	1	&	1	&	0	\\
75	&	LR	&	Civilian Unemployment Rate: 16 yr + 	&	Percent	&	M	&	BLS	&	0	&	0	&	0	&	0	\\
76	&	LU0	&	Civilians Unemployed for Less Than 5 Weeks	&	1000--P	&	M	&	BLS	&	0	&	1	&	0	&	0	\\
77	&	LU5	&	Civilians Unemployed for 5-14 Weeks 	&	1000--P	&	M	&	BLS	&	0	&	1	&	0	&	1	\\
78	&	LU15	&	Civilians Unemployed for 15-26 Weeks	&	1000--P	&	M	&	BLS	&	0	&	1	&	0	&	1	\\
79	&	LUT27	&	Civilians Unemployed for 27 Weeks and Over 	&	1000--P	&	M	&	BLS	&	0	&	1	&	0	&	1	\\\hline
80	&	LUAD	&	Average [Mean] Duration of Unemployment	&	Weeks	&	M	&	BLS	&	0	&	1	&	0	&	0	\\
81	&	LANAGRA	&	All Employees: Total Nonfarm 	&	1000--P	&	M	&	BLS	&	0	&	1	&	1	&	1	\\
82	&	LAPRIVA	&	All Employees: Total Private Industries 	&	1000--P	&	M	&	BLS	&	0	&	1	&	1	&	0	\\
83	&	LANTRMA	&	All Employees: Mining and Logging 	&	1000--P	&	M	&	BLS	&	0	&	1	&	1	&	1	\\
84	&	LACONSA	&	All Employees: Construction 	&	1000--P	&	M	&	BLS	&	0	&	1	&	1	&	1	\\
85	&	LAMANUA	&	All Employees: Manufacturing 	&	1000--P	&	M	&	BLS	&	0	&	1	&	1	&	0	\\
86	&	LATTULA	&	All Employees: Trade, Transportation \& Utilities 	&	1000--P	&	M	&	BLS	&	0	&	1	&	1	&	1	\\
87	&	LAINFOA	&	All Employees: Information Services 	&	1000--P	&	M	&	BLS	&	0	&	1	&	1	&	1	\\
88	&	LAFIREA	&	All Employees: Financial Activities 	&	1000--P	&	M	&	BLS	&	0	&	1	&	1	&	1	\\
89	&	LAPBSVA	&	All Employees: Professional \& Business Services 	&	1000--P	&	M	&	BLS	&	0	&	1	&	1	&	1	\\
90	&	LAEDUHA	&	All Employees: Education \& Health Services 	&	1000--P	&	M	&	BLS	&	0	&	1	&	1	&	1	\\
91	&	LALEIHA	&	All Employees: Leisure \& Hospitality 	&	1000--P	&	M	&	BLS	&	0	&	1	&	1	&	1	\\
92	&	LASRVOA	&	All Employees: Other Services 	&	1000--P	&	M	&	BLS	&	0	&	1	&	1	&	1	\\
93	&	LAGOVTA	&	All Employees: Government 	&	1000--P	&	M	&	BLS	&	0	&	1	&	1	&	0	\\
94	&	LAFGOVA	&	All Employees: Federal Government 	&	1000--P	&	M	&	BLS	&	0	&	1	&	1	&	1	\\
95	&	LASGOVA	&	All Employees: State Government 	&	1000--P	&	M	&	BLS	&	0	&	1	&	1	&	0	\\
96	&	LALGOVA	&	All Employees: Local Government 	&	1000--P	&	M	&	BLS	&	0	&	1	&	1	&	0	\\\hline
97	&	PETEXA	&	West Texas Intermediate Spot Price FOB$^\ast$ 	&	\$--B	&	M	&	EIA	&	0	&	2	&	0	&	0	\\\hline
98	&	NAPMNI	&	ISM Mfg: New Orders Index 	&	Index	&	M	&	ISM	&	1	&	0	&	0	&	1	\\
99	&	NAPMOI	&	ISM Mfg: Production Index 	&	Index	&	M	&	ISM	&	1	&	0	&	0	&	1	\\
100	&	NAPMEI	&	ISM Mfg: Employment Index 	&	Index	&	M	&	ISM	&	1	&	0	&	0	&	1	\\
101	&	NAPMVDI	&	ISM Mfg: Supplier Deliveries Index 	&	Index	&	M	&	ISM	&	1	&	0	&	0	&	0	\\
102	&	NAPMII	&	ISM Mfg: Inventories Index 	&	Index	&	M	&	ISM	&	1	&	0	&	0	&	0	\\\hline
103	&	SP500	&	Standard \& Poor's 500 Stock Price Index 	&	41-43=10	&	D	&	WSJ	&	0	&	1	&	1	&	0	\\\hline
\end{tabular}																			

\begin{tabular}{p{\textwidth}}
\scriptsize $^\ast$ Cushing, Oklahoma \\
\end{tabular}	

\bigskip\bigskip
\begin{tabular}{
p{.125\textwidth}															
p{.4\textwidth}															
p{.1\textwidth}															
p{.02\textwidth}															
p{.1\textwidth}															
p{.02\textwidth}															
p{.01\textwidth}		
}\hline\hline
\bf Series ID	&	\bf Definition	&	\bf Unit	&	\bf F	&	\bf Source	\\\hline
GDPPOTHQ	&	Real Potential Gross Domestic Product &	BoC 2009\$	&	Q	&	CBO	\\
NAIRUQ	&	Natural Rate of Unemployment	&	percent &	Q	&	CBO	\\
GDPPLUS	&	US GDPplus 	&	percent	&	Q	&	FRBPHIL	\\\hline
\end{tabular}

\end{document}